\documentclass[floats,floatfix,amssymb,prd,superscriptaddress,nofootinbib,twocolumn,aps]{revtex4-1}
\usepackage{graphicx,amssymb,amsmath,amsthm,amsfonts,epsfig,epsf}
\usepackage{amsmath,amsfonts}
\usepackage{bm} 
\usepackage[caption=false]{subfig}
\usepackage[usenames]{color}
\usepackage{multirow}
\usepackage[colorlinks=true,urlcolor=blue]{hyperref}%
\newcommand{\C}[1]{\mathcal{#1}}
\begin{document}

\title{Extreme mass ratio inspirals on the equatorial plane in the adiabatic order} 
 
\author{Ryuichi Fujita}
\affiliation{Center for Gravitational Physics, Yukawa Institute for Theoretical
  Physics, Kyoto University, Kyoto 606-8502, Japan~}
\affiliation{{Institute of Liberal Arts, Otemon Gakuin University, Osaka 567-8502, Japan~}}
\author{Masaru Shibata} 
\affiliation{Max Planck Institute for
  Gravitational Physics (Albert Einstein Institute), Am Mühlenberg 1,
  Potsdam-Golm 14476, Germany}
\affiliation{Center for Gravitational Physics, Yukawa Institute for Theoretical
  Physics, Kyoto University, Kyoto 606-8502, Japan~}

\date{\today} 

\begin{abstract}
We compute gravitational waves from inspiraling stellar-mass
compact objects on the equatorial plane of a massive spinning black
hole (BH).  Our inspiral orbits are computed by taking into account
the adiabatic change of orbital parameters due to gravitational
radiation in the lowest order in mass ratio.  We employ an
interpolation method to compute the adiabatic change at arbitrary
points inside the region of orbital parameter space computed in
advance.  Using the obtained inspiral orbits and associated
gravitational waves, we compute power spectra of gravitational waves
and the signal-to-noise ratio (SNR) for several values of the BH spin,
the masses of the binary, and the initial orbital eccentricity during
a hypothetical three-year Laser Interferometer Space Antenna observation before final plunge. 
We find that (i) the SNR increases as the BH spin and the mass of the compact object
increase for the BH mass $M \agt 10^6M_\odot$, (ii) the SNR has a
maximum for $M \approx 10^6M_\odot$, 
and (iii) the SNR increases as the initial eccentricity increases 
for $M=10^6M_\odot$. 
We also show that incorporating the contribution from the higher
multipole modes of gravitational waves is crucial for enhancing the
detection rate.
\end{abstract}


\maketitle
\section{Introduction}
The inspirals of stellar-mass compact objects of mass $\mu\sim
1-100M_\odot$ into supermassive black holes (SMBHs) of mass $M\sim
10^5-10^7M_\odot$ are among the key sources for the future space-based
gravitational-wave detector Laser Interferometer Space Antenna (LISA)~\cite{Audley:2017drz} (see, e.g.,
  Refs.~\cite{Luo:2015ght,Guo:2018npi} for other future space-based
  detectors in the LISA band).  Such extreme-mass-ratio inspirals
(EMRIs) are expected to have typically $\sim 10^5$ orbital cycles
($\sim 10^6$\,rad in gravitational-wave phase) during a few years of 
observation by LISA.  The observation of gravitational waves from
EMRIs will provide an opportunity of precision probes of general
relativity and information in the vicinity of SMBHs (see, e.g.,
Refs.~\cite{Babak:2017tow,Berry:2019wgg}). However, for these research
purposes, one has to prepare accurate models of gravitational
waveforms suitable for the data analysis of gravitational waves from
EMRIs.

Since the mass ratio $\eta\equiv \mu/M$ is $\lesssim 10^{-3}$, EMRIs
can be modeled by using black hole (BH) perturbation theory (see,
e.g., Refs.~\cite{Mino:1997bx,Sasaki:2003xr}).  In the limit of the
test mass, $\eta\rightarrow 0$, the compact object follows timelike
geodesic orbits in background Kerr spacetime.  At higher order in the
mass ratio, however, the orbit deviates slightly from geodesic orbits
due to the interaction with its own gravitational field, gravitational
self-force (GSF) (see, e.g.,
Refs.~\cite{Mino:1996nk,Quinn:1996am,Poisson:2011nh,Barack:2009ux,Barack:2018yvs}
and references therein). 
Using the two-time-scale expansion method in
Ref.~\cite{Hinderer:2008dm}, the orbital phase can be expanded with
respect to $\eta$ as
\begin{align}
\Phi =\frac{1}{\eta}(\Phi^{(0)}+\eta \Phi^{(1)}+{\cal O}(\eta^2)), 
\end{align}
where $\Phi^{(0)}$ and $\Phi^{(1)}$ are quantities of order unity, and
resonances~\cite{Flanagan:2010cd} of ${\cal O}(\eta^{1/2})$ are
neglected.  $\Phi^{(0)}/\eta$ denotes the orbital phase determined by the
time-averaged dissipative part of the first-order GSF, that
corresponds to the adiabatic change of the constants of motion of the
geodesics. $\Phi^{(1)}$ denotes the remaining parts of the self-force.
$\Phi^{(0)}$ must be computed much more precisely than any others
because $\Phi^{(0)}/\eta$ is the dominant part of the orbital phase.

In order to determine $\Phi^{(0)}$, one has to compute orbital
inspirals by incorporating the adiabatic change of the constants of
motion, $dI^i/dt$, due to the gravitational-wave emission, where $I^i$
denotes three constants of motion (see Sec.~\ref{sec:formulation} for
details). Many numerical results of $dI^i/dt$ have been derived
for spherical
orbits~\cite{Shibata:1993yf,Hughes:1999bq,Hughes:2001jr},
eccentric-equatorial
orbits~\cite{Tanaka:1993pu,Cutler:1994pb,Glampedakis:2002ya}, and
eccentric-inclined
orbits~\cite{Drasco:2005kz,Drasco:2007gn,Fujita:2009us}.  The orbital
phase $\Phi^{(0)}/\eta$ is also computed for spherical orbits in
Ref.~\cite{Hughes:2001jr}, but the accuracy in $\Phi^{(0)}/\eta$ is worse
than 1\,rad, which is the minimum accuracy required for the
gravitational-wave modeling suitable for the data analysis of EMRIs. 

In this paper, we derive adiabatic orbital inspirals on the equatorial
plane of the Kerr BH focusing on the time-averaged dissipative part of
the first-order GSF (i.e., the lowest-order part in $\eta$).  We
compute the adiabatic evolution of the inspiral orbits using the
osculating geodesics method~\cite{Pound:2007th,Gair:2010iv}, in which
a sequence of geodesic orbits is assumed to be tangent to the true
inspiral orbit at each moment (see
Refs.~\cite{Warburton:2011fk,Osburn:2015duj} for inspiral orbits
including the conservative part of the first-order GSF in
Schwarzschild spacetime).  Our adiabatic inspiral orbits are computed
by taking into account the adiabatic change of the constants of motion
due to the emission of gravitational waves at each geodesic orbit.

The issue in this line of the study is that 
numerical computation for the adiabatic change of $I^i$ for each of
  $\sim 10^5$ inspiral orbits is extremely costly, even if we restrict
  our attention to equatorial inspirals. Thus, we employ the following
  alternative strategy.  First, we compute the adiabatic change of
  $I^i$ for a number of data points in the semilatus rectum, $p$, and
  the orbital eccentricity, $e$.  Then, we use an interpolation
method to obtain $dI^i/dt$ at arbitrary points in the phase space
  of $(p,e)$, for which $dI^i/dt$ are computed in advance. Using
this strategy, we obtain inspiral orbits and associated gravitational
waves with an inexpensive computational cost.

This paper is organized as follows.  In Sec.~\ref{sec:formulation}, we
review formulations necessary to compute the geodesic motion in Kerr
spacetime and the adiabatic change of the constants of motion due to
the emission of gravitational waves.  In Sec.~\ref{sec:method}, we
first summarize our approach to obtain the adiabatic inspiral orbits
using the osculating geodesic method with interpolated fluxes. Then,
we describe the accuracy for the adiabatic change of
constants of motion numerically derived. The issues to improve the
accuracy are also discussed.  Several representative inspiral
  orbits and associated gravitational waves are presented in
  Sec.~\ref{sec:results}, paying particular attention to
  gravitational-wave spectra. We show the dependence of the
  gravitational-wave spectra on the mass and spin of SMBHs and orbital
  eccentricity of compact objects.  We summarize this paper in
Sec.~\ref{sec:summary}.  Throughout this paper we use the geometrical
units with $c=G=1$ where $c$ and $G$ are the speed of light and
gravitational constant, respectively.

\section{Formulation}
\label{sec:formulation} 
The purpose of this paper is to explore inspiral orbits of a
stellar-mass compact object of mass $\mu$ around a Kerr BH of mass $M\gg \mu$.  
Specifically, we derive gravitational waves emitted by the
orbiting object using the BH perturbation theory and consider the
adiabatic evolution of the orbit due to the gravitational-wave
emission.  We use the methods presented in
Refs.~\cite{Fujita:2004rb,Fujita:2009uz,Fujita:2009us}, based on the
formalism developed in Refs.~\cite{MST,MST_RW,MSTR,Sasaki:2003xr}, to
numerically compute gravitational-wave fluxes by a stellar-mass
object with bound orbits around a Kerr BH of spin $a$ for large sets
of orbital parameters.  Then, one can obtain inspiral orbits by
incorporating the adiabatic change of orbital parameters due to the
gravitational-wave emission.  In this paper, we focus only on the
inspirals on the equatorial plane of the BH as a first step.
\subsection{Bound geodesics}
\label{sec:geodesics} 
First, we summarize the method to determine the generic geodesic
  orbit in Kerr spacetime.  Using Boyer-Lindquist coordinates for
the Kerr solution, $(t, r, \theta,\phi)$, and Mino
time~\cite{Mino:2003yg} $\lambda=\int d\tau/(r^2+a^2\cos^2\theta)$,
the geodesic equations are written as

\begin{align}
\left(\frac{dr}{d\lambda}\right)^2 &= R(r),
\label{eq:geodesic_r}\\
\left(\frac{d\cos\theta}{d\lambda}\right)^2 &=\Theta(\cos\theta), 
\label{eq:geodesic_theta}
\end{align}
\begin{align}
\frac{d\phi}{d\lambda} &= \Phi_r(r)+\Phi_\theta(\theta), 
\quad \quad \label{eq:geodesic_phi}\\
\frac{dt}{d\lambda} &= T_r(r)+T_\theta(\theta), 
\label{eq:geodesic_t}
\end{align}
where 
\begin{align}
R(r)&=\left[P(r)\right]^2 - \Delta[r^2+(\C{L}_{z}-a \C{E})^2+\C{C}],\\
P(r)&=(r^2+a^2)\C{E}-a \C{L}_{z},\\
\Theta(\cos\theta)&
= \C{C} - (\C{C}+a^2(1-\C{E}^2)+\C{L}_{z}^2)\cos^2\theta\cr
                  &\quad + a^2(1-\C{E}^2)\cos^4\theta, \\
\Phi_r(r)&=\frac{a}{\Delta} P(r), \\
\Phi_\theta(\theta) &= \frac{\C{L}_{z}}{1-\cos^{2}\theta} - a\C{E}, \\
T_r(r) &= \frac{r^{2}+a^{2}}{\Delta} P(r),\\
T_\theta(\theta) &=- a^2 \C{E}(1-\cos^{2}\theta) + a\C{L}_{z}, 
\end{align}
and $\Delta=r^2-2Mr+a^2$.  $\C{E}$, $\C{L}_{z}$, and $\C{C}$ are
constants that denote the specific energy, the $z$ component of the
specific angular momentum, and the Carter constant of a stellar-mass
compact object, respectively.  The geodesic orbits in Kerr
spacetime can be characterized by these three constants of motion
$(\C{E},\C{L}_{z},\C{C}$).  In the following, we often refer to these
three constants in a vector form as $I^i$.

One can also use another set of three parameters, the semilatus
rectum $p$, the orbital eccentricity $e$, and the inclination angle
$\theta_{\textrm{inc}}$, to characterize the geodesics for bound
orbits.  These parameters are related to turning points of the radial
motion, $r_\textrm{max}$ and $r_\textrm{min}$, and the polar motion,
$\theta_\textrm{min}$, via
\begin{align}
p&=\frac{2 r_\textrm{max}r_\textrm{min}}{r_\textrm{max}+r_\textrm{min}},\quad 
e=\frac{r_\textrm{max}-r_\textrm{min}}{r_\textrm{max}+r_\textrm{min}},
\end{align}
and $\theta_{\textrm{inc}}=\pi/2-({\textrm{sgn}} \C{L}_{z})
  \theta_\textrm{min}$.  Note that $r_\textrm{min}$ is written as
$p/(1+e)$ and the minimum value of $r_{\textrm{min}}$ is written as
$2M-a+2M^{1/2}\sqrt{M-a}$, which is realized for the marginally bound
orbit~\cite{Bardeen:1972fi}.  $(\C{E},\C{L}_{z},\C{C}$) are written as
functions of $(p,e,\theta_{\textrm{inc}})$ using the method in
Ref.~\cite{Schmidt:2002qk}. In the following, we refer to
$(p,e,\theta_{\textrm{inc}})$ as $J^i$.

Using $J^i$, the radial and polar motions can be parametrized by 
\begin{align}
r(\lambda)&=\frac{p}{1+e\cos[\chi(\lambda)-\chi_0]},\cr
\cos\theta(\lambda)&=\cos\theta_{\textrm{inc}} \cos[\psi(\lambda)-\psi_0],
\end{align}
where $\chi(\lambda)$ and $\psi(\lambda)$ are monotonic parameters that run 
from $0$ to $2\pi$ over one radial and polar cycle, respectively. 
$\chi_0$ and $\psi_0$ take 
the values of $\chi$ and $\psi$ at $r=r_\textrm{min}$ and $\theta=\theta_\textrm{min}$, 
respectively. 
In the osculating geodesic method, the inspiral orbit under the GSF 
is described by the evolution of $p$, $e$, $\theta_{\textrm{inc}}$, 
$\chi_0$, and $\psi_0$. 
The principal orbital elements $p$, $e$, and $\theta_{\textrm{inc}}$ evolve due to 
the dissipative part of the GSF, while 
the positional orbital elements $\chi_0$ and $\psi_0$ evolve due to 
the conservative part of the GSF. 

Since the equations of radial and polar motion are decoupled in
Eqs.~(\ref{eq:geodesic_r}) and (\ref{eq:geodesic_theta}), for the bound
orbits, $r(\lambda)$ and $\theta(\lambda)$ become periodic functions
that are independent of each other.  The fundamental periods for the
radial and polar motion, $\Lambda_r$ and $\Lambda_\theta$, are
calculated by
\begin{align}
\Lambda_r=2\int_{\textrm{r}_{\textrm{min}}}^{\textrm{r}_{\textrm{max}}}\frac{dr}{\sqrt{R(r)}},\quad
\Lambda_\theta=4\int_{0}^{\cos\theta_\textrm{min}}
\frac{d\cos\theta}{\sqrt{\Theta(\cos\theta)}},
\end{align}
and thus, the angular frequencies of the radial and polar motion become
\begin{equation}
\Upsilon_r=\frac{2\pi}{\Lambda_r},\quad\quad
\Upsilon_\theta=\frac{2\pi}{\Lambda_\theta}. 
\end{equation}
$\Upsilon_r$ and $\Upsilon_\theta$ can be expressed in 
complete elliptic integrals of the first kind; 
see, e.g., Refs.~\cite{Fujita:2009us,Fujita:2009bp}. 

We define the angle variables as $w_r=\Upsilon_r\lambda$ and $w_\theta=\Upsilon_\theta\lambda$. 
Then, any functions that depend only on $r$ or $\theta$ become
periodic with respect to $w_r$ or $w_\theta$, respectively,
with the period of $2\pi$. 

To solve Eqs.~(\ref{eq:geodesic_phi}) and (\ref{eq:geodesic_t}), 
we expand their right-hand sides into Fourier series as~\cite{Drasco:2003ky}
\begin{align}
\frac{dt}{d\lambda} &=\sum_{k,n}T_{k,n}
e^{-ik\Upsilon_r\lambda}e^{-in\Upsilon_\theta\lambda}, \\
\frac{d\phi}{d\lambda} &=\sum_{k,n}\Phi_{k,n}
e^{-ik\Upsilon_r\lambda}e^{-in\Upsilon_\theta\lambda},
\end{align}
where
\begin{align}
T_{k,n}&=\int_0^{2\pi} \frac{dw_r}{2\pi} \int_0^{2\pi}\frac{dw_\theta}{2\pi}
(T_r(r)+T_\theta(\theta)) 
e^{i k w_r}e^{i n w_\theta}, \\
\Phi_{k,n}&=\int_0^{2\pi} \frac{dw_r}{2\pi} \int_0^{2\pi}\frac{dw_\theta}{2\pi} 
(\Phi_r(r)+\Phi_\theta(\theta)) 
e^{i k w_r}e^{i n w_\theta}. 
\end{align}
Since $T_{k,n}=0$ and $\Phi_{k,n}=0$ for $k\neq 0$ and $n\neq 0$,
we have
\begin{align}
\frac{dt}{d\lambda} &=
\Gamma+\sum_{k\neq 0}T_{k,0}e^{-ikw_r}+\sum_{n\neq 0}T_{0,n}e^{-inw_\theta}, 
\label{eq:dtdlam}\\
\frac{d\phi}{d\lambda} &=
\Upsilon_\phi+\sum_{k\neq 0}\Phi_{k,0}e^{-ikw_r}+\sum_{n\neq 0}\Phi_{0,n}e^{-inw_\theta}, 
\label{eq:dphidlam}
\end{align}
where
\begin{align}
\Gamma&\equiv T_{00} =\Upsilon_{t^{(r)}}+\Upsilon_{t^{(\theta)}},\cr 
\Upsilon_\phi&\equiv \Phi_{00} =\Upsilon_{\phi^{(r)}}+\Upsilon_{\phi^{(\theta)}},\cr 
\Upsilon_{t^{(r)}}&=\frac{1}{2\pi}\int_0^{2\pi}dw_r T_r,\,\,\,
\Upsilon_{t^{(\theta)}}=\frac{1}{2\pi}\int_0^{2\pi}dw_\theta T_\theta,\cr
\Upsilon_{\phi^{(r)}}&=\frac{1}{2\pi}\int_0^{2\pi}dw_r \Phi_r,\,\,\,
\Upsilon_{\phi^{(\theta)}}=\frac{1}{2\pi}\int_0^{2\pi}dw_\theta \Phi_\theta.
\end{align}
Then, we obtain the functions $t(\lambda)$ and $\phi(\lambda)$ 
by integrating Eqs.~(\ref{eq:dtdlam}) and (\ref{eq:dphidlam}) in the following forms: 
\begin{align}
t(\lambda)&=\Gamma\lambda
+\sum_{k\neq 0}\frac{iT_{k,0}}{k\Upsilon_r}e^{-ikw_r}
+\sum_{n\neq 0}\frac{iT_{0,n}}{n\Upsilon_\theta}e^{-inw_\theta}, 
\\
\phi(\lambda)&=\Upsilon_\phi\lambda
+\sum_{k\neq 0}\frac{i\Phi_{k,0}}{k\Upsilon_r}e^{-ikw_r}
+\sum_{n\neq 0}\frac{i\Phi_{0,n}}{n\Upsilon_\theta}e^{-inw_\theta}.
\end{align}
Here, the two variables, $\Gamma$ and $\Upsilon_\phi$, denote 
the average rates of change of $t$ and $\phi$ as functions of $\lambda$, 
respectively. 

\subsection{Secular evolution of orbital parameters}
\label{sec:dIdt2dJdt} 
In the Teukolsky formalism~\cite{Teukolsky:1973ha}, 
the gravitational perturbation on Kerr spacetime is described in terms of 
the Newman-Penrose variables, $\Psi_0$ and $\Psi_4$, which satisfy a master equation.
The Weyl scalar $\Psi_4$
is related to gravitational waves at infinity as
\begin{align}
\Psi_4\rightarrow\frac{1}{2}(\ddot{h}_{+}-i\,\ddot{h}_{\times }).
\end{align}
The master equation for $\Psi_4$
can be separated into radial and angular parts if we
expand $\Psi_4$ in harmonic modes as
\begin{align}
\rho^{-4} \Psi_4=\displaystyle \sum_{\ell m}\int_{-\infty}^{\infty} d\omega 
e^{-i\omega t+i m \varphi} \ _{-2}S_{\ell m}^{a\omega}(\theta)
R_{\ell m\omega}(r),
\end{align}
where $\rho=(r-i a \cos\theta)^{-1}$, 
and $_{-2}S_{\ell m}^{a\omega}(\theta)$ 
is the spin-weighted spheroidal harmonics with spin $s=-2$. 
The radial function $R_{\ell m\omega}(r)$ satisfies 
the so-called Teukolsky equation, 
\begin{align}
\Delta^2\frac{d}{dr}\left(\frac{1}{\Delta}\frac{dR_{\ell m\omega}}{dr}
\right)-V(r) R_{\ell m\omega}=T_{\ell m\omega},
\label{eq:radial-teukolsky}
\end{align}
where the potential term $V(r)$ is 
\begin{align}
V(r) = -\frac{K^2 + 4i(r-M)K}{\Delta} + 8i\omega r + \lambda_{\ell m\omega},
\end{align}
with $K=(r^2+a^2)\omega-ma$ and $\lambda_{\ell m\omega}$ the
eigenvalue of $_{-2}S_{\ell m}^{a\omega}(\theta)$.

The asymptotic behavior of the solution at the horizon and infinity is written,
respectively, as
\begin{align}
R_{\ell m \omega}(r\rightarrow r_+)
&\equiv Z^\textrm{H}_{\ell m \omega}\Delta^2e^{-iPr^*},
\label{eq:ZH}
\end{align}
and
\begin{align}
R_{\ell m \omega}(r\rightarrow \infty)
\equiv Z^\infty_{\ell m \omega}r^3e^{i\omega r^*},
\label{eq:Zinf}
\end{align}
where $r_+=M+\sqrt{M^2-a^2}$, $P=\omega - ma/2Mr_+$, and 
$r^*$ is the tortoise coordinate. 

For the bound orbits of a stellar-mass object, the amplitude of the
partial wave $Z^{\infty/\textrm{H}}_{\ell m\omega}$, defined in
Eqs.~(\ref{eq:ZH}) and (\ref{eq:Zinf}), can be expanded as
\begin{align}
Z^{\infty,\textrm{H}}_{\ell m\omega}
\equiv \sum_{kn} \tilde{Z}^{\infty,\textrm{H}}_{\ell mkn}
\delta(\omega - \omega_{mkn})\;,
\end{align}
where 
\begin{equation} \label{eq:omega_mkn}
\omega_{mkn}\equiv (m\Upsilon_\phi + k\Upsilon_\theta + n\Upsilon_r)/\Gamma\;.
\end{equation}

Using these functions, gravitational waves at infinity are expressed as
\begin{align}
 h_{+}-ih_{\times}=-\frac{2}{r}\sum_{\ell mkn}
\frac{\tilde{Z}^{\infty}_{\ell mkn}}{\omega_{mkn}^2}
\frac{{}_{-2}S^{a\omega_{mkn}}_{\ell m}(\theta)}{\sqrt{2\pi}}
e^{i\omega_{mkn}(r^*-t)+im\phi}.
\end{align}
In addition, 
the adiabatic change for $({\cal E},{\cal L}_z,{\cal C})$
due to the emission of gravitational waves are
expressed as~\cite{Mino:2003yg,Sago:2005gd,Sago:2005fn}
\begin{align}
 \left<\frac{d\C{E}}{dt}\right>
&= -\mu^2\sum_{\ell mkn}
\frac{1}{4\pi\omega^2_{mkn}}\left(
\left|\tilde{Z}^{\infty}_{\ell mkn}\right|^2
+\alpha_{\ell mkn}
\left|\tilde{Z}^{\textrm{H}}_{\ell mkn}\right|^2
\right), \label{eq:dEdt} \\
 \left<\frac{d\C{L}_{z}}{dt}\right> &= -\mu^2\sum_{\ell mkn}
\frac{m}{4\pi\omega^3_{mkn}}
\left(
\left|\tilde{Z}^{\infty}_{\ell mkn}\right|^2
+\alpha_{\ell mkn}
\left|\tilde{Z}^{\textrm{H}}_{\ell mkn}\right|^2
\right),\label{eq:dLzdt} \\
\left<\frac{d\C{C}}{dt}\right>
&= \left<\frac{d\C{Q}}{dt}\right>-2(a\C{E}-\C{L}_{z})\left(
a\left<\frac{d\C{E}}{dt}\right>-\left<\frac{d\C{L}_{z}}{dt}\right>
\right), \label{eq:dCdt} \\
\left<\frac{d\C{Q}}{dt}\right> &=
2\Upsilon_{t^{(r)}}
\left<\frac{d\C{E}}{dt}\right>
-2\Upsilon_{\phi^{(r)}}
\left<\frac{d\C{L}_{z}}{dt}\right>\cr
&+\mu^3\sum_{\ell mkn}\frac{n\Upsilon_r}
{2\pi\omega^3_{mkn}}\left(
\left|\tilde{Z}^{\infty}_{\ell mkn}\right|^2
+\alpha_{\ell mkn}
\left|\tilde{Z}^{\textrm{H}}_{\ell mkn}\right|^2
\right),\cr \label{eq:dQdt}
\end{align}
where 
\begin{align}
 \alpha_{\ell mkn} =
 \frac{256(2Mr_+)^5 P (P^2+4\epsilon^2)(P^2+16\epsilon^2)\omega_{mkn}^3}
{C^{\textrm{TS}}_{\ell mkn}},
\label{eq:TS_const}
\end{align}
$\epsilon=\sqrt{M^2-a^2}/4Mr_{+}$, and $C^{\textrm{TS}}_{\ell mkn}$ is
the Teukolsky-Starobinsky constant~\cite{Teukolsky:1974yv} (see
Ref.~\cite{Drasco:2005is} for the scalar case).  Here
$\left<\cdots\right>$ denotes the time average.  We note that
$\tilde{Z}^{\infty}_{\ell mkn}$ and $\tilde{Z}^{\textrm{H}}_{\ell
  mkn}$ in Eqs.~(\ref{eq:dEdt})--(\ref{eq:dQdt}) denote fluxes at
infinity and the horizon, respectively.

Once we obtain the adiabatic change of $I^i$, 
we can derive the adiabatic change of $J^i$ using 
\begin{align}
\left\langle\frac{dJ^i}{dt}\right\rangle = 
\left (G^{-1}\right)^{i}_{j}\left\langle\frac{dI^j}{dt}\right\rangle,
\label{eq:dJdt}
\end{align}
where $G^{i}_{j}=\partial I^i/\partial J^j$. 
In this paper, we consider the case of $\C{C}=0$, and thus, 
$d\C{C}/dt=0$ ($d\theta_{\textrm{inc}}/dt=0$).

\section{Our method to obtain inspiral orbits}
\label{sec:method} 
The purpose of this paper is to derive adiabatic inspiral orbits and
associated gravitational waves.  We ignore the change in the mass and
spin of the BH due to the absorption of gravitational waves because
they are small effects.  In order to obtain the adiabatic inspiral
orbits, we construct a sequence of the osculating
orbits~\cite{Pound:2007th,Gair:2010iv}, which are assumed to be
tangent to the true inspiral orbit at each instance.  We ignore the
evolution of the positional orbital elements, which is one of the
higher-order effects in the mass ratio (see
Refs.~\cite{Pound:2007th,Gair:2010iv} for a method to include the
evolution of the positional orbital elements).  We incorporate the
adiabatic change of the constants of motion due to the emission of
gravitational waves for each geodesic orbit.  Then, the error in our
inspiral orbit from the true inspiral orbit is of ${\cal O}(\eta)$,
which is caused by higher-order effects ignored in this paper. The
evolution of $(p, e)$ is calculated by determining $(\langle dp/dt
\rangle, \langle de/dt \rangle)$ from Eq.~(\ref{eq:dJdt}) for fixed
values of $M$ and $q(=a/M)$: dimensionless spin parameter. In the
following we refer to $q$ simply as the BH spin.

\subsection{Procedure for determining inspiral orbits}
\label{sec:inspiral}

For the numerical evolution of $p$, i.e., $p(t)$, using interpolated
gravitational-wave fluxes, we take a Euler step as $p(t+\Delta
t)=p(t)+\langle dp/dt \rangle\Delta t$, where $\Delta t$ is a time step 
and $\langle dp/dt \rangle$ is computed from Eq.~(\ref{eq:dJdt}).  
In this paper, we choose $\Delta t = p/\langle dp/dt \rangle \epsilon_t$, 
where $\epsilon_t \approx 10^{-4}$.
In order to estimate the relative error of $p(t)$, 
we compare $p(t)$ by setting $\epsilon_t=10^{-4}$ with a reference solution 
for $p(t)$ obtained by setting $\epsilon_t =10^{-6}$. 
We find that the relative error in $p(t)$ by setting $\epsilon_t=10^{-4}$ 
is about ${\cal O}(\epsilon_t)$, i.e., $10^{-4}$. 
In Sec.~\ref{sec:results}, we find that the power spectra and 
the signal-to-noise ratio (SNR)
for typical EMRIs span about a few orders of magnitude.
The error of $10^{-4}$ in
the inspiral orbits is acceptable for computing power spectra of
gravitational waves and the SNR within the 
error of $10^{-3}$, although the 
error size in the adiabatic change needs to be better than
$10^{-6}$ to suppress the error in the total orbital phase less than
1\,rad (see discussion below).  
We note that the above procedure can be
straightforwardly extended to the higher-order BH perturbation theory
in $\eta$.

\begin{figure*}[htbp] 
\centering 
\includegraphics[width=59mm]{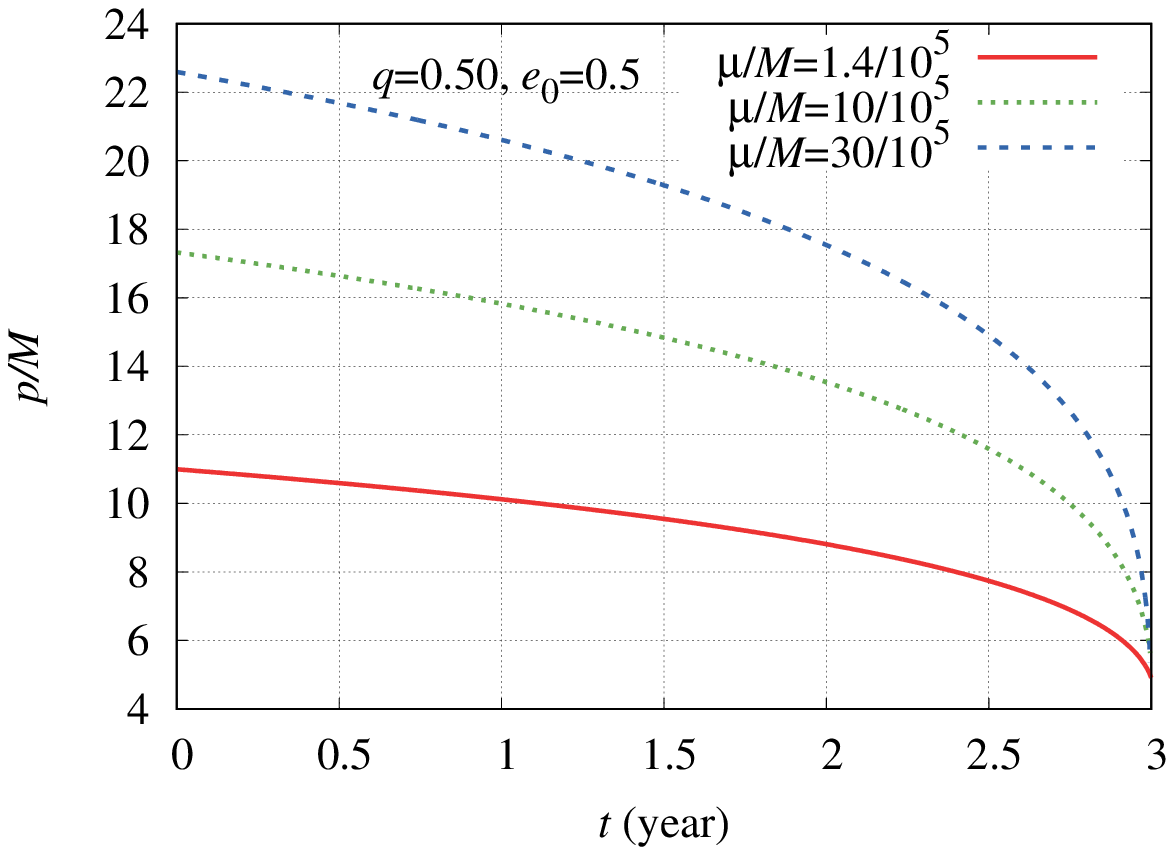}
\includegraphics[width=59mm]{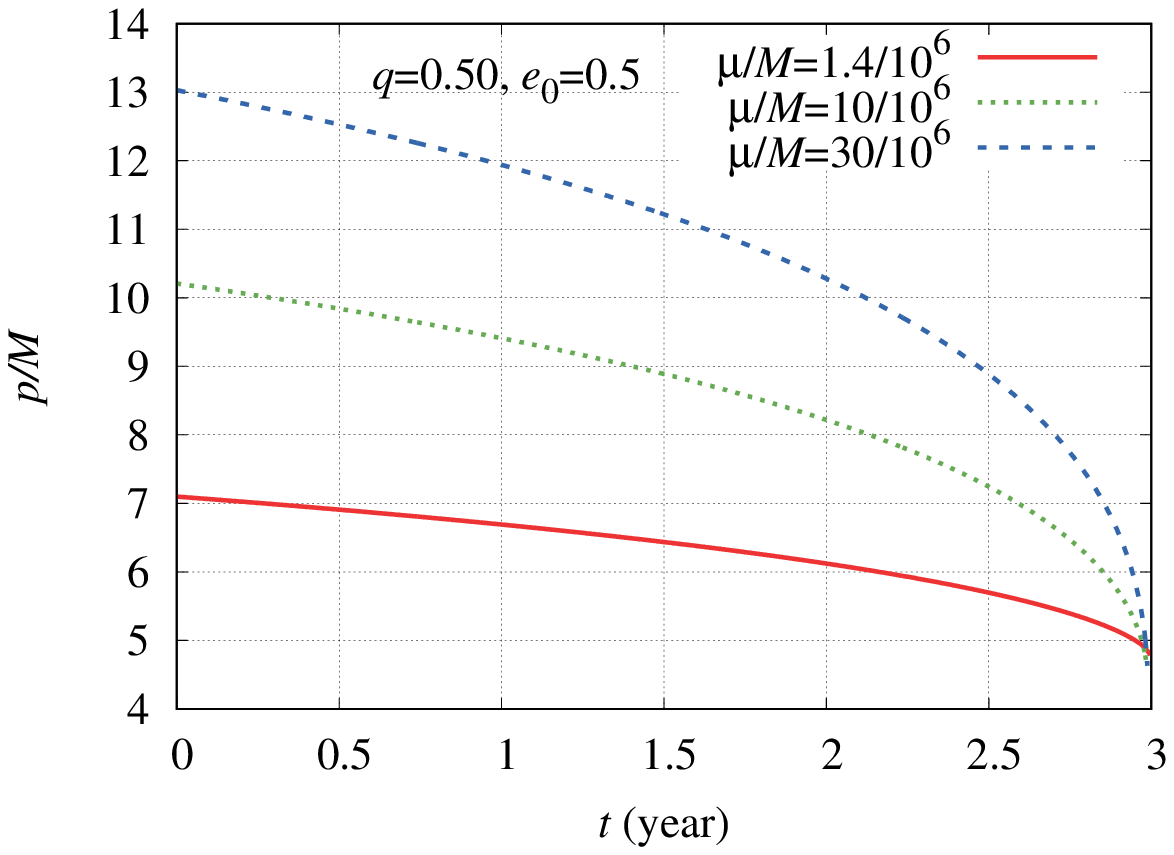}%
\includegraphics[width=59mm]{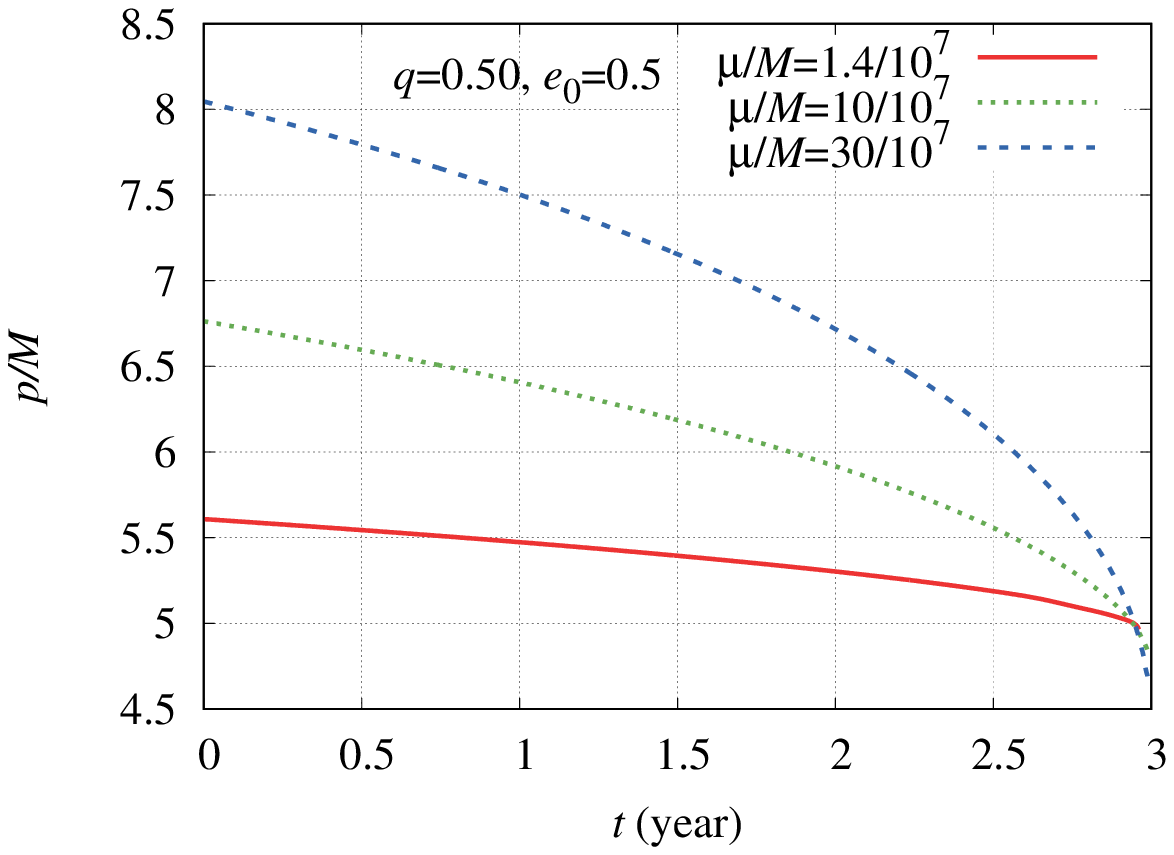}\\
\includegraphics[width=59mm]{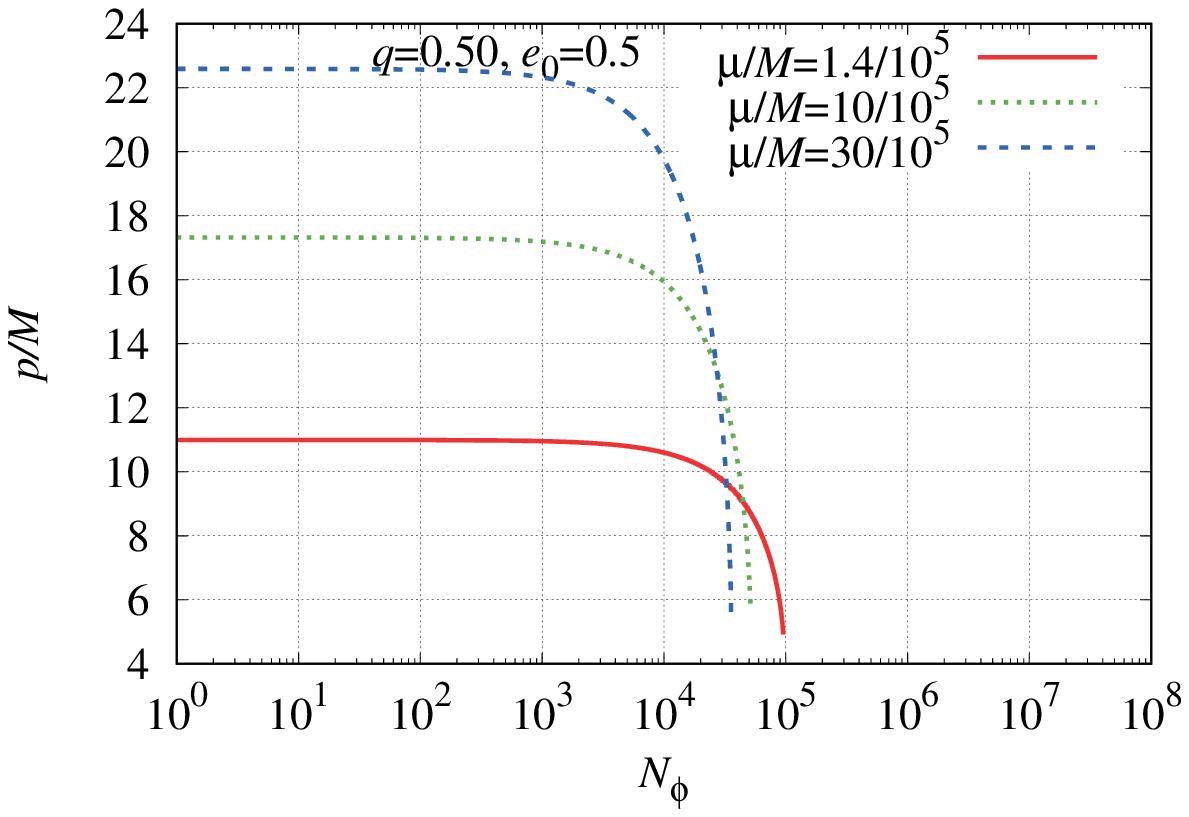}
\includegraphics[width=59mm]{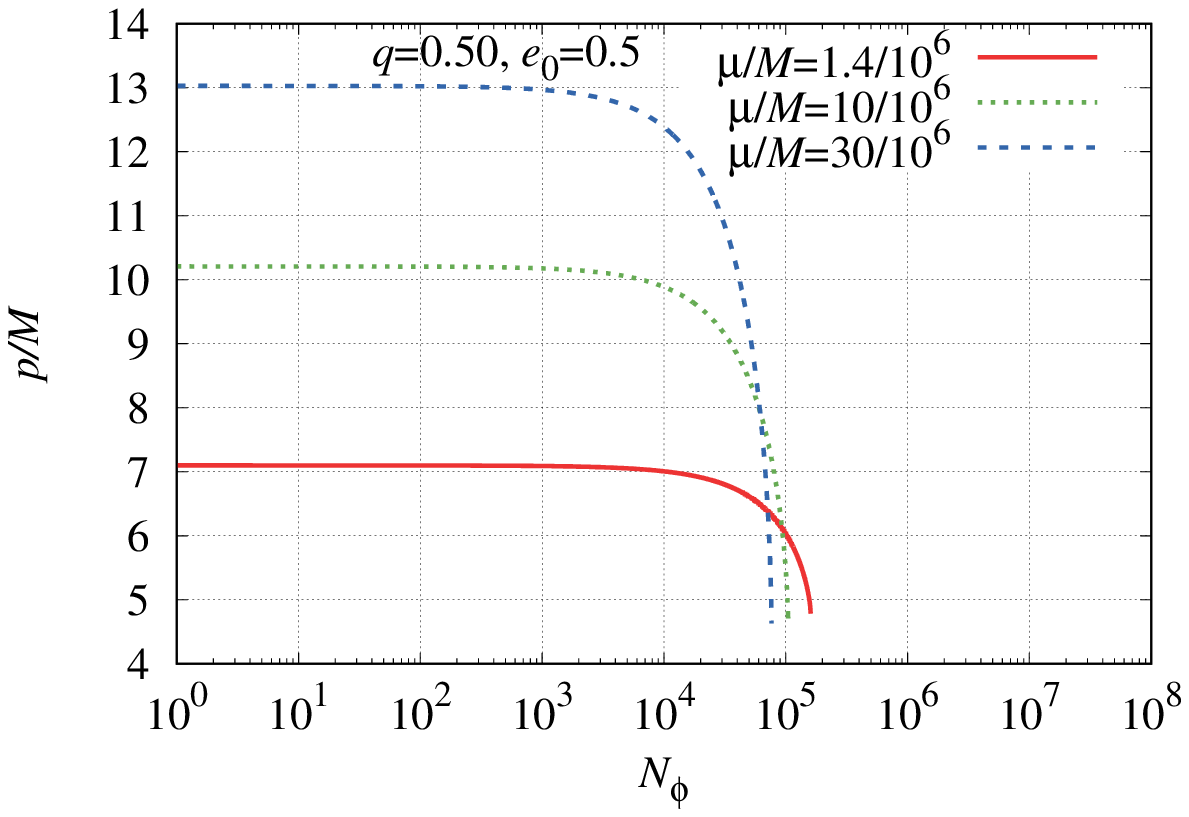}%
\includegraphics[width=59mm]{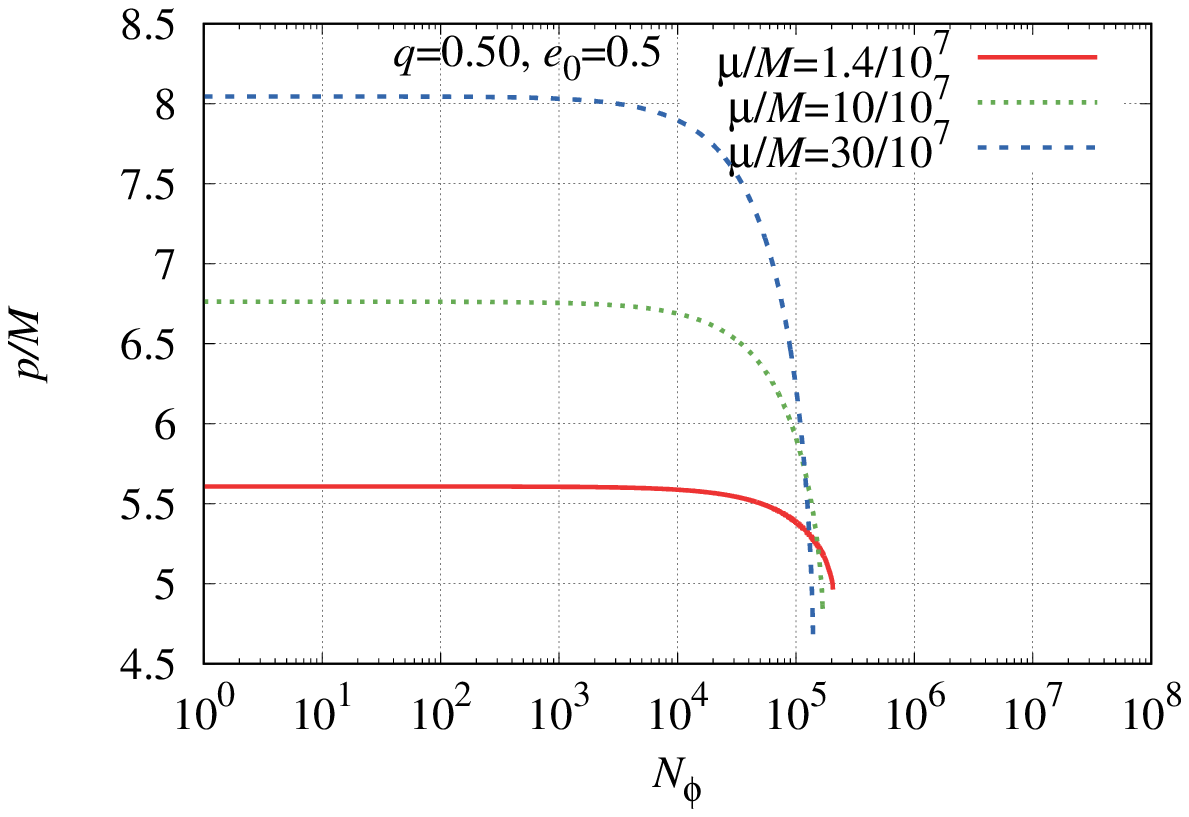}
\caption{Top panels show the evolution of the semilatus rectum $p$
  as a function of time for the last 3-year inspirals before plunge
  with $q=0.5$, $M=10^5M_\odot$ (left), $10^6M_\odot$ (middle), and
  $10^7M_\odot$ (right), and the initial orbital eccentricity
  $e_0=0.5$.  Bottom panels show $p$ as a function of orbital cycles
  for the last 3-year inspirals before plunge with the same
  parameters $(q,e_0,\mu,M)$ as in the top panels.  }
\label{fig:cycle}
\end{figure*}

It is feasible to numerically calculate the adiabatic change of the
constants of motion only for a restricted number of parameter sets of
$(p, e)$ in reasonable computational time.  Thus, we first compute
them for the restricted data points and use an interpolation method to
obtain $(\langle dp/dt \rangle, \langle de/dt \rangle)$ at arbitrary
points inside the region of the defined parameter space of $(p,e)$.  Then, we obtain
an inspiral orbit using the interpolated values of $(\langle dp/dt
\rangle, \langle de/dt \rangle)$.  We employ a fifth-order Lagrange
interpolation~\cite{Recipes}, both in $p$ and $e$, for this.  We note
that the similar method is employed, e.g., in
Ref.~\cite{Osburn:2015duj} for the local fitting of gravitational
self-force in Schwarzschild spacetime.

In this paper, we compute $(\langle dp/dt \rangle, \langle de/dt
\rangle)$ for $\approx 18000$ data points in the $(p,e)$ space for
each value of $q$.  The number of the sampling points for $p$ is $256$
for the range of $p_\textrm{ISO}\le p\le 30M$ with a log-even
spaced grid. Here, $p_\textrm{ISO}$ denotes $p$ at the innermost
stable orbit for each value of $q$.  For $e$, the sampling point is
chosen to be $e=0.005$, $0.01$, and $0.0125\le e\le 0.925$ with the
grid spacing $\Delta e=0.0125$ (the total number is 76).   The
  total computational cost for determining the gravitational-wave
  fluxes with these sampling points and $|q|=0$, $0.1$, $0.3$, $0.5$,
  $0.7$, and $0.9$ is about $20$~d using $\sim 400$~processors with $\sim
  2.6$~GHz clock speed.  

Figure~\ref{fig:cycle} illustrates the results for inspiral orbits
determined by the above procedure for $q=0.5$ and $e_0=0.5$ with
$M=(10^5, 10^6, 10^7M_\odot)$ and $\mu=(1.4, 10, 30M_\odot)$. Here,
$e_0$ denotes the initial orbital eccentricity. We plot the evolution
of $p$ and the number of orbital cycles $N_\phi$ for the last 3 years
before plunge of a stellar-mass compact object into SMBHs.  This
figure shows that the lifetime of the EMRIs becomes 3 years for the
cases that $p/M \approx 5$--20, depending on the masses of the SMBH
and compact star. It also shows that the typical total cycles of the
orbit is $N_\phi \sim 10^5$ (i.e., the typical total phase of
gravitational waves is $\sim 10^6$\,rad) for the last 3-year inspiral
orbits before plunge.  Thus, if we require that the error in a
gravitational-wave phase model is smaller than 1\,rad, the error of
the gravitational-wave fluxes has to be within $10^{-6}$. 

\subsection{Accuracy of gravitational-wave fluxes}

The adiabatic change of the orbital parameters is computed from
gravitational-wave fluxes, i.e., $\tilde{Z}^{\infty/\textrm{H}}_{\ell
  mn}$, where we omit the $k$-mode because we focus on the equatorial
orbits. For the computation of $\tilde{Z}^{\infty/\textrm{H}}_{\ell
  mn}$, one needs to integrate $_{-2}S_{\ell m}^{a\omega}(\theta)$ and
$R_{\ell m\omega}(r)$ with the source term $T_{\ell m \omega}$ along a
geodesic orbit.  We use the numerical methods developed in
Refs.~\cite{Fujita:2004rb,Fujita:2009uz} to compute $_{-2}S_{\ell
  m}^{a\omega}(\theta)$ and $R_{\ell m\omega}(r)$.  One can compute
them with the machine precision in most cases. However, in some cases,
the accuracy of $R_{\ell m\omega}(r)$ is limited by that of the
so-called renormalized angular momentum $\nu$ introduced in
Refs.~\cite{MST,MST_RW,MSTR}: it is infeasible to accurately
determine $\nu$ for large values of $M\omega$, typically $M\omega>3$
for $(\ell,m)=(2,2)$ in double precision
calculation~\cite{Fujita:2004rb,Fujita:2009uz} (see, however,
  Refs.~\cite{Shah:2013uya,Shah:2014tka,Shah:2015nva,Hopper:2015jxa}
  which use \textit{Mathematica} codes to determine $\nu$ in high
  precision, $\sim 100$ decimal places).
Since the high-frequency modes play an important role, it becomes
challenging to accurately compute gravitational-wave fluxes for
compact orbits (with small values of $r_{\textrm{min}}$), in particular
for the high BH spin of $q\agt 0.9$.

We use the trapezium rule, which has an excellent convergence property
to integrate periodic functions, to derive 
$\tilde{Z}^{\infty/\textrm{H}}_{\ell mn}$~\cite{Fujita:2009us,Hopper:2015jxa}. 
We choose the maximum number of the grid points in the trapezium rule
as $2^{14}+1$ to save computational time.  In
Ref.~\cite{Fujita:2009us}, with this number of the grid points, it is
found that one can compute $dI^i/dt$ with the accuracy of $10^{-10}$
for $q=0$, $p=10M$, and $e=0.9$.  However, the numerical accuracy is
not as good as this level for a high value of $q \agt 0.9$ as we find
in the present work.  To summarize, the numerical accuracy in
$\tilde{Z}^{\infty/\textrm{H}}_{\ell mn}$ is currently limited by that
of $\nu$ and the number of the grid points used in the trapezium rule
for a high value of $q \agt 0.9$. Improving the accuracy for this 
special case is the issue left for the future work.

The numerical accuracy in the gravitational-wave fluxes is also
limited by truncating the mode summation in Eqs.~(\ref{eq:dEdt}) and
(\ref{eq:dLzdt}). In the present work, the mode summation in the
fluxes, Eqs.~(\ref{eq:dEdt}) and (\ref{eq:dLzdt}), is performed until 
the error becomes smaller than $10^{-6}$ at least for $p\geq 6M$
and $e\le 0.8$.  We choose this error size because the total cycle of
gravitational waves during a few years LISA observation is of the
order of $10^{5}$ as already illustrated in Fig.~\ref{fig:cycle}.

The mode summation in Eqs.~(\ref{eq:dEdt}) and (\ref{eq:dLzdt}) is
expressed as
\begin{align}
F &=\sum_{\ell=2}^{\infty} F_{\ell},\label{eq:sum_Fl}
\end{align}
\begin{align}
F_\ell &=\sum_{m=-\ell}^{\ell} F_{\ell m},\label{eq:sum_Flm}
\\
F_{\ell m} &=2 \sum_{n=n_i}^{\infty} F_{\ell m n},\label{eq:sum_Flmn}
\end{align}
where $n_i$ is the minimum integer which satisfies $m\Upsilon_\phi+n_i \Upsilon_r>0$, 
and 
\begin{align}
F=\left<\frac{d\C{E}}{dt}\right>\quad \textrm{or} \quad \left<\frac{d\C{L}_z}{dt}\right>. 
\end{align}
In Eq.~(\ref{eq:sum_Flmn}), we used the relation of $ F_{\ell m n}=
F_{\ell -m -n}$ to take into account the modes of $M\omega<0$.

We truncate the $\ell$-summation in Eq.~(\ref{eq:sum_Fl}) by choosing a
maximum value of $\ell$ as $\ell_{\textrm{max}}$.  Then, $F$ is
written as
\begin{align}
F =\sum_{\ell=2}^{\ell_{\textrm{max}}} F_{\ell}+\delta F_{\ell_{\textrm{max}}}, 
\end{align}
where $\delta F_{\ell_{\textrm{max}}}$ is the error due to
restricting the $\ell$-summation up to
$\ell=\ell_{\textrm{max}}$. The reason that we set the maximum 
value of $\ell$ is that for very high values of $\ell \agt \ell_{\textrm{max}}$, 
the value of $\nu$ cannot be numerically calculated
  accurately. As mentioned above, this problem could be fixed if we
  can improve the precision for the numerical calculation of $\nu$.

\begin{figure}[t] 
\centering 
\includegraphics[width=88mm]{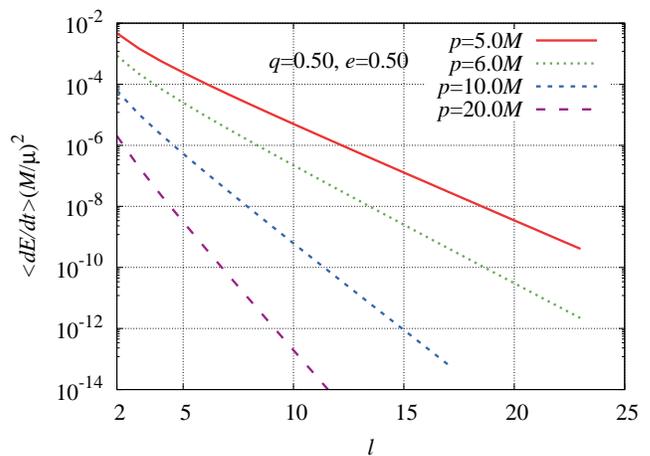}
\caption{Gravitational-wave energy flux ($F_\ell$) as a function of
  $\ell$ for $q=0.5$, $e=0.5$, and $p=5.0M$, $6.0M$, $10.0M$, and
  $20.0M$.}
\label{fig:dEdt_ell_q050_e050}
\end{figure}

Figure~\ref{fig:dEdt_ell_q050_e050} shows the energy flux of
gravitational waves, $F_\ell$, as a function of $\ell$ for $q=0.5$ and
$e=0.5$.  $F_\ell$ decreases approximately exponentially with the
increase of $\ell$, but the decrease rate becomes less steep for
smaller values of $p$.  If we assume the exponential decrease of
$F_\ell$ in $\ell$, the error size by the truncation of the
higher-$\ell$ modes, $\delta F_{\ell_{\textrm{max}}}$, can be
estimated as
\begin{align}
\delta F_{\ell_{\textrm{max}}}
=F\sum_{\ell=\ell_{\textrm{max}}+1}^{\infty} {\textrm{e}}^{-\alpha\ell}
=F \frac{{\textrm{e}}^{-\alpha\ell_{\textrm{max}}}}{{\textrm{e}}^{\alpha}-1},
\end{align}
where $\alpha$ is a positive constant.  

\begin{figure*}[htbp]
\centering 
\includegraphics[width=59mm]{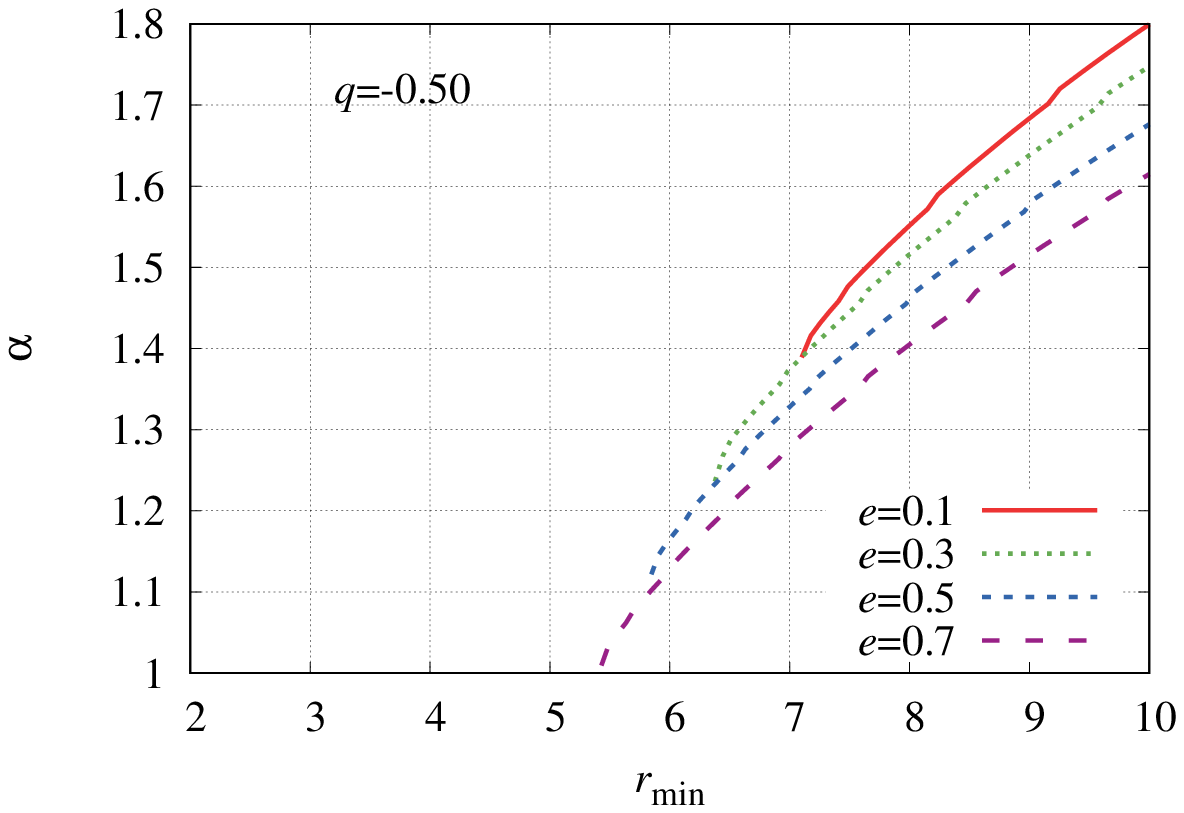}%
\includegraphics[width=59mm]{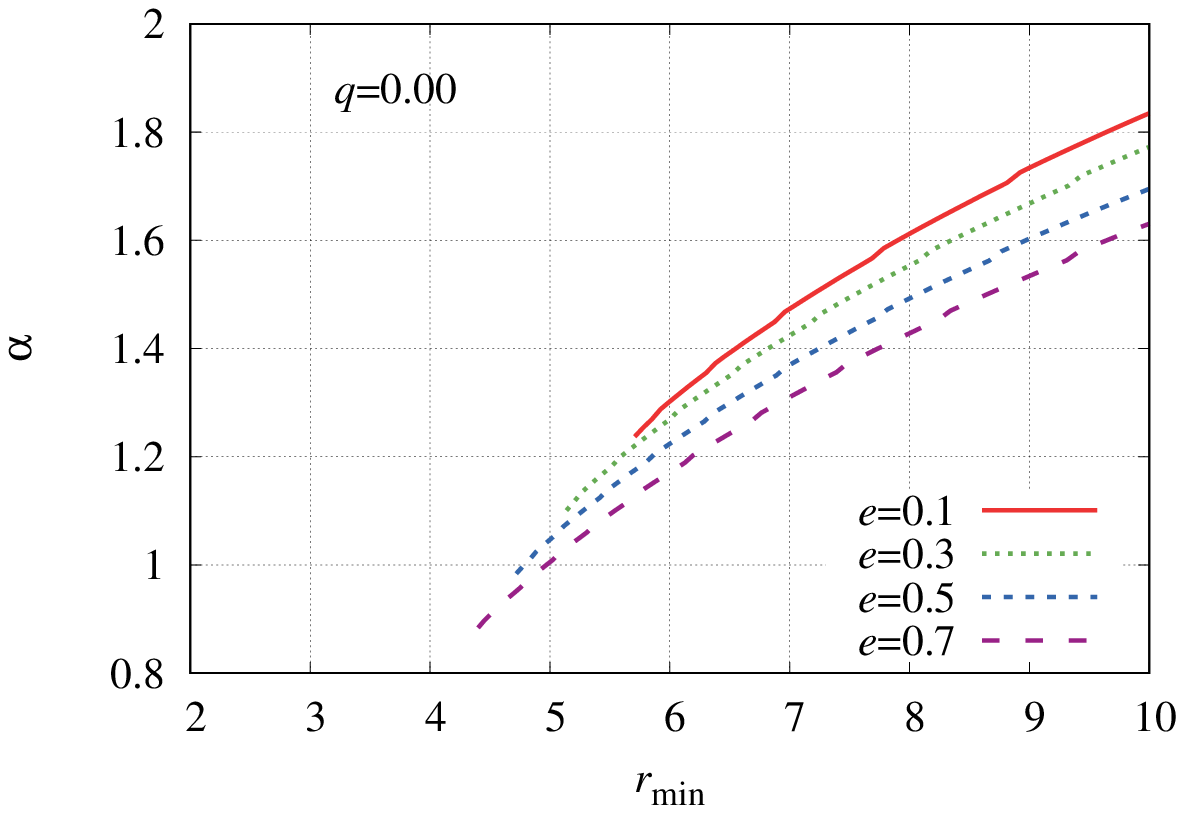}%
\includegraphics[width=59mm]{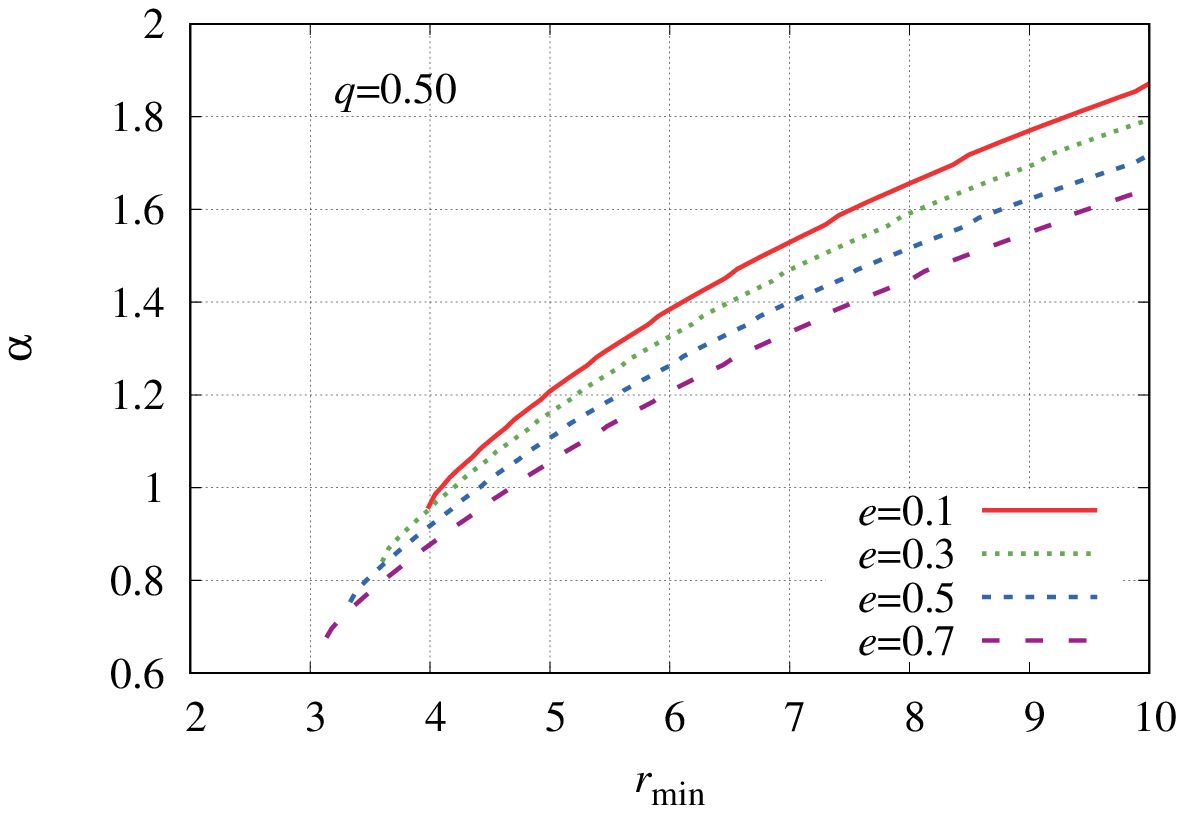}
\caption{$\alpha$ as a function of $r_{\textrm{min}}$ for $q=-0.5$ (left), 
$0$ (middle), and $0.5$ (right). 
}
\label{fig:dEdt_ell_alpha}
\end{figure*}

In this paper the maximum value of $\ell$ is set to be
$\ell_{\textrm{max}}=25$.  This implies that $\delta
F_{\ell_{\textrm{max}}}/F$ is less than $10^{-6}$ for $\alpha\gtrsim
0.6$.  Figure~\ref{fig:dEdt_ell_alpha} shows $\alpha$ as a function of
$r_{\textrm{min}}$ for $q=-0.5$ (left), $0$ (middle), and $0.5$
(right) with several values of $e$.  We find that the value of
$\alpha$ is larger than $0.6$ for any stable orbits with $q\alt 0.5$
for which the minimum value of the orbital radius, $r_{\textrm{min}}$,
is larger than $\sim 3M$.  Hence, we conclude that the
error due to restricting the $\ell$-summation up to
$\ell_{\textrm{max}}=25$ in the energy dissipation rate for $q\lesssim
0.5$ is less than $10^{-6}$.  However, $\ell_{\textrm{max}}=25$ is not
large enough to achieve the required error size for orbits close to
the separatrix with $q\gtrsim 0.6$.

The summation over the $n$-modes in Eq.~(\ref{eq:sum_Flmn}) is
approximated as
\begin{align}
F_{\ell m} =2 \sum_{n=n_i}^{n_f} F_{\ell m n},
\end{align}
where $n_f$ is determined by $F_{\ell m n_f}<10^{-6}F$.  We note that
the values of $n$ for the dominant modes of $F_{\ell m n}$ shifts to
larger values of $n$ for larger values of $\ell$ and $e$ (see
Refs.~\cite{Drasco:2005kz,Fujita:2009us}), and $F_{\ell m n}$
decreases exponentially with the increase of $n$ after the dominant
mode of $F_{\ell m n}$ is
reached~\cite{Drasco:2005kz,Fujita:2009us}.  In this paper, the
maximum value of $n_f$ is set to be $1000$. This choice is large
enough for the orbits with $e \alt 0.8$.

\begin{figure}[htbp] 
\centering 
\includegraphics[width=88mm]{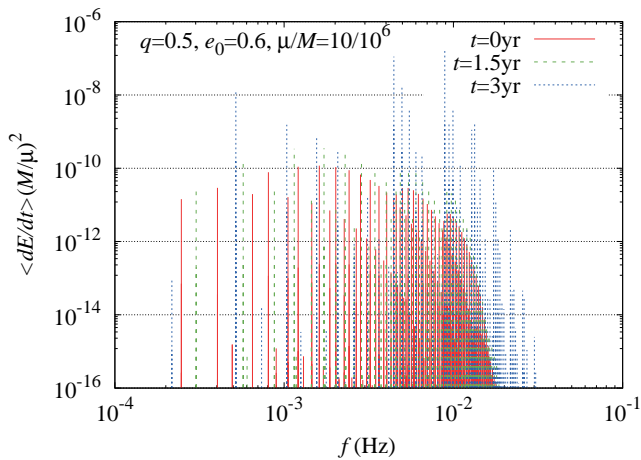}
\caption{ The energy spectrum during the 3-year inspiral before plunge
  for $q=0.5$ and $e_0=0.6$ with $M=10^6M_\odot$ and $\mu=10M_\odot$.
  The values of $(p,e)$ take $(10.1M,0.60)$ at $t=0$~year, $(8.9M,0.50)$
  at $t=1.5$~years, and $(4.6M,0.24)$ at $t=3$~years, respectively.  }
\label{fig:dEdt_f_q050_e060}
\end{figure}

\begin{figure*}[htbp]
\centering
\includegraphics[width=59mm]{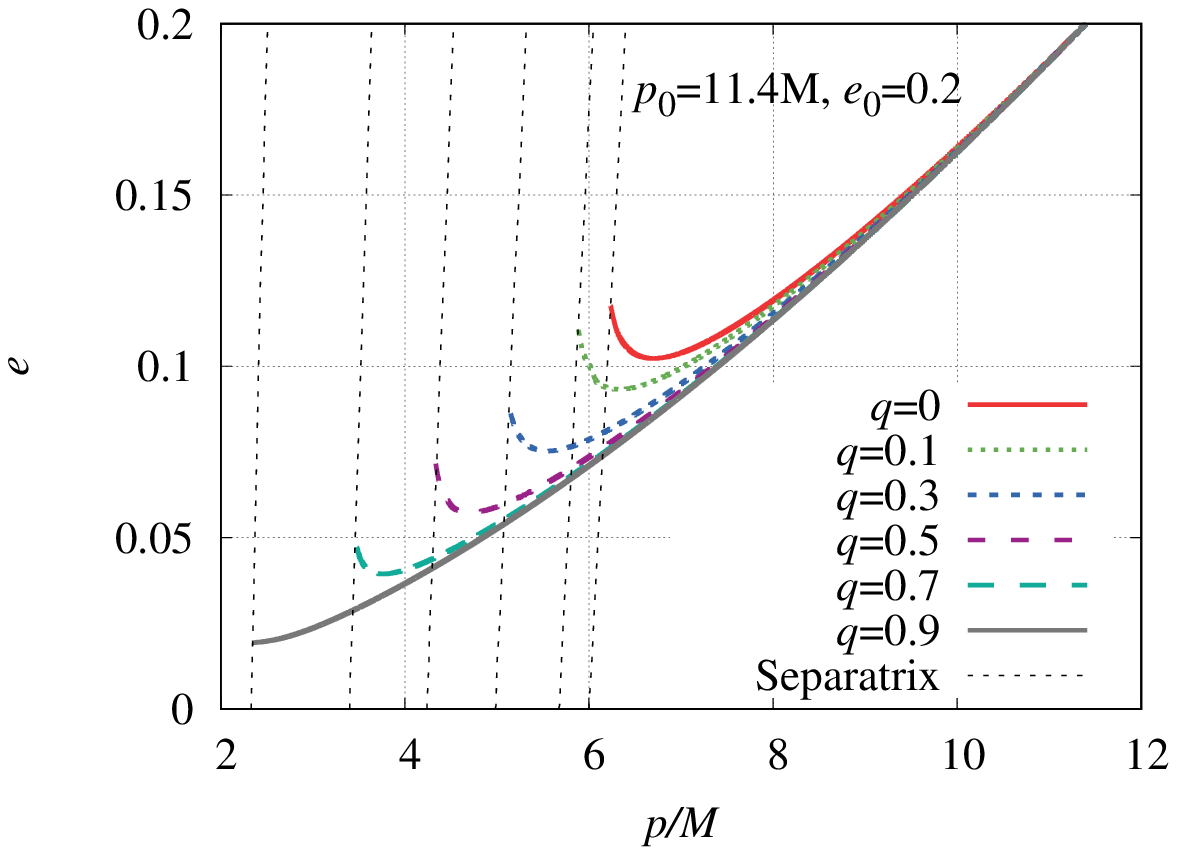}
\includegraphics[width=59mm]{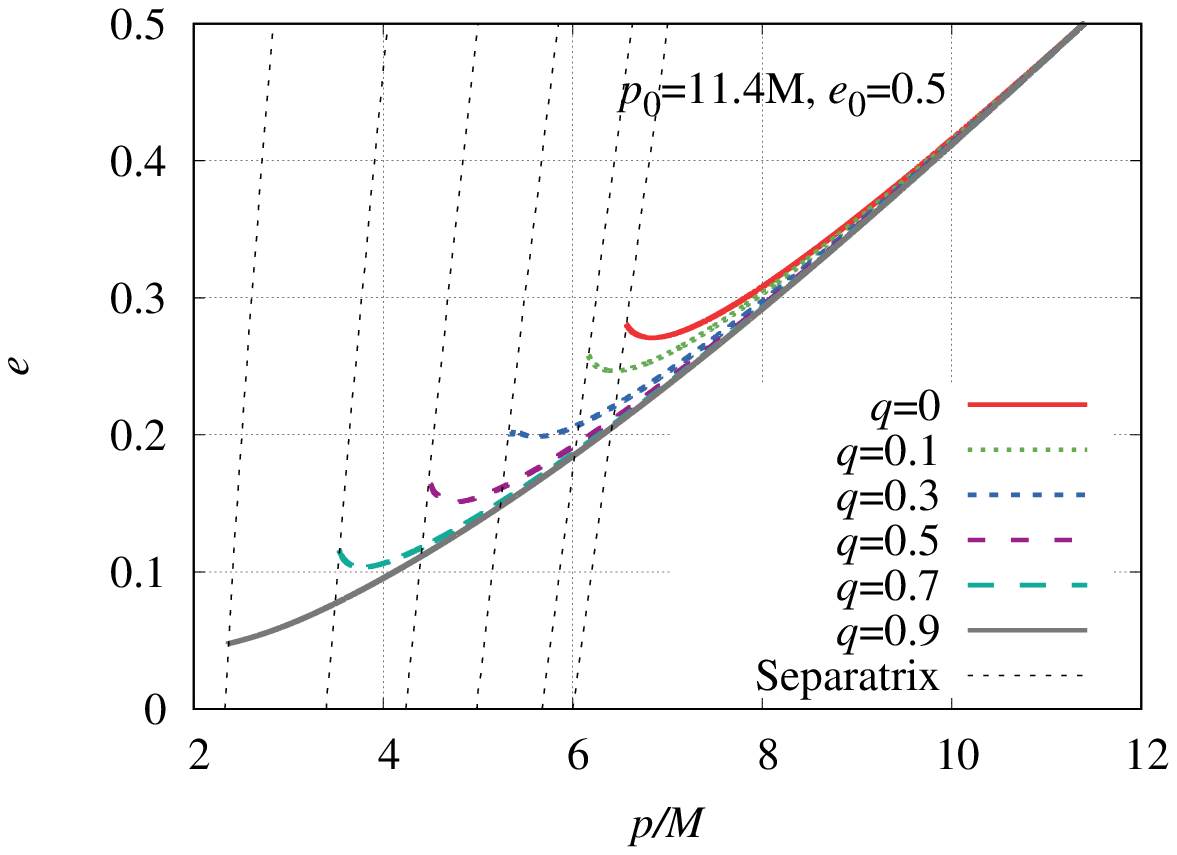}
\includegraphics[width=59mm]{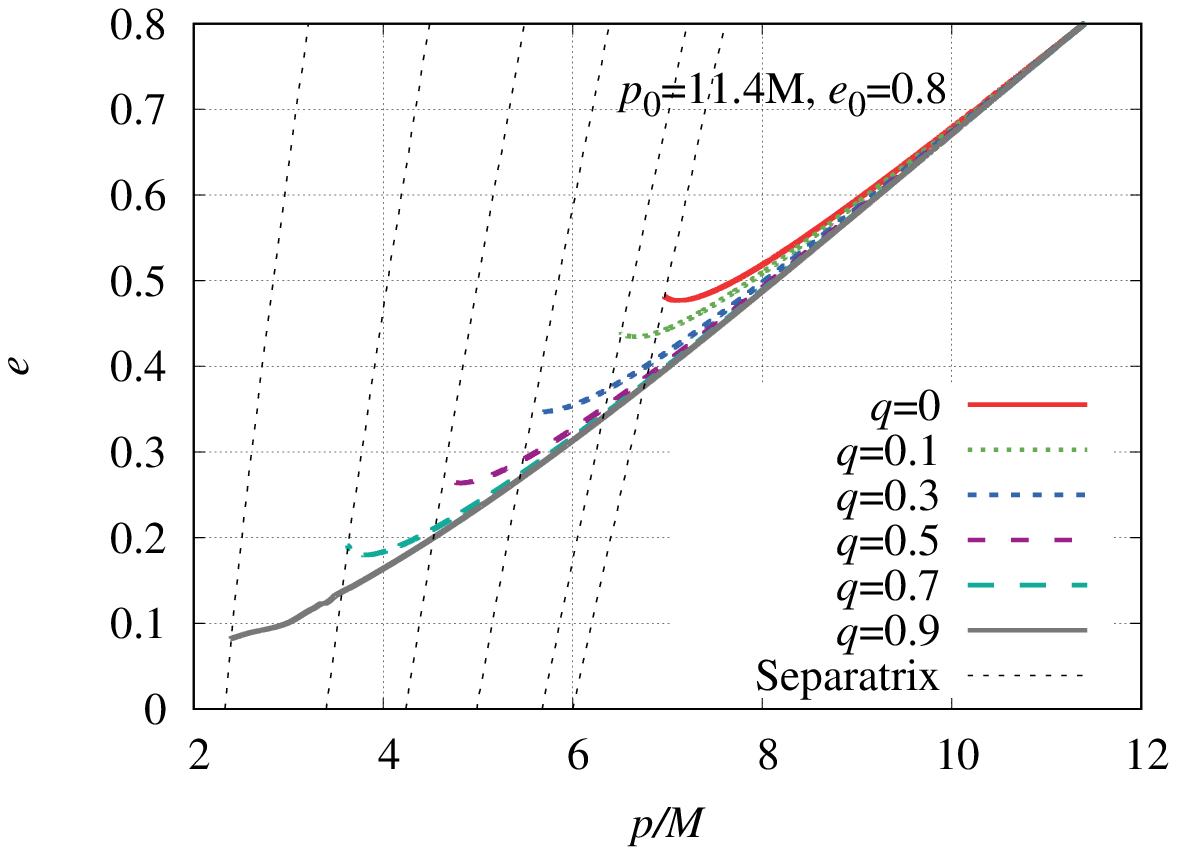}
\caption{The orbital eccentricity $e$ as a function of the semilatus
  rectum $p$ for inspiral orbits from $p_0=11.4M$ to the separatrix
  with $q=0$, $0.1$, $0.3$, $0.5$, $0.7$, and $0.9$.  The initial
  orbital eccentricity is chosen to be $e_0=0.2$ (left), $0.5$
  (middle), and $0.8$ (right).  The dashed curves show the separatrix
  of stable orbits.  Note that it takes $\sim 100M^2/\mu$ for
  $q=-0.5$, $\sim 190M^2/\mu$ for $q=0$ and $\sim 300M^2/\mu$ until
  plunge for $q=0.5$ with $e_0=0.2$ (see also Fig.~\ref{fig:orbit_pn}).
  The inspiral time becomes longer for the larger BH spin with
  fixed values of $p_0$ and $e_0$ because $p$ at the separatrix
  becomes smaller.  We note that the orbits near the separatrix are
  not obtained for $q=0.9$ as accurately as for other values of $q$
  (see a discussion in Sec.~\ref{sec:method}).}
\label{fig:orbit_pe_e0}
\end{figure*}

\begin{figure*}[htbp]
\centering
\includegraphics[width=59mm]{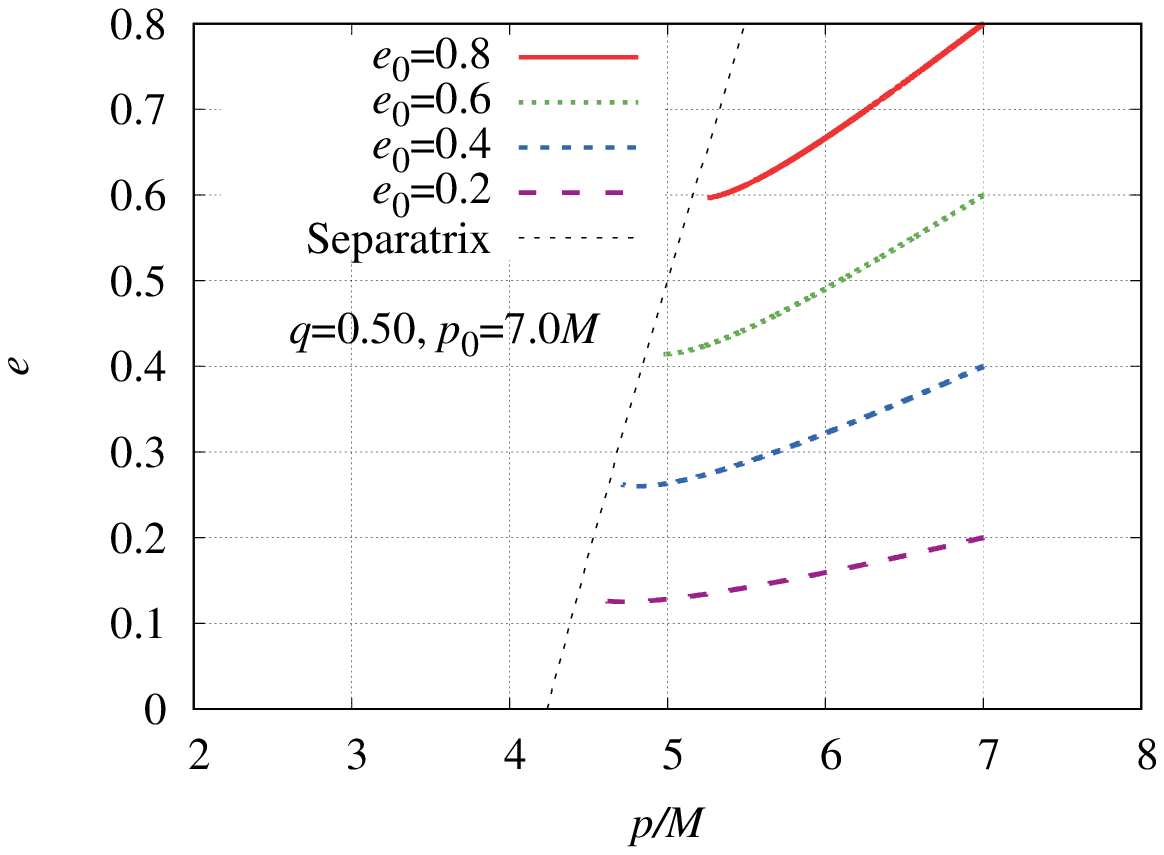} 
\includegraphics[width=59mm]{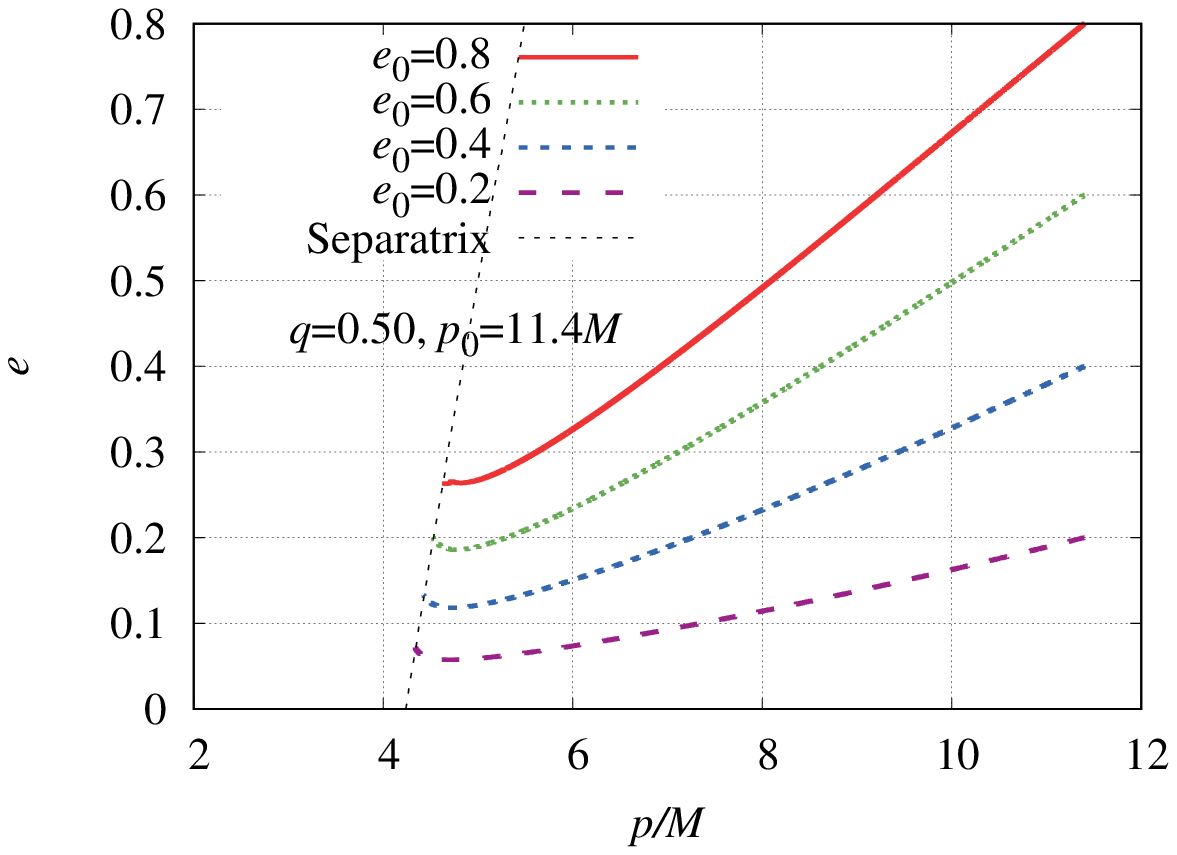} 
\includegraphics[width=59mm]{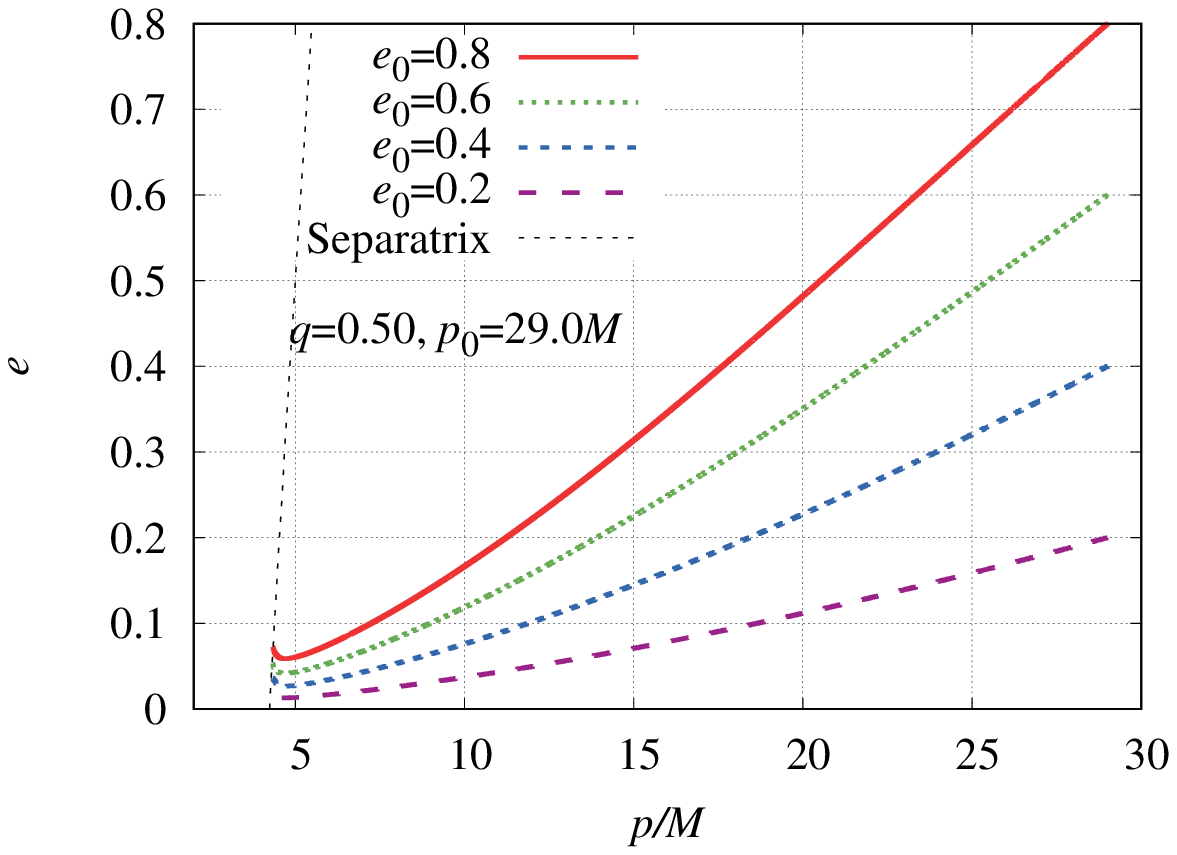}
\caption{The same as Fig.~\ref{fig:orbit_pe_e0}, but for inspiral
  orbits from $p_0=7.0M$ (left), $11.4M$ (middle), and $29.0M$ (right)
  to the separatrix for $q=0.5$ with $e_0=0.8$, $0.6$, $0.4$, and
  $0.2$.  The dashed curves denote the separatrix of stable orbits for
  $q=0.5$.}
\label{fig:orbit_pe_p0}
\end{figure*}

Figure~\ref{fig:dEdt_f_q050_e060} shows the energy spectrum during the
3-year inspiral before plunge for $q=0.5$ and $e_0=0.6$ with
$M=10^6M_\odot$ and $\mu=10M_\odot$.  The values of $(p,e)$ take
$(10.1M,0.60)$ at the beginning, $(8.9M,0.50)$ at $1.5$~years, and
$(4.6M,0.24)$ at the plunge, respectively.  We note that the number of
the $n$-modes necessary for the required accuracy for fixed values of
$(\ell,m)$ decreases as approaching the separatrix because of the
circularization of the orbital eccentricity (see, e.g.,
Figs.~\ref{fig:orbit_pe_e0} and \ref{fig:orbit_pe_p0}). By contrast,
the number of the $\ell$-modes necessary for the required accuracy
increases with the orbital evolution, because the value of $r_{\textrm{min}}$ 
decreases and relativistic effects are enhanced with the orbital evolution.
To achieve the relative error in the energy dissipation rate 
within $10^{-6}$, the maximum values of $(\ell,n)$ become $(18,109)$,
$(19,76)$, and $(24,42)$ at the beginning, $1.5$~years, and the plunge,
respectively. 

\begin{figure*}[htbp] 
\centering 
\includegraphics[width=59mm]{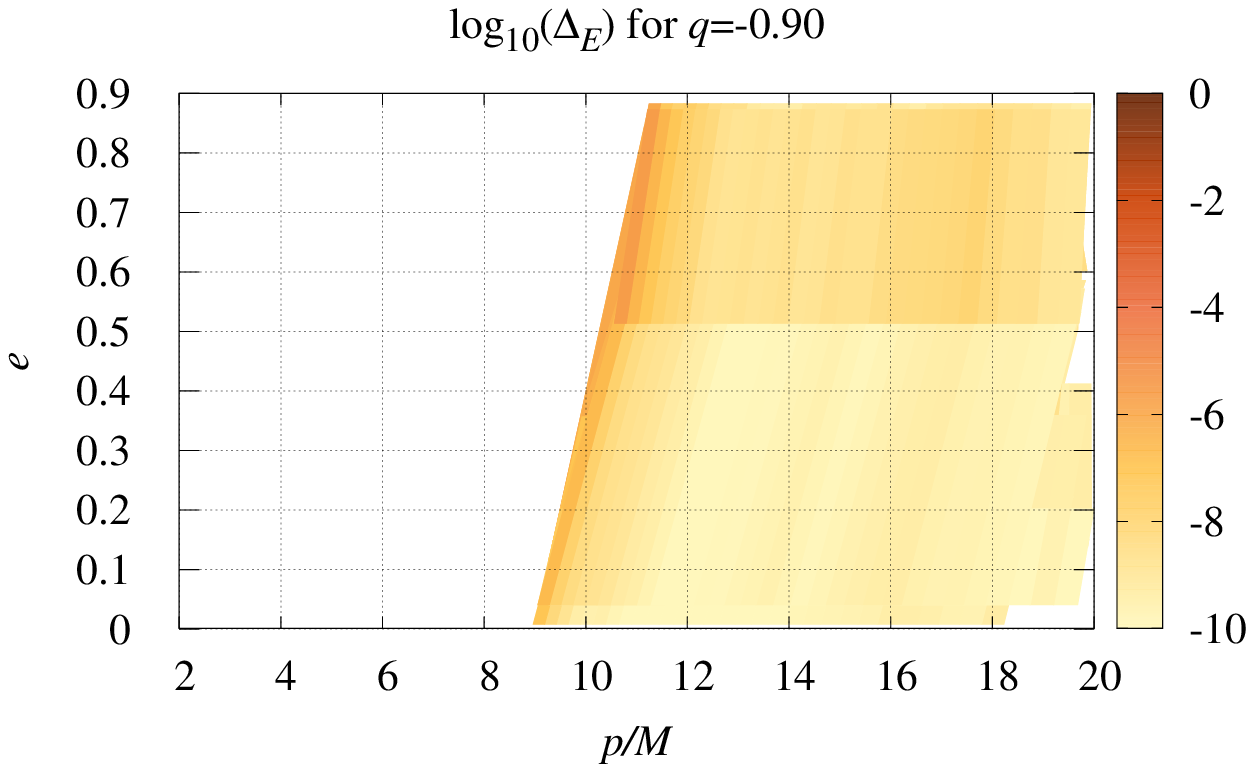}
\includegraphics[width=59mm]{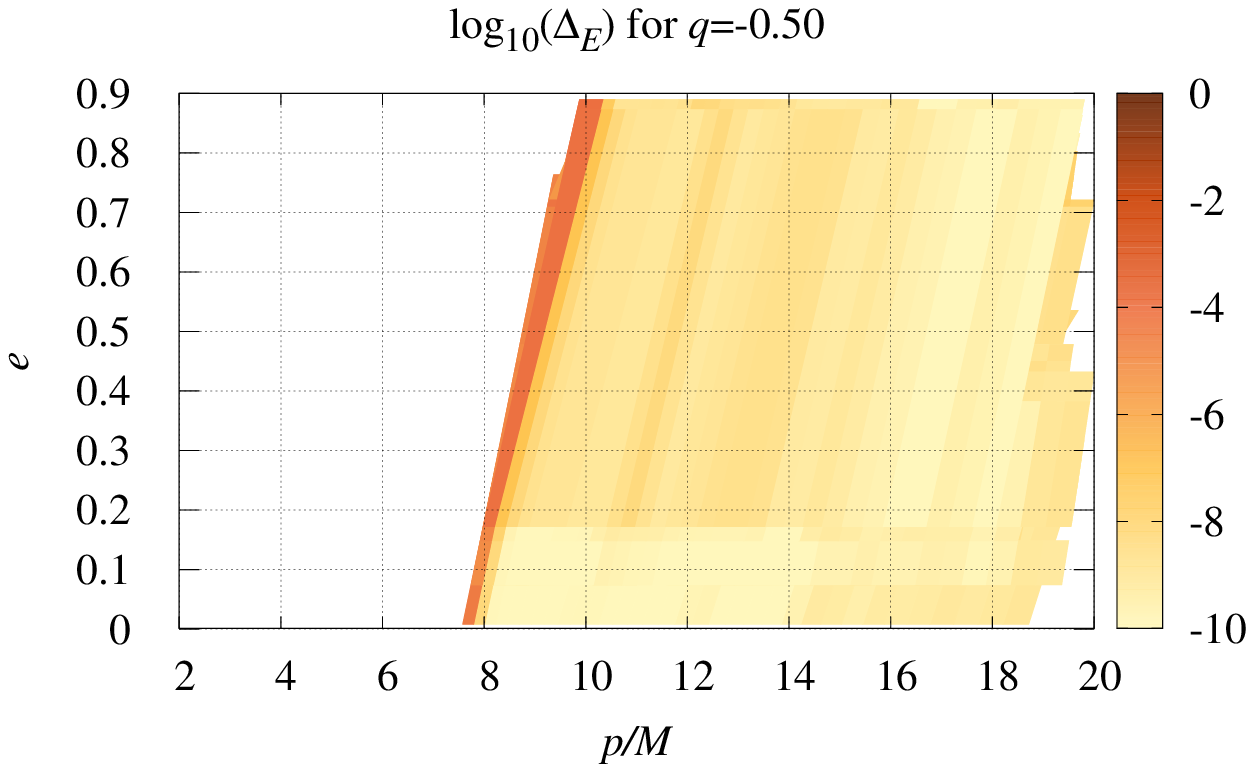}
\includegraphics[width=59mm]{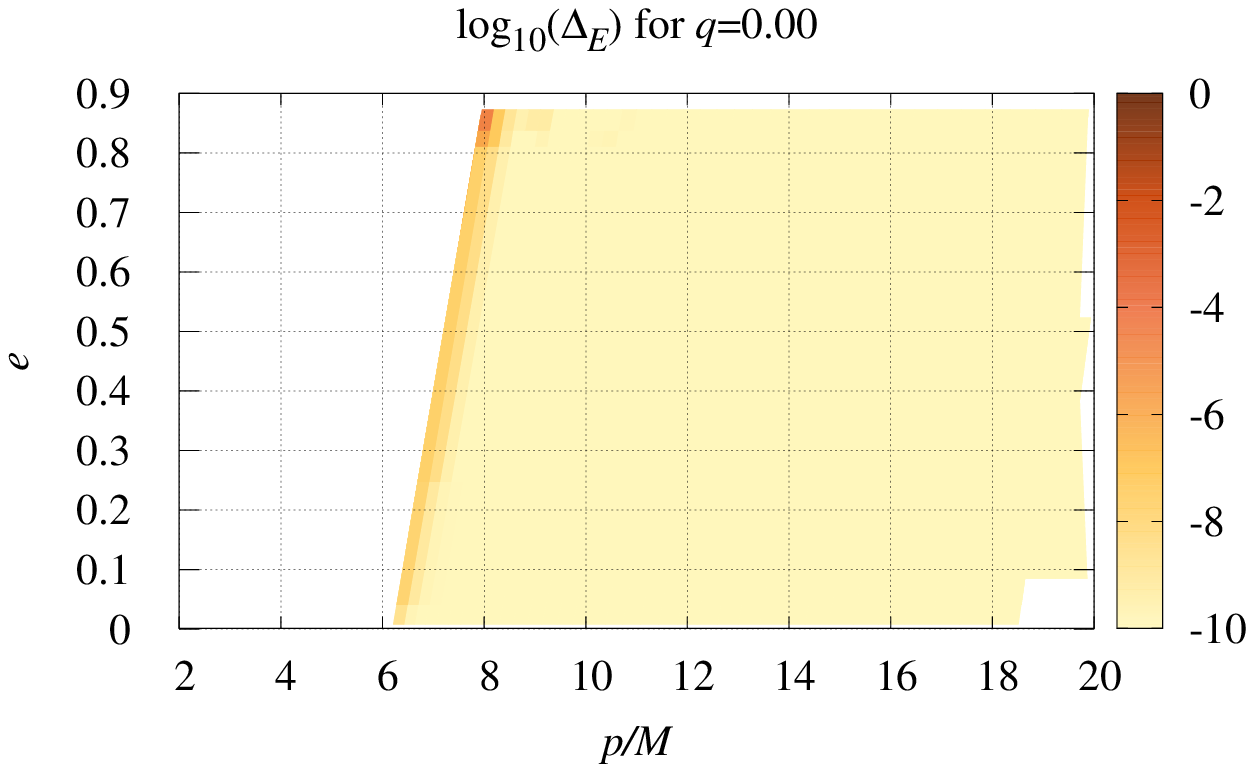}\\
\includegraphics[width=59mm]{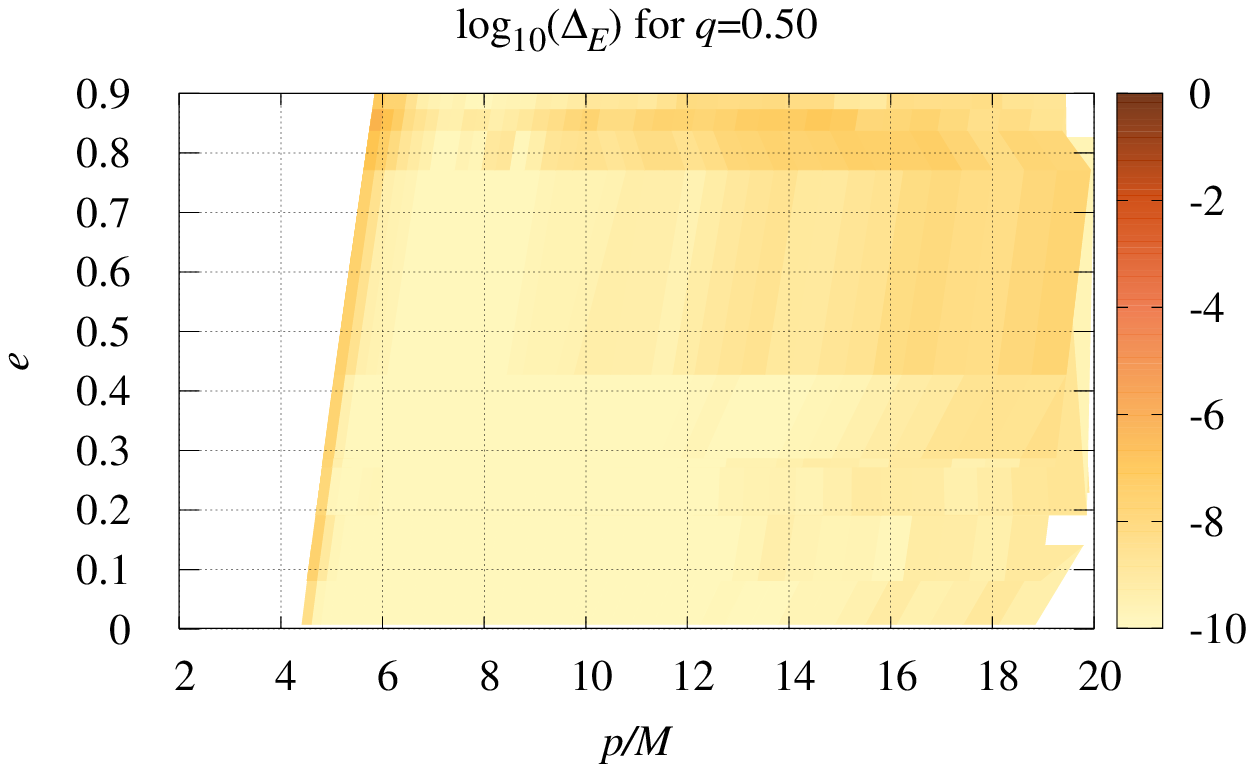}
\includegraphics[width=59mm]{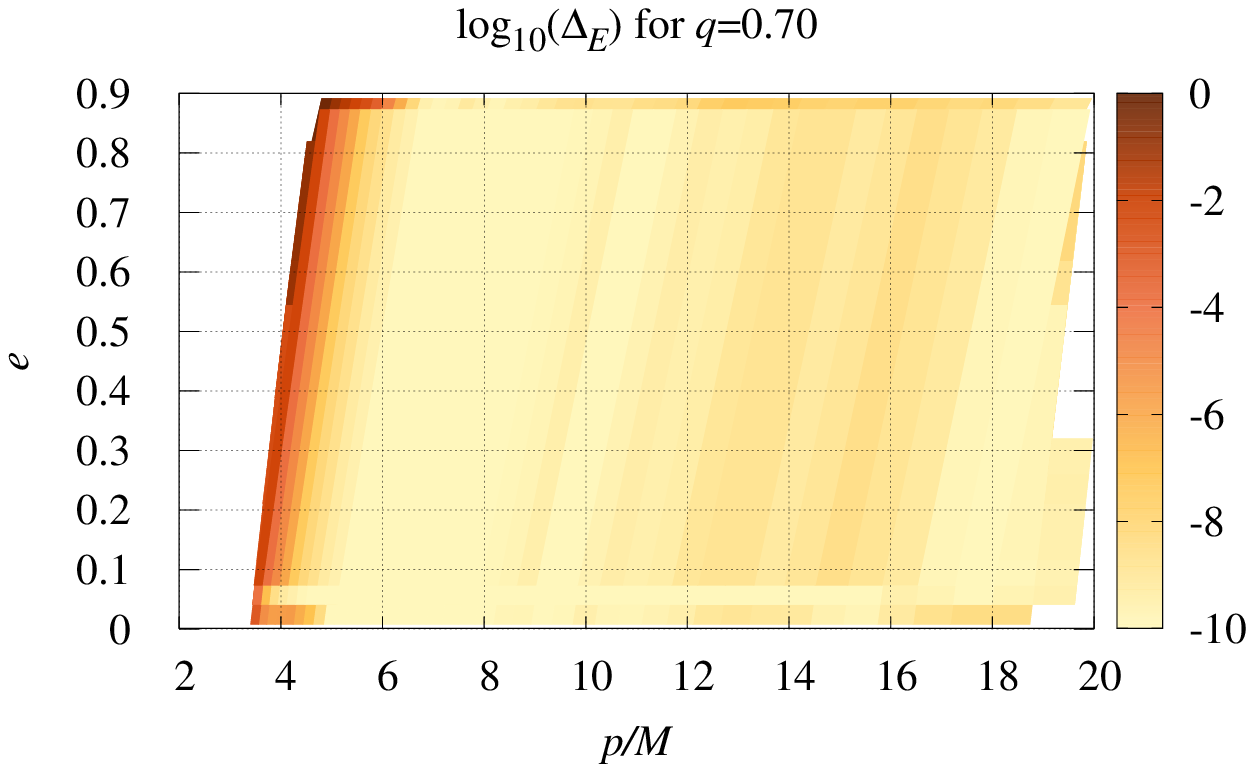}
\includegraphics[width=59mm]{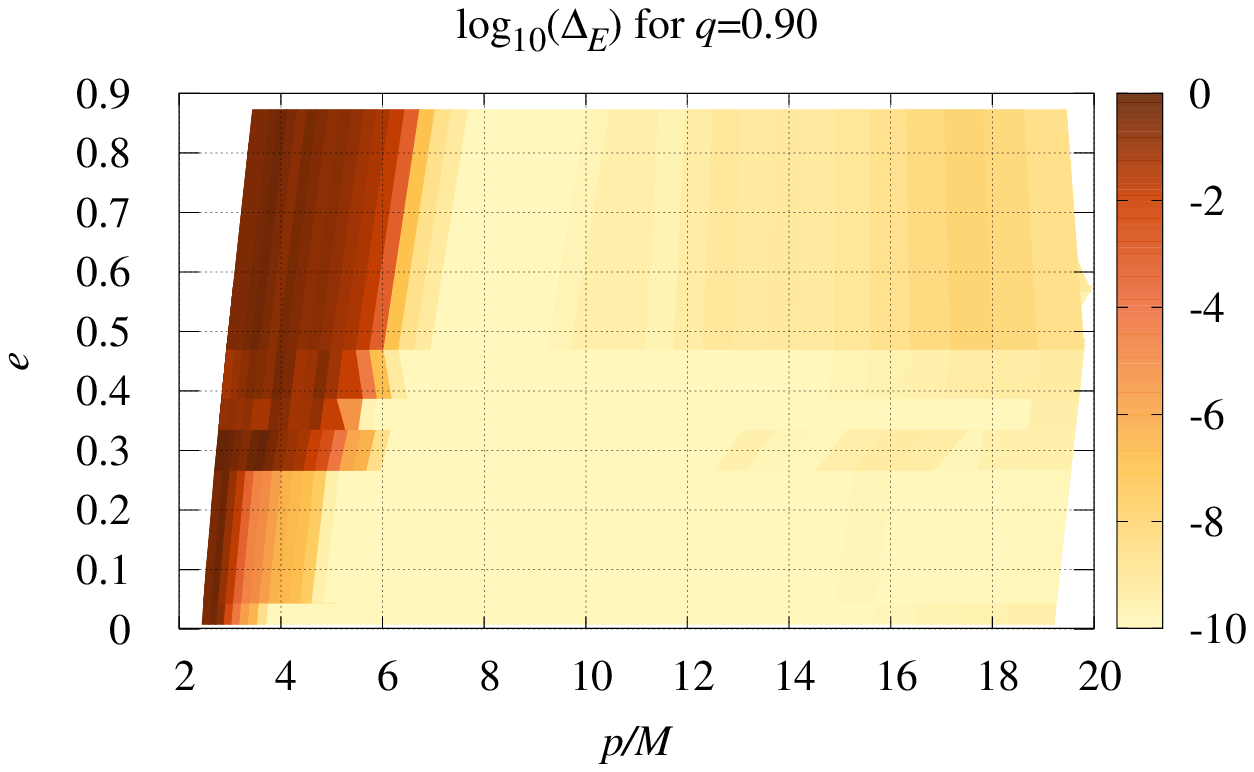} 
\caption{Relative error in the interpolated energy flux,
  $\Delta_{\textrm{E}}$, for $q=-0.9$, $-0.5$, $0.0$, $0.5$, $0.7$,
  and $0.9$.  The error is estimated by comparison with numerical data
  independent of those used for the interpolation.  The error is
  smaller than $10^{-6}$ for $r_{\textrm{min}}=p/(1+e)\gtrsim 3M$.}
\label{fig:flux_err}
\end{figure*}

To confirm the validity of the interpolation, the relative error in
the interpolated energy flux is estimated by comparison with numerical
data independent of those used for the interpolation, and the results
are shown in Fig.~\ref{fig:flux_err}.  This shows that the error is
smaller than $10^{-6}$ for $r_{\textrm{min}}=p/(1+e)\agt 3M$.
Thus, the required accuracy is always achieved for $q\le 0.5$.  As
  already mentioned, the accuracy is also limited by the accuracy of
  $\nu$, the number of the grid points used in the trapezium rule, and
  the truncation with respect to the $\ell$-summation.  By these
  limitations, the accuracy with the error less than $10^{-6}$ is not
  achieved for the compact orbits of $r_{\textrm{min}} \alt 3M$.
  Figure~\ref{fig:flux_err} shows the similar feature for the
  magnitude of the error associated with the interpolation. This
  suggests that the accuracy would be limited by that for the
  individual data set, not by the interpolation.  To conclude,
  currently, for the case that the value of $r_{\textrm{min}}$ is
  smaller than $3M$ (i.e., for $q\agt 0.6$), the accuracy of $10^{-6}$
  is not achieved due to the error of the individual data set.

Here, we should note the following point: the lifetime of the binaries
with an orbit near the separatrix to plunge is so short that the total
cycle of the orbits is at most $10^4$ (see Fig.~\ref{fig:cycle}). This
indicates that for such compact orbits, the accuracy of $\lesssim
10^{-5}$ would be acceptable.  Thus in this paper, we believe that the
accuracy of our numerical results for the inspiral orbits is
acceptable for $q\leq 0.7$. However, for $q\geq 0.9$, we should keep
in mind that the accuracy is not sufficient.  Improving the accuracy
for the case of $q$ close to unity is the issue to be solved in the
future work.

\section{Results}
\label{sec:results}
In this section, we present inspiral orbits and corresponding
gravitational-wave spectra for the typical EMRIs as sources of LISA
using the procedure described in Sec.~\ref{sec:method}. We also
calculate the SNR of gravitational waves for
such EMRIs using the LISA's designed sensitivity curve.

\subsection{Inspiral orbits}
\label{sec:p_e}
Figures~\ref{fig:orbit_pe_e0} and \ref{fig:orbit_pe_p0} illustrate the
inspiral orbits showing the orbital eccentricity $e$ as a function of
the semilatus rectum $p$ for several values of $(q,p_0,e_0)$.  Here,
$p_0$ and $e_0$ are the initial semilatus rectum and the initial
orbital eccentricity, respectively.  In Fig.~\ref{fig:orbit_pe_e0},
the inspirals start from $p_0=11.4M$ to the separatrix with $e_0=0.2$
(left), $0.5$ (middle), and $0.8$ (right) for $q=0$ to $0.9$.  We note
that in our assumption imposed in this paper, the curves $e(p)$ do not
depend on $M$ and $\mu$~\cite{Peters:1964zz}.  It takes $\sim
100M^2/\mu$ for $q=-0.5$, $\sim 190M^2/\mu$ for $q=0$, and $\sim
300M^2/\mu$ for $q=0.5$ from $p_0=11.4M$ to the plunge with $e_0=0.2$
(cf. also Fig.~\ref{fig:orbit_pn}).  Here, $\sim 300M^2/\mu$ is
  $\sim 5$ years for $M=10^6M_\odot$ and $\mu=10M_\odot$.  We note that 
the inspiral time becomes longer for the larger BH spin with fixed
values of $p_0$ and $e_0$ because $p$ at the separatrix becomes
smaller.  The inspiral time also becomes longer for the larger values of
$e_0$ with a fixed value of $p_0$, because the adiabatic change of
  $p$ (i.e., $\langle dp/dt \rangle$) becomes smaller for larger
  values of $e$ with a fixed value of $p$~\cite{Peters:1964zz}.
Figure~\ref{fig:orbit_pe_e0} shows that the circularization of the
orbital eccentricity occurs for the most stages of the inspiral, but
the eccentricity slightly increases near the separatrix as already
found in
Refs.~\cite{Cutler:1994pb,Glampedakis:2002ya,Fujita:2009us,Warburton:2011fk,Osburn:2015duj}.

Figure~\ref{fig:orbit_pe_p0} compares the inspiral orbits with
$p_0=7.0M$ (left), $11.4M$ (middle), and $29.0M$ (right), fixing
$q=0.5$, but varying $e_0$ from $0.8$ to $0.2$.  This also shows that
the orbits always circularize in the early inspiral, but the
eccentricity increases as approaching the separatrix.  For the larger
value of $p_0$, the plunge occurs for a small value of $p$, which is
approximately equal to the radius of the innermost stable circular
orbit.  Thus, for the case that $p_0/M$ is fairly large $\sim 30$, the
circularization occurs significantly even for $e_0=0.8$ and results in
the nearly circular orbits just before the plunge. For relatively
small SMBH mass, e.g., $M \sim 10^5M_\odot$, the inspiral proceeds from
$p_0 \approx 30M$ to the plunge in a few years for $\mu \sim
10M_\odot$. For such a case, the final orbit is likely to be nearly
circular even if $e_0$ is initially high as $e_0=0.8$.  By contrast,
for a higher value of $M$, the inspiral time of stellar-mass object is
a few years even if $p_0$ is smaller than $10M$. For the small value of
$p_0$, the eccentricity does not change significantly until the plunge
orbit is reached.  Thus such a plunge orbit could have a large
eccentricity if $e_0$ at $p<10M$ is so.
We note that the two-body relaxation in star clusters of the galactic centers could produce highly eccentric EMRIs 
with $p\le 10M$ in the LISA band that do not plunge immediately because 
the value of $p$ at the separatrix becomes smaller than that of a Schwarzschild BH 
if one takes into account the BH spin~\cite{AmaroSeoane:2012cr}.

\subsection{Gravitational-wave spectra and SNR} 
\label{sec:snr}
Following Ref.~\cite{Finn:2000sy}, 
the squared SNR averaged over all source directions is defined by 
\begin{align}
\textrm{SNR}^2= 4\int_{0}^{\infty} d(\log f) \left[\frac{h_\textrm{eff}(f)}{h_\textrm{n}(f)}\right]^2,
\label{eq:snr}
\end{align}
where $h_\textrm{n}(f)$ is the noise amplitude and $h_\textrm{eff}(f)$
is the power spectrum defined below.  The noise amplitude is
defined by $h_\textrm{n}(f)=\sqrt{f S_\textrm{n}(f)}$
\cite{Moore:2014lga}, where $S_\textrm{n}(f)$ is the one-sided noise
power spectral density. In this paper, $S_\textrm{n}(f)$ is taken to
be the analytic form of the LISA's designed sky-averaged 
sensitivity presented in Ref.~\cite{Babak:2017tow}.

The power spectrum is defined by the summation of 
the power spectrum for $\ell$, $h_\textrm{eff}^{\ell}$, as
\begin{align}
h_\textrm{eff}(f)=\sum_{\ell} h_\textrm{eff}^{\ell}(f),
\label{eq:heff}
\end{align}
where 
\begin{align}
h_\textrm{eff}^{\ell}(f)=\sum_{m n} h_\textrm{eff}^{\ell m n}(f),
\label{eq:hl}
\end{align}
and $h_\textrm{eff}^{\ell m n}(f)$ is estimated by~\cite{Finn:2000sy}
\begin{align}
h_\textrm{eff}^{\ell m n}(f_{\ell m n})=\frac{1}{\pi D}\sqrt{\frac{2\dot{E}_{\ell m n}}{\dot{f}_{\ell m n}}}. 
\label{eq:hlmn}
\end{align}
Here, the dot denotes the time derivative, $D$ is the distance to the
source, $\dot{E}_{\ell m n}$ is the energy flux to infinity due to the
emission of gravitational waves at frequency $f_{\ell m n}$, which is
defined as
\begin{align}
f_{\ell m n}=\frac{m\Upsilon_\phi + n\Upsilon_r}{2\pi\Gamma}
\equiv m f_\phi+n f_r, 
\end{align}
where Eq.~(\ref{eq:omega_mkn}) is used.  $f_\phi$ and $f_r$ are the
frequencies of the azimuthal and radial motion, respectively.  In our
inspiral orbits, $f_\phi$ and $f_r$ can be computed at each time step
by using $p(t)$ and $e(t)$.  We then compute $\dot{f}_{\ell m n}$
from $\Delta f_{\ell m n}/\Delta t$, where $\Delta t$ is the time step
for evolving the orbital motion and $\Delta f_{\ell m n}=f_{\ell m
  n}(t+\Delta t)-f_{\ell m n}(t)$ (see Sec.~\ref{sec:inspiral} for our
choice of the time step).  We use $\approx 100$ frequency bins to
smooth the modal power spectrum $h_\textrm{eff}^{\ell m n}(f)$.  We
compute $h_\textrm{eff}^{\ell m n}(f)$ for the modes of $\ell=2$--$4$,
$-\ell \le m\le l$, and $n_0\le n\le n_0+45$ where $n_0=-m$.  
We choose this value of $n_0$ in order to compute the SNR 
with the relative error of $\lesssim$ 10\%. We check the error by varying 
$n_0$ from 10 to 60. For $e_0=0.4$ and $\ell=2$, 
$n_0=10$ is sufficient to compute the SNR with such accuracy. 
However, we need a larger value of $n_0$ for higher values of 
$e_0$ and $\ell$~\cite{Drasco:2005kz,Fujita:2009us}. 
For $e_0=0.8$ and $\ell=2$ ($\ell=4$), we need 
$n_0=30$ ($n_0=45$) to compute the SNR with the error of $\lesssim$ 10\%.
The power spectrum $h_\textrm{eff}(f)$ is computed by summing all the
modes of $h_\textrm{eff}^{\ell m n}(f)$ at each frequency bin.

\begin{figure}[t]
\centering
\includegraphics[width=88mm]{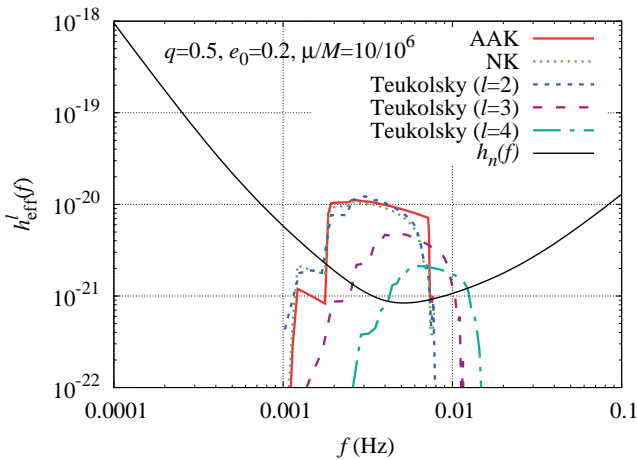}
\caption{Power spectra with $\ell=2$ for numerical kludge
  (NK)~\cite{Gair:2005ih,Babak:2006uv}, augmented analytic kludge
  (AAK)~\cite{Chua:2017ujo}, and Teukolsky (this work) models for a
  $10M_\odot$ compact object inspiraling around a $10^6M_\odot$ SMBH of
  spin $q=0.5$ at $D=1$\,Gpc during the last 3-year inspiral before
  plunge. We consider the inspiral with $p_0=10.3M$ and $e_0=0.2$.
  The amplitudes in the Teukolsky model for $\ell=3$ and $4$ modes are
  also shown.  The curve $h_{\textrm{n}}(f)$ shows the LISA's
  designed sky-averaged sensitivity~\cite{Babak:2017tow}. }
\label{fig:heff_kludge}
\end{figure}

\begin{figure*}[htbp]
\centering
\includegraphics[width=89mm]{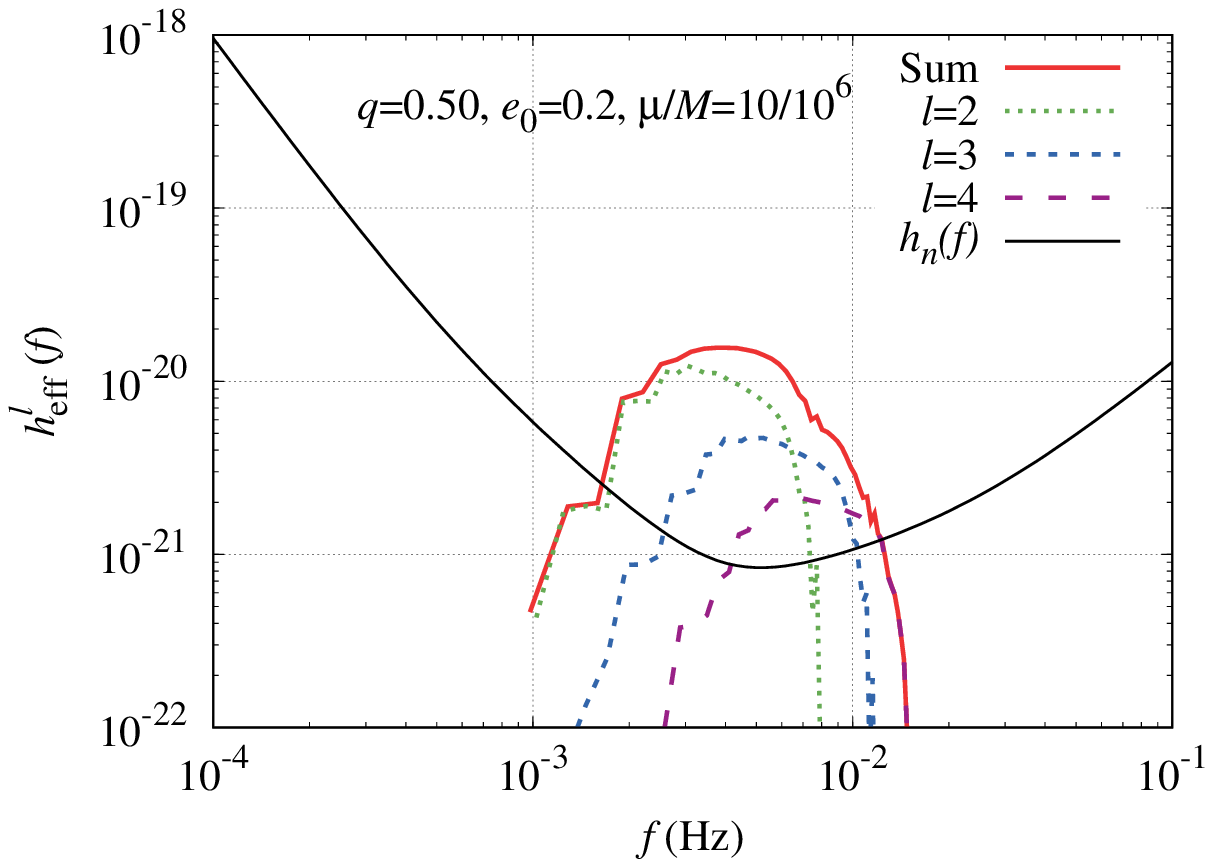}%
\includegraphics[width=89mm]{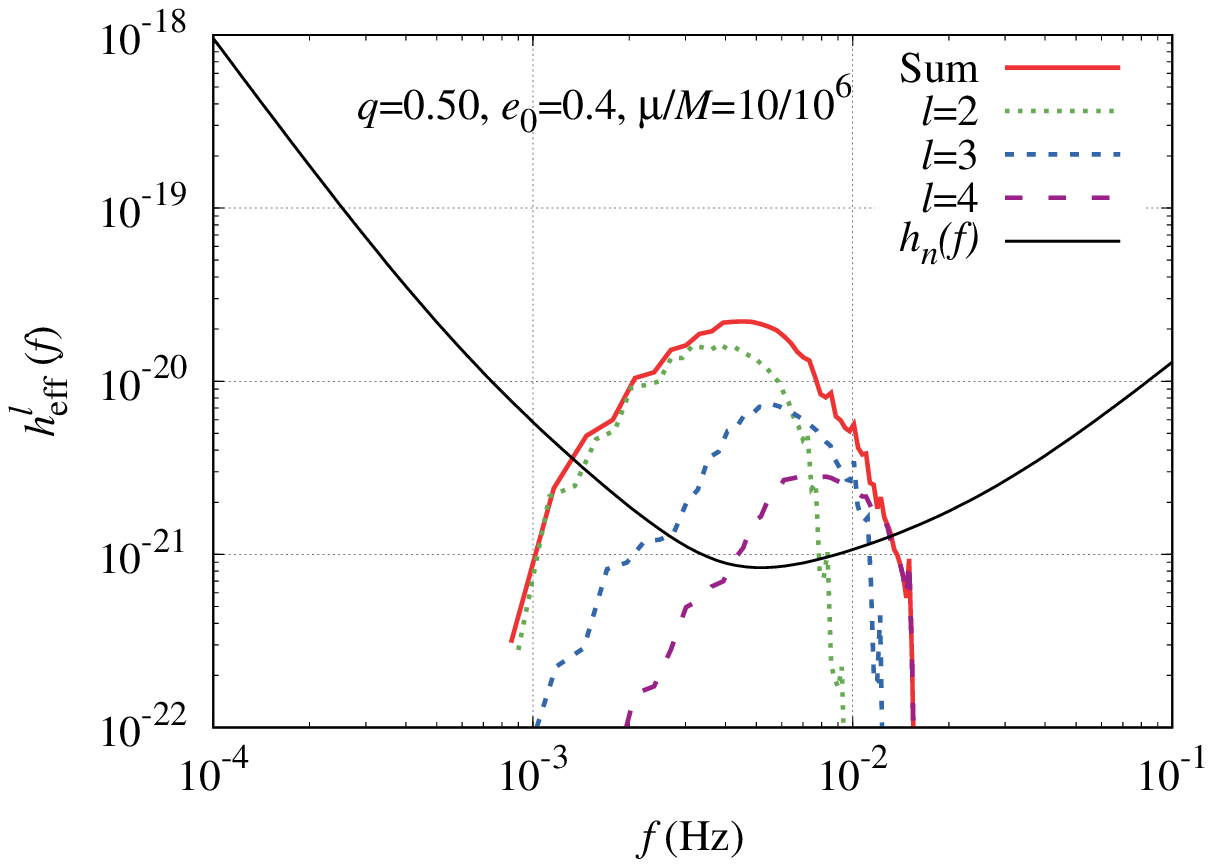}\\%
\includegraphics[width=89mm]{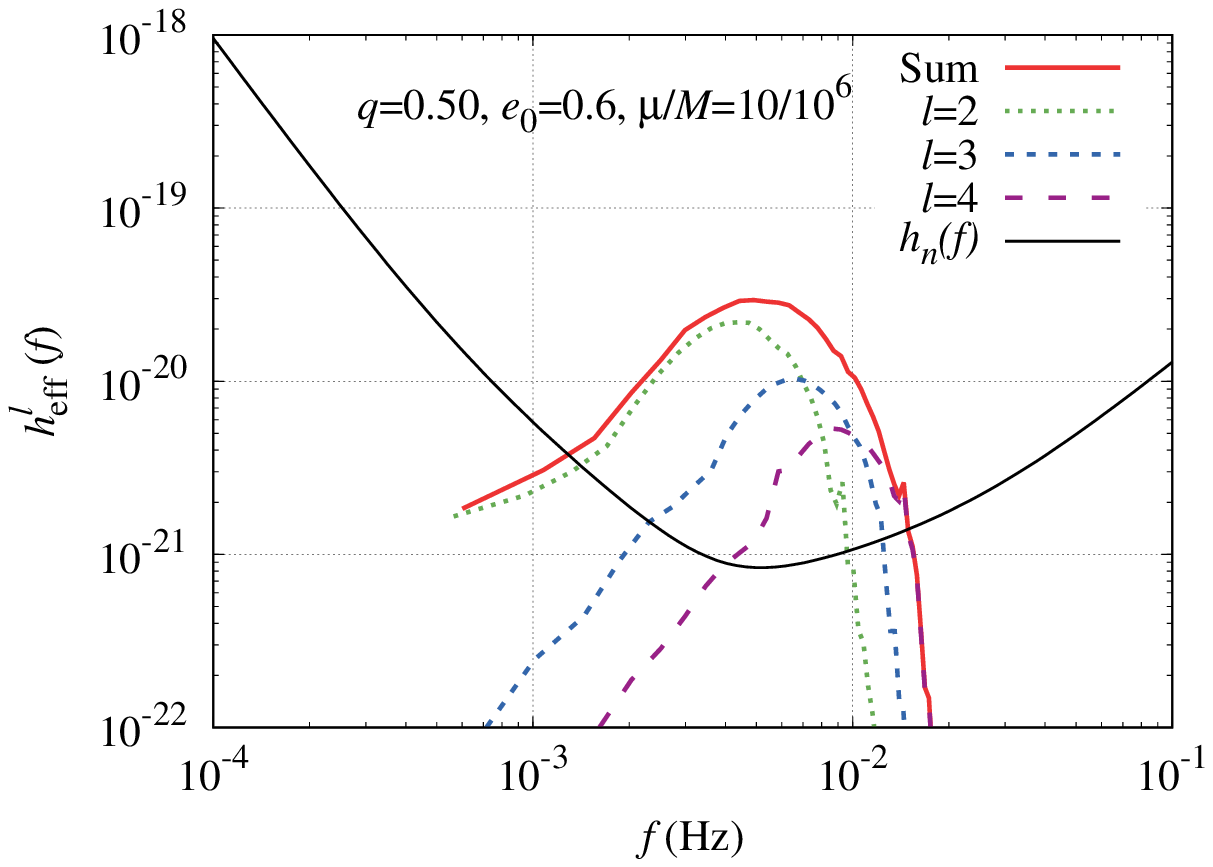}%
\includegraphics[width=89mm]{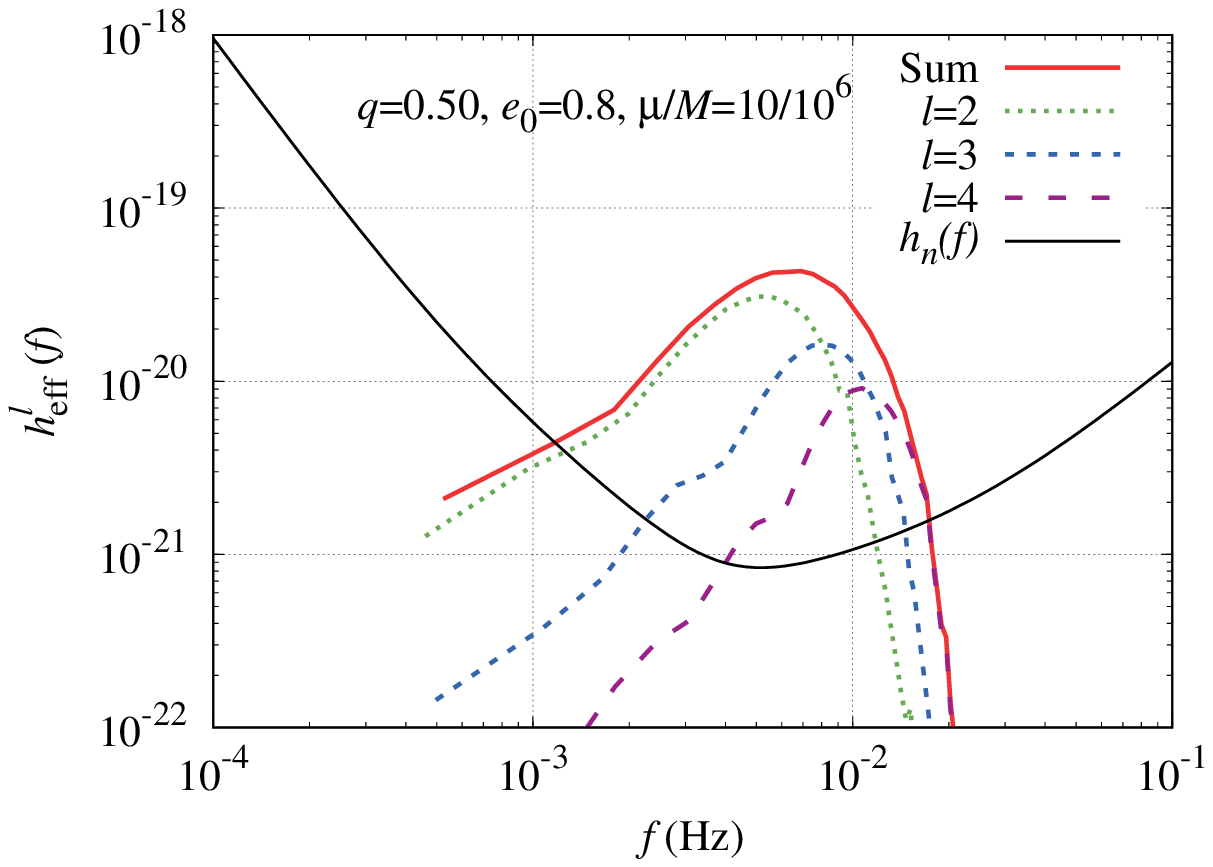}%
\caption{Power spectra for $\ell=2$, $3$, and $4$ induced by a
  $10M_\odot$ compact object inspiraling around a $10^6M_\odot$
  SMBH of spin $q=0.5$ at $D=1$\,Gpc for the last 3-year inspiral
  before plunge.  The initial orbital eccentricity is chosen to be
  $e_0=0.2$ (top left), $0.4$ (top right), $0.6$ (bottom left), and
  $0.8$ (bottom right).  The values of $p_0$ take $10.33M$, $10.28M$,
  $10.11M$, and $9.58M$ for $e_0=0.2$, $0.4$, $0.6$, and $0.8$,
  respectively.  Complicated structures in the amplitude can be
  understood by noting contributions from higher radial modes ($n$-modes) to
  gravitational waves (see, e.g., Ref.~\cite{Barack:2003fp}).  The
  curve $h_{\textrm{n}}(f)$ shows LISA's designed sky-averaged
  sensitivity~\cite{Babak:2017tow}.  }
\label{fig:heff_q050_p11.4}
\end{figure*}

\begin{figure*}[htbp]
\centering
\includegraphics[width=59mm]{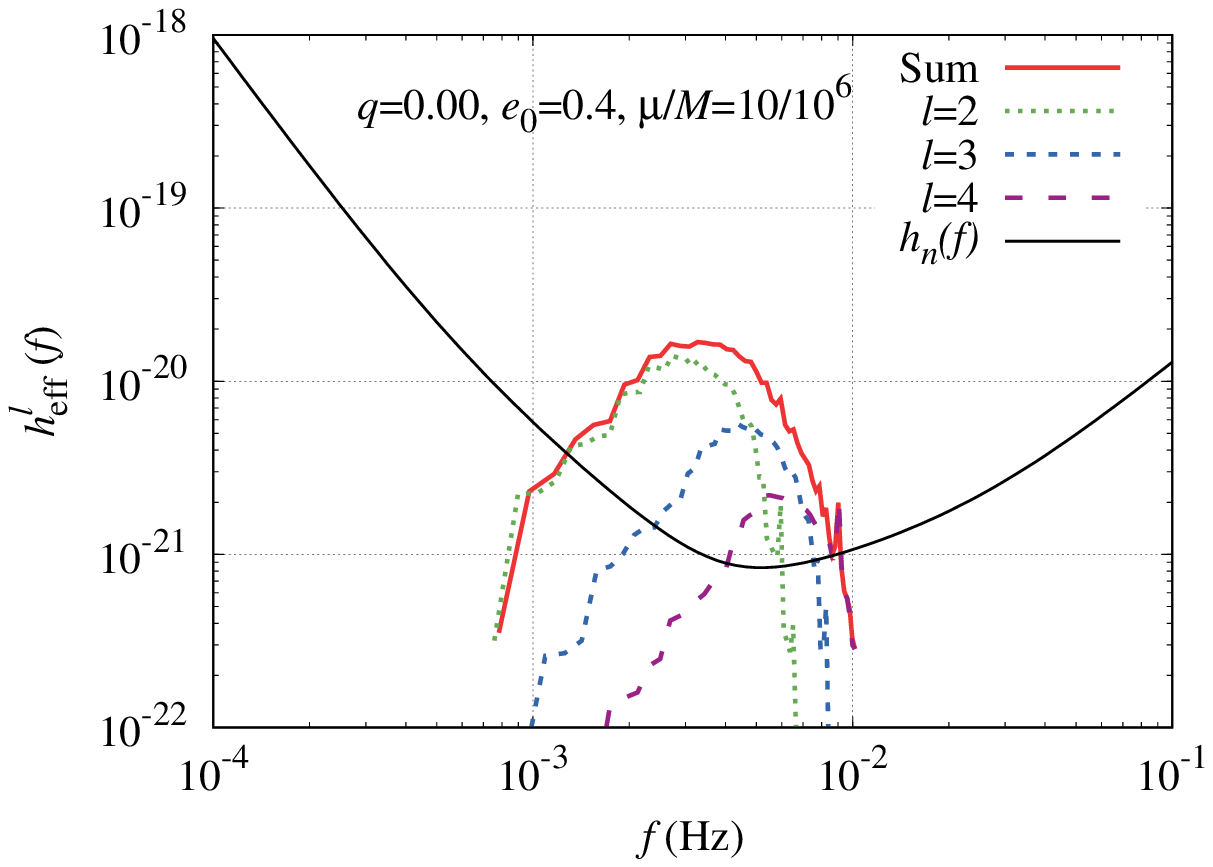}%
\includegraphics[width=59mm]{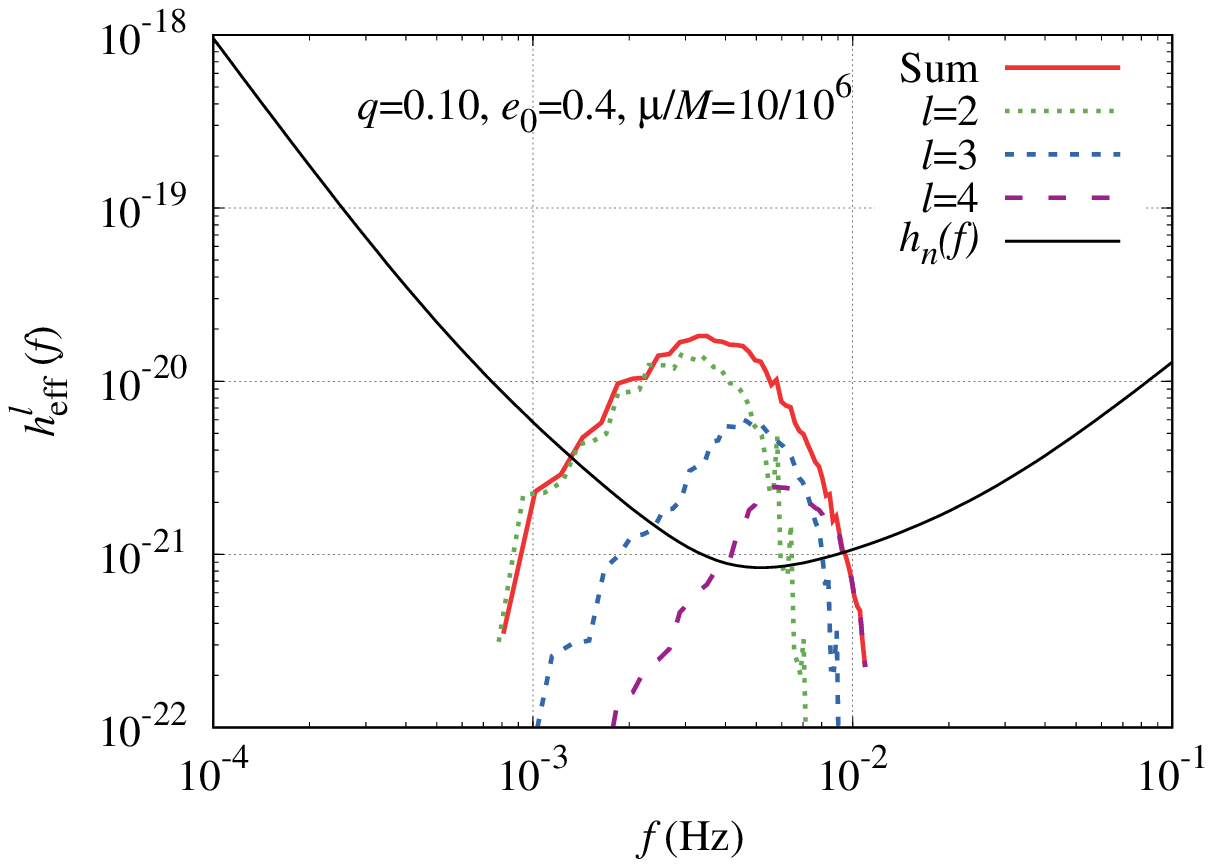}%
\includegraphics[width=59mm]{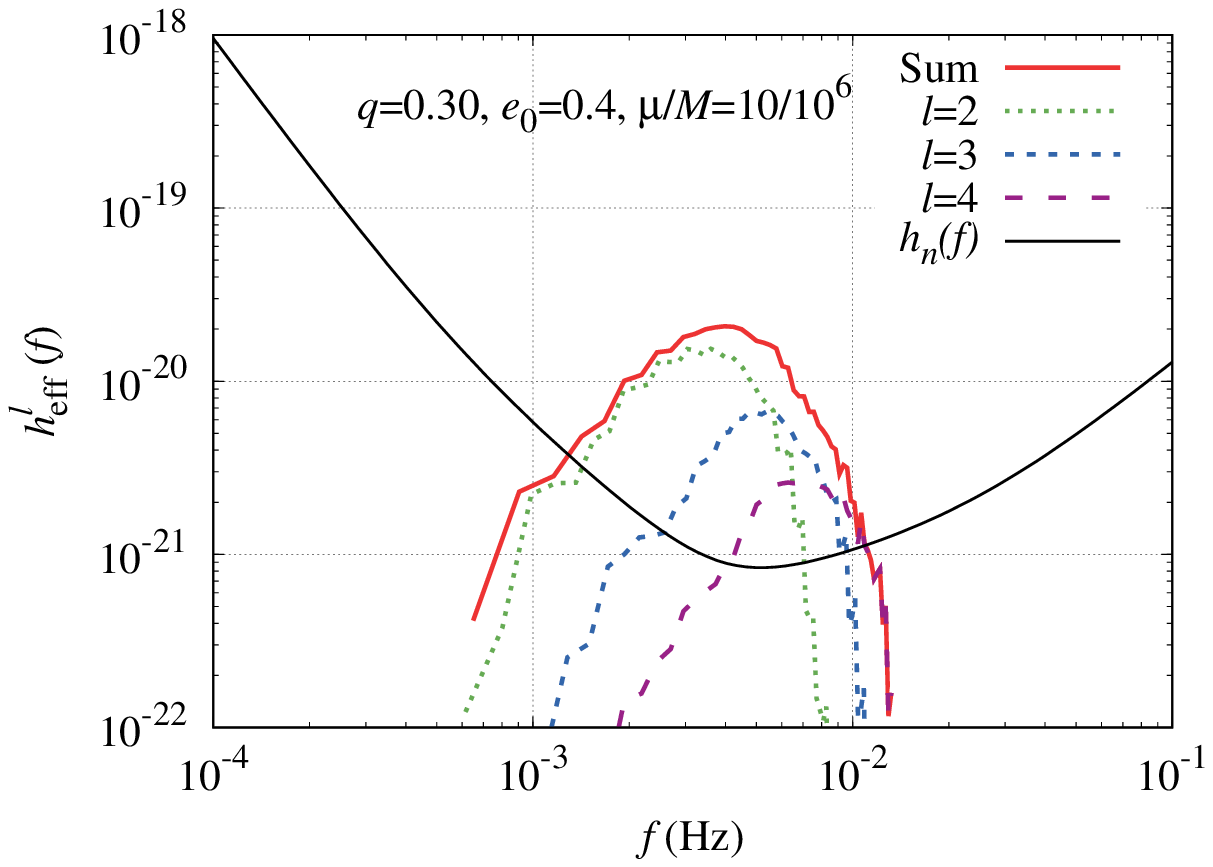}\\%
\includegraphics[width=59mm]{./fig/heff_q0.50e0.4mu10M1.e+6Tobs3.eps}%
\includegraphics[width=59mm]{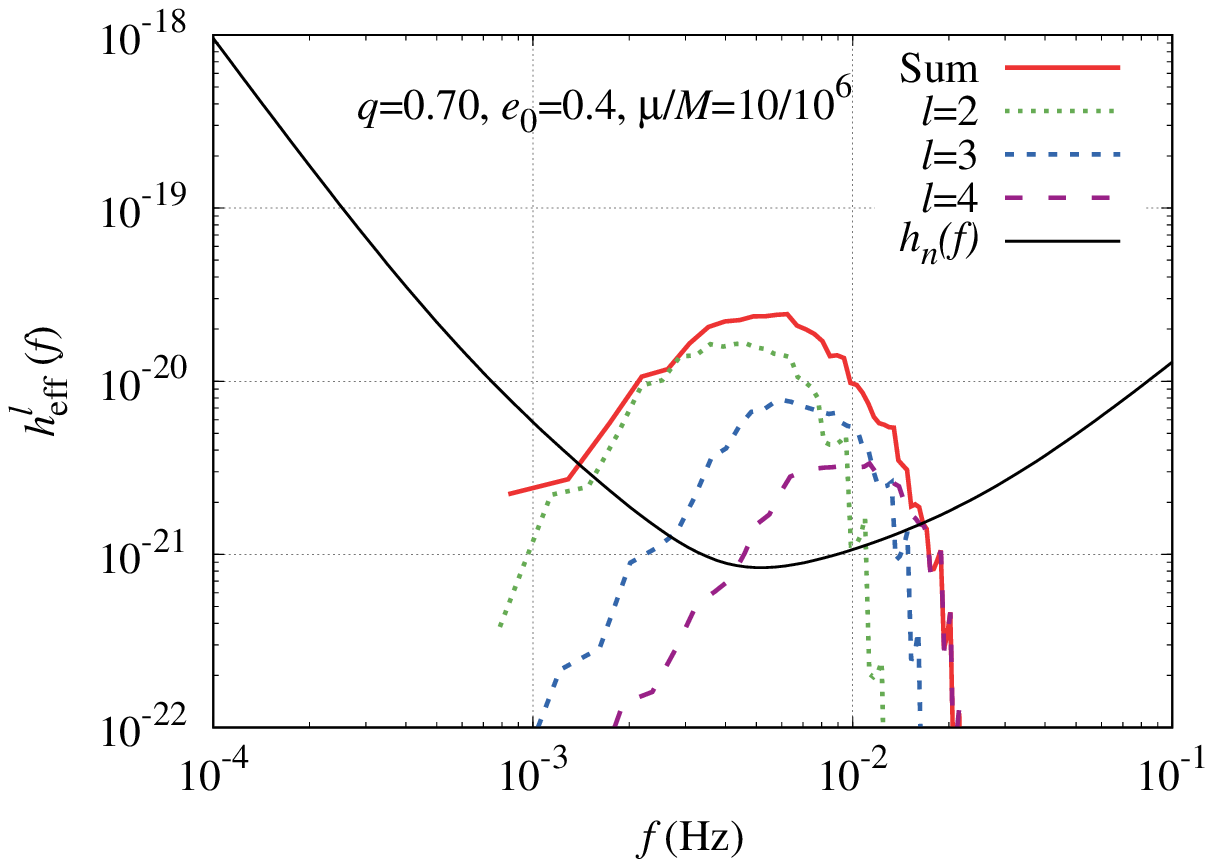}%
\includegraphics[width=59mm]{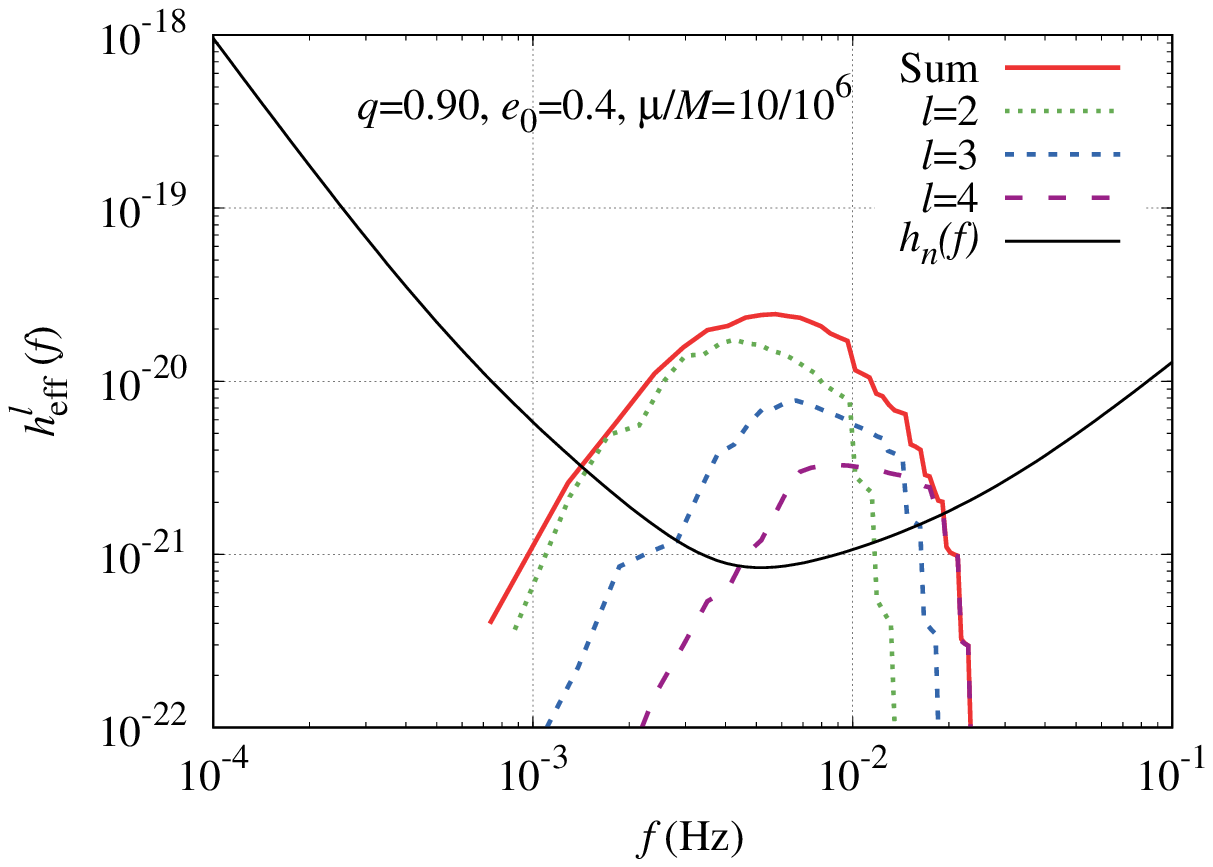}
\caption{The same as Fig.~\ref{fig:heff_q050_p11.4}, but for
  $e_0=0.4$, and $q=0$ (top left), $0.1$ (top middle), $0.3$ (top
  right), $0.5$ (bottom left), $0.7$ (bottom middle), and $0.9$
  (bottom right).  The values of $p_0$ take $11.38M$, $11.15M$,
  $10.71M$, $10.28M$, $9.83M$, and $9.41M$ for $q=0$, $0.1$, $0.3$,
  $0.5$, $0.7$, and $0.9$, respectively.  }
\label{fig:heff_p11.4_e0.4}
\end{figure*}

\begin{figure*}[t]
\includegraphics[width=59mm]{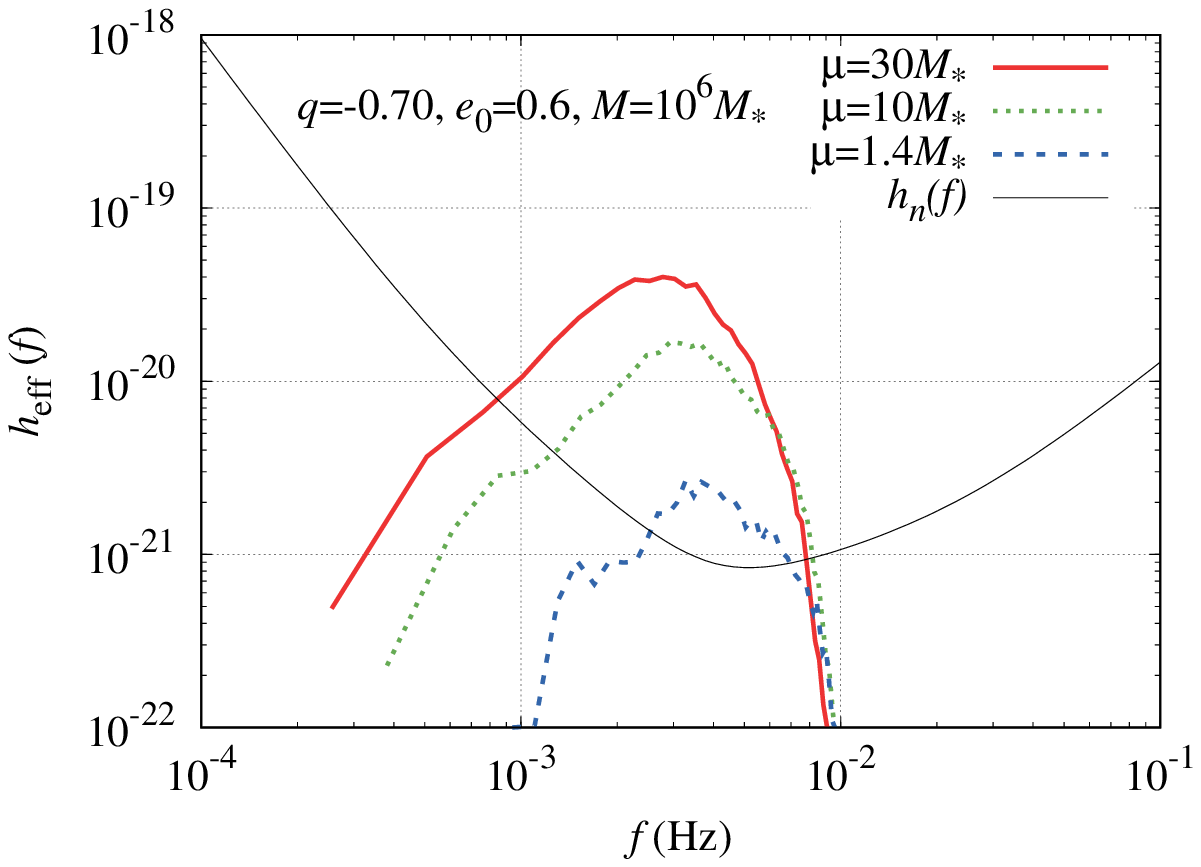}%
\includegraphics[width=59mm]{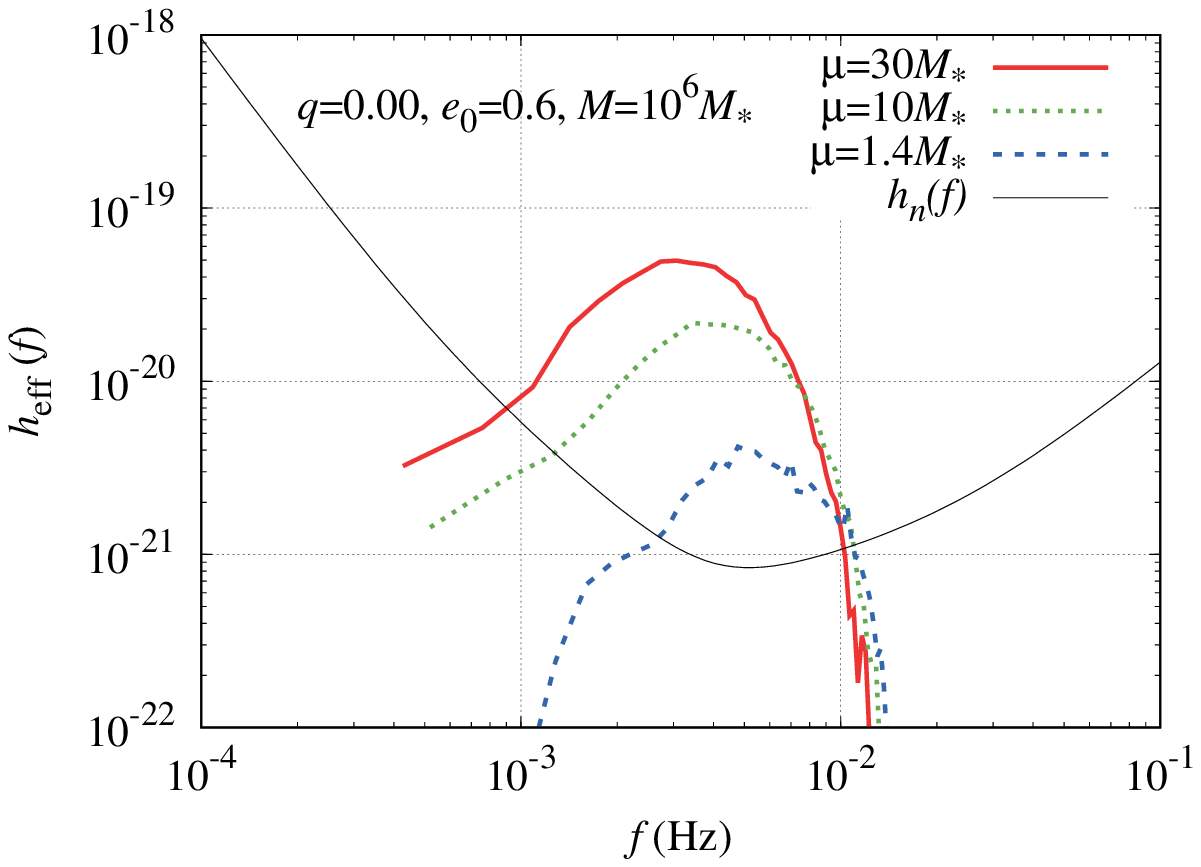}%
\includegraphics[width=59mm]{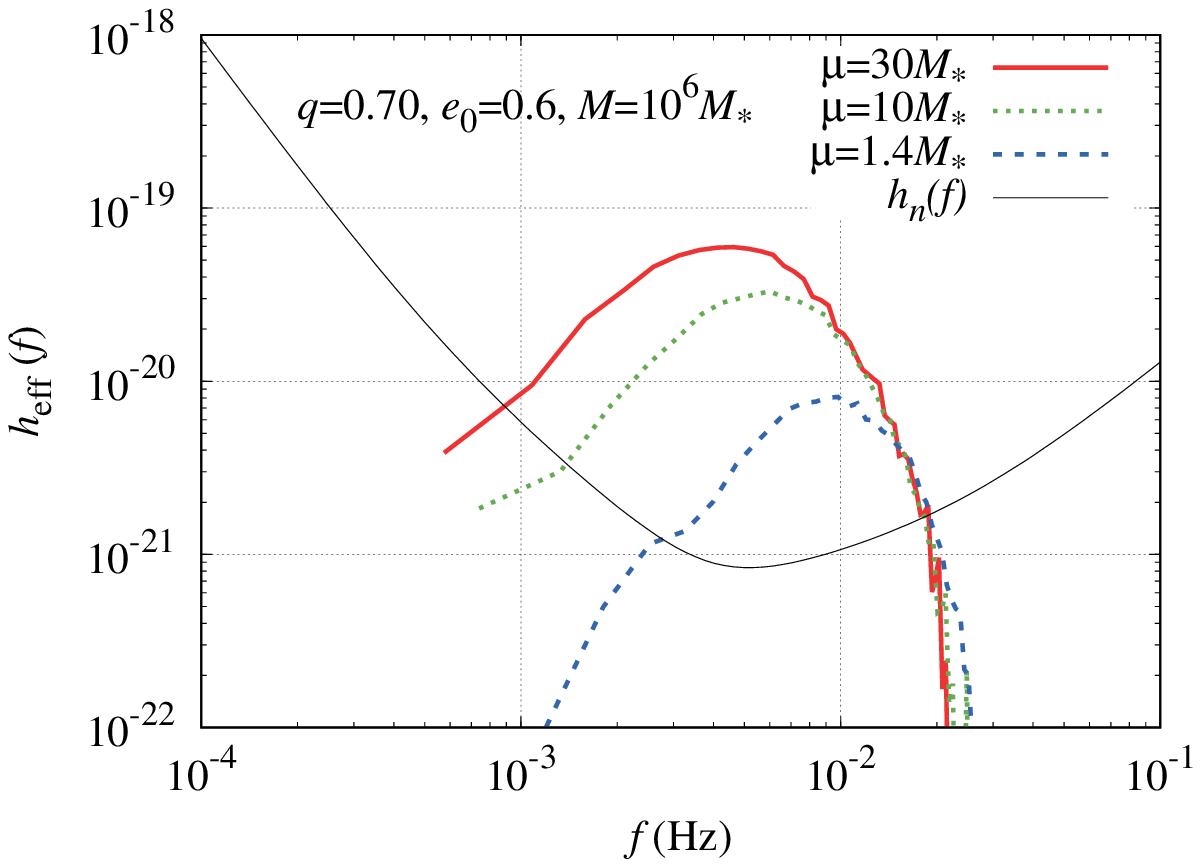}
\caption{ Power spectrum summed over the $\ell=2$--4 modes for a
  compact object of mass $\mu=(1.4,10,30)M_\odot$ inspiraling around an
  SMBH of $M=10^6M_\odot$ with $q=-0.7$ (left), $0$ (middle), and
  $0.7$ (right) at $D=1$\,Gpc during the last 3-year inspiral with
  $e_0=0.6$.  }
\label{fig:heff_M6}
\end{figure*}

\begin{figure*}[t]
\includegraphics[width=59mm]{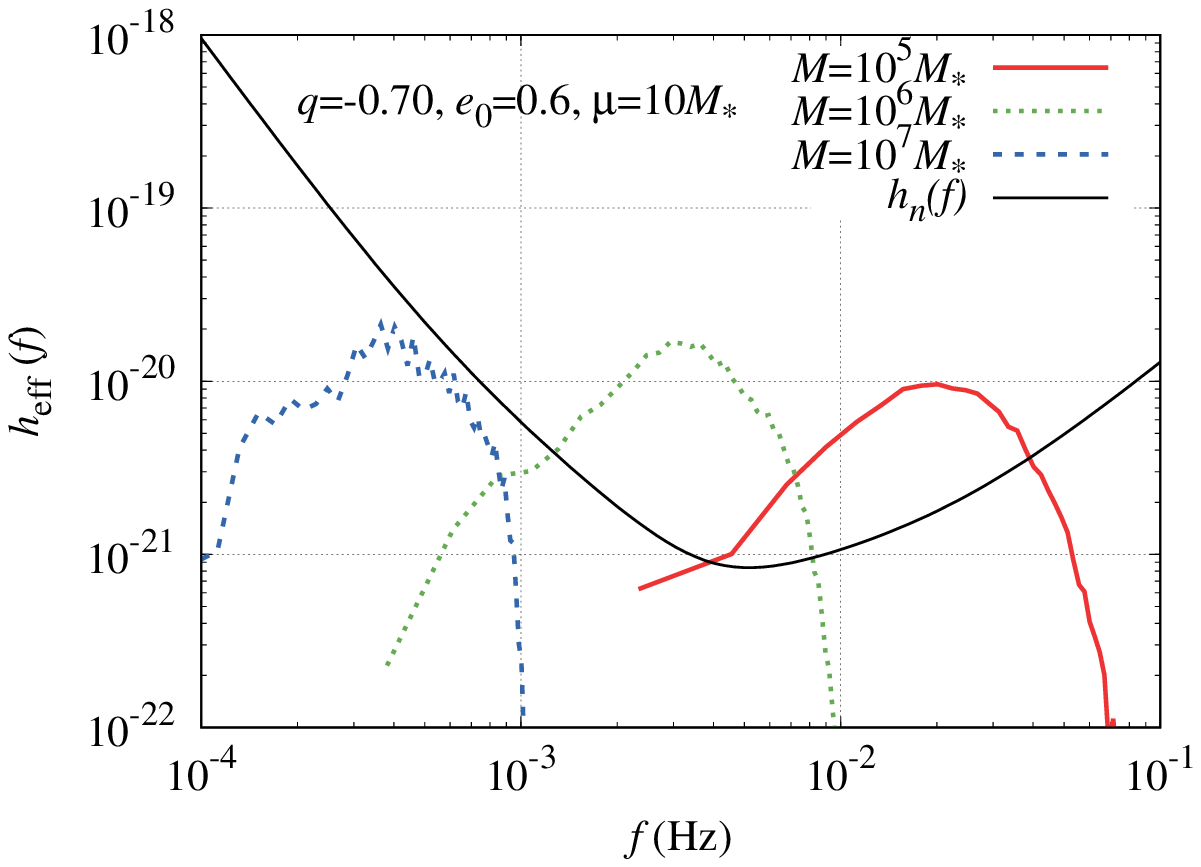}%
\includegraphics[width=59mm]{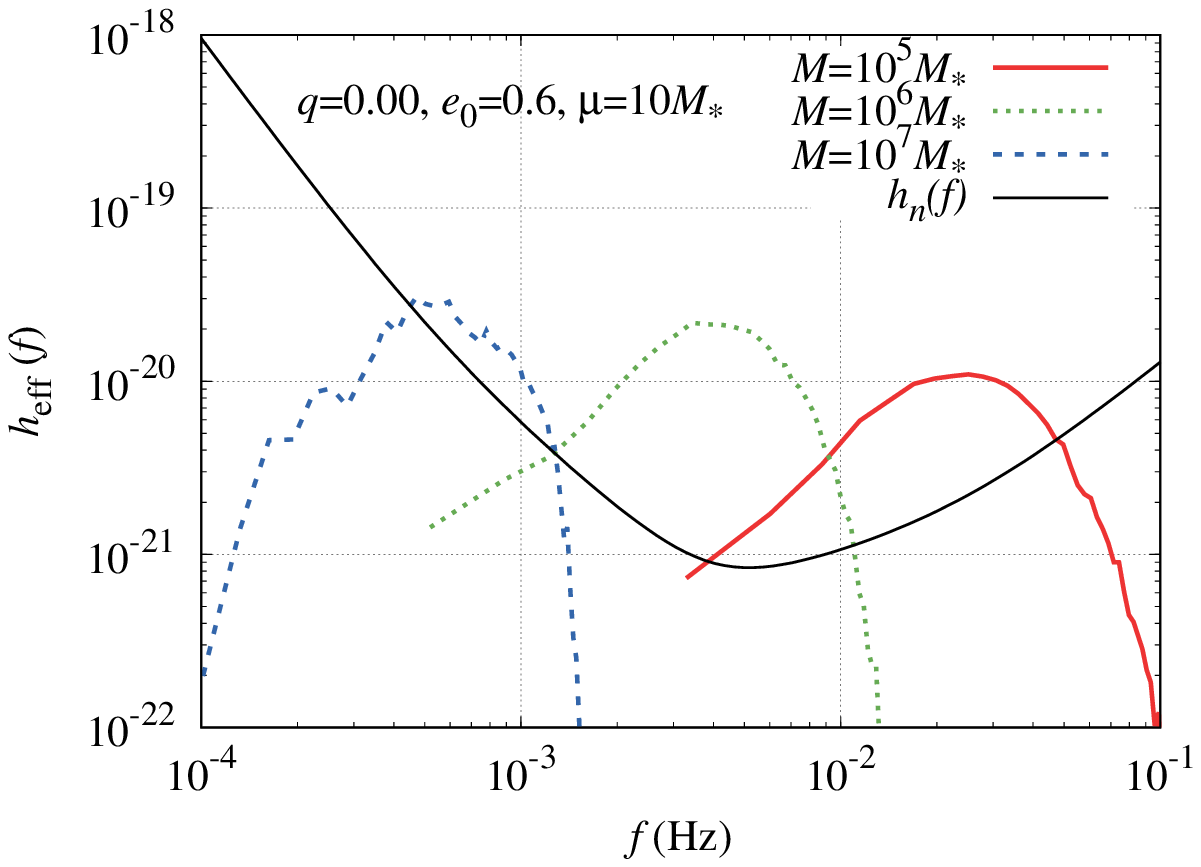}%
\includegraphics[width=59mm]{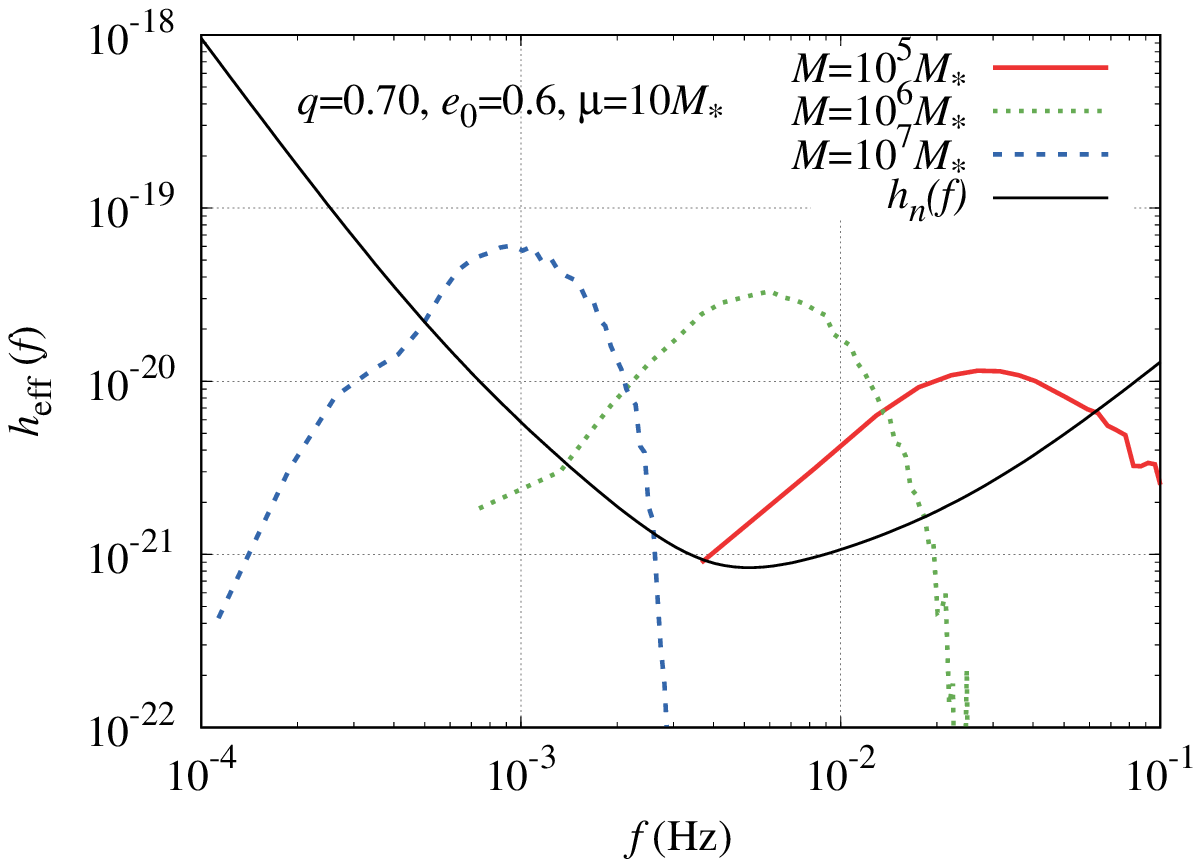}
\caption{
The same as Fig.~\ref{fig:heff_M6}, but for 
$\mu=10M_\odot$ and $M=(10^5,10^6,10^7)M_\odot$. 
}
\label{fig:heff_mu10}
\end{figure*}

\begin{figure*}[htbp]
\centering
\includegraphics[width=89mm]{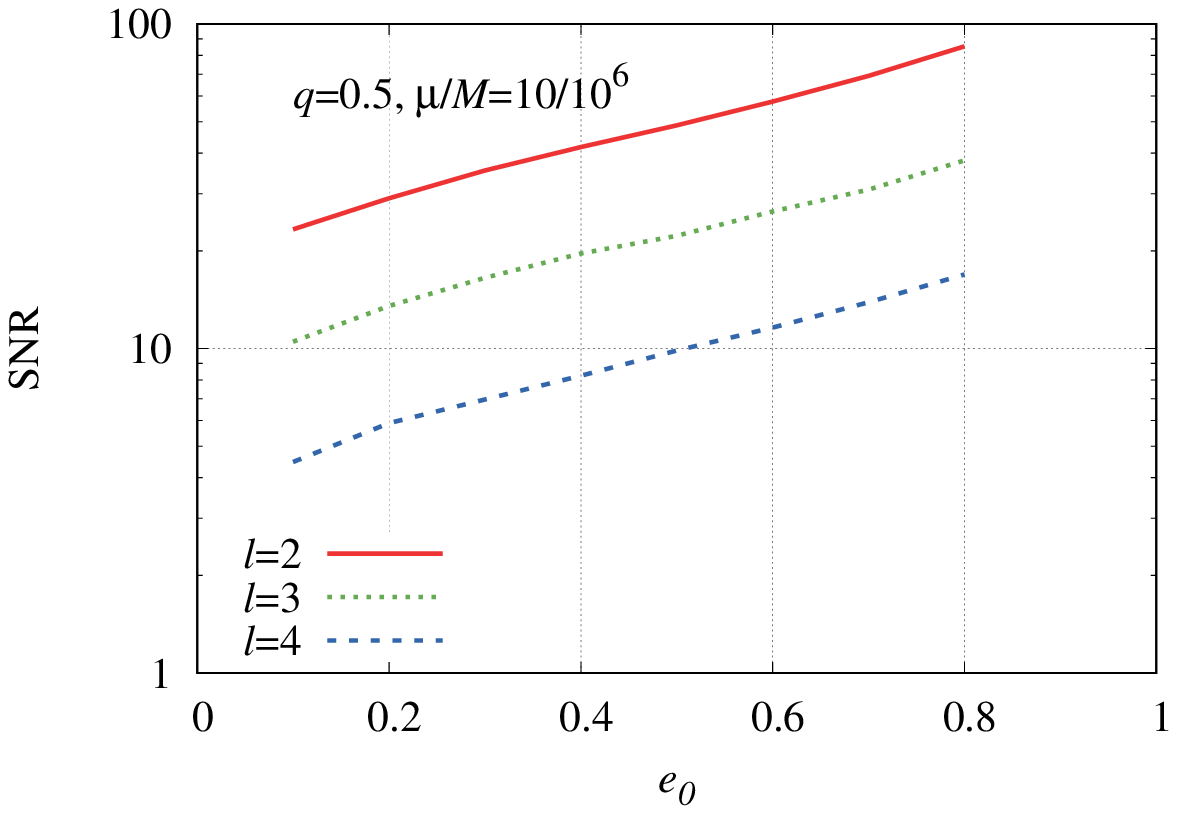}%
\includegraphics[width=89mm]{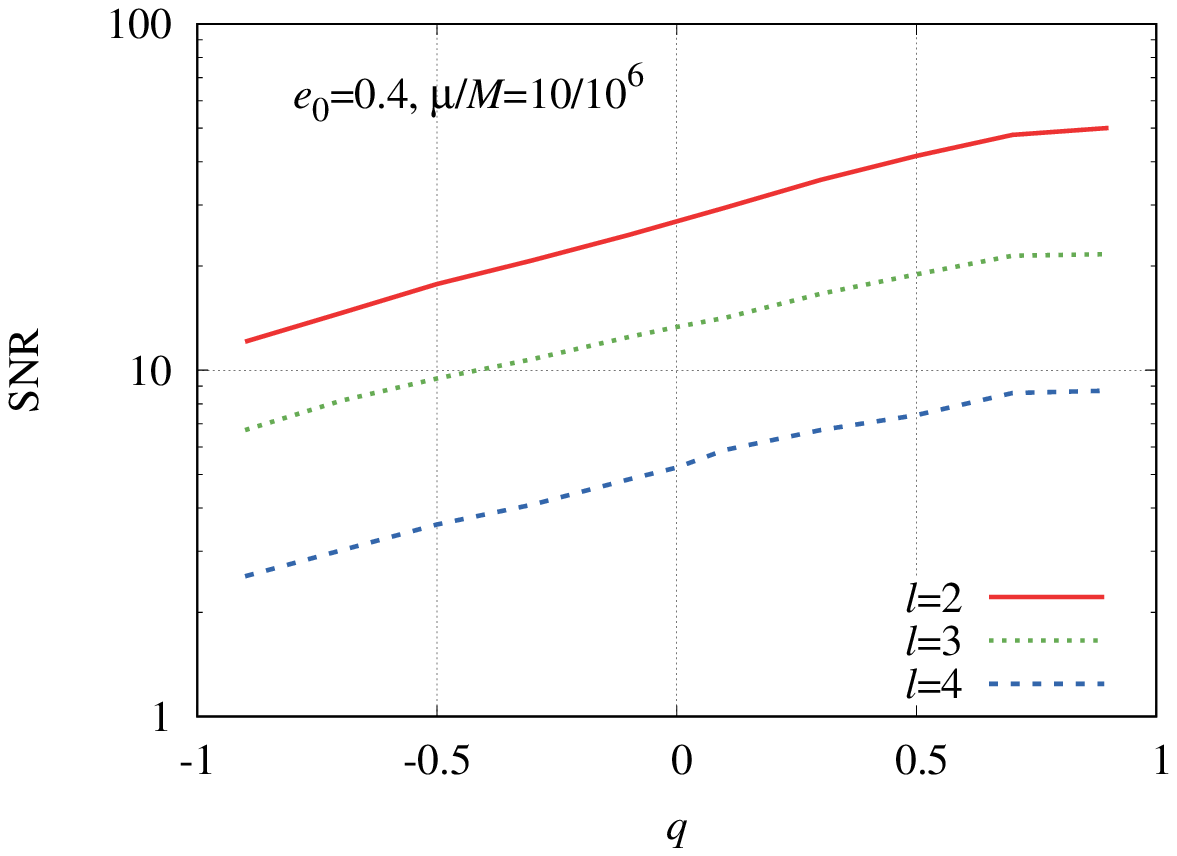}%
\caption{ Left: SNR associated with different $\ell$-modes from
  $\ell=2$ to $4$ as functions of $e_0$ for a $10M_\odot$ compact
  object inspiraling around a $10^6M_\odot$ SMBH of spin $q=0.5$
  at $D=1$\,Gpc for the last 3-year inspiral before plunge.  Right: 
  SNR as functions of $q$ for $M=10^6M_\odot$, $\mu=10M_\odot$, and
  $e_0=0.4$.  }
\label{fig:snr_q050_p11.4}
\end{figure*}

\begin{figure*}[t]
\includegraphics[width=59mm]{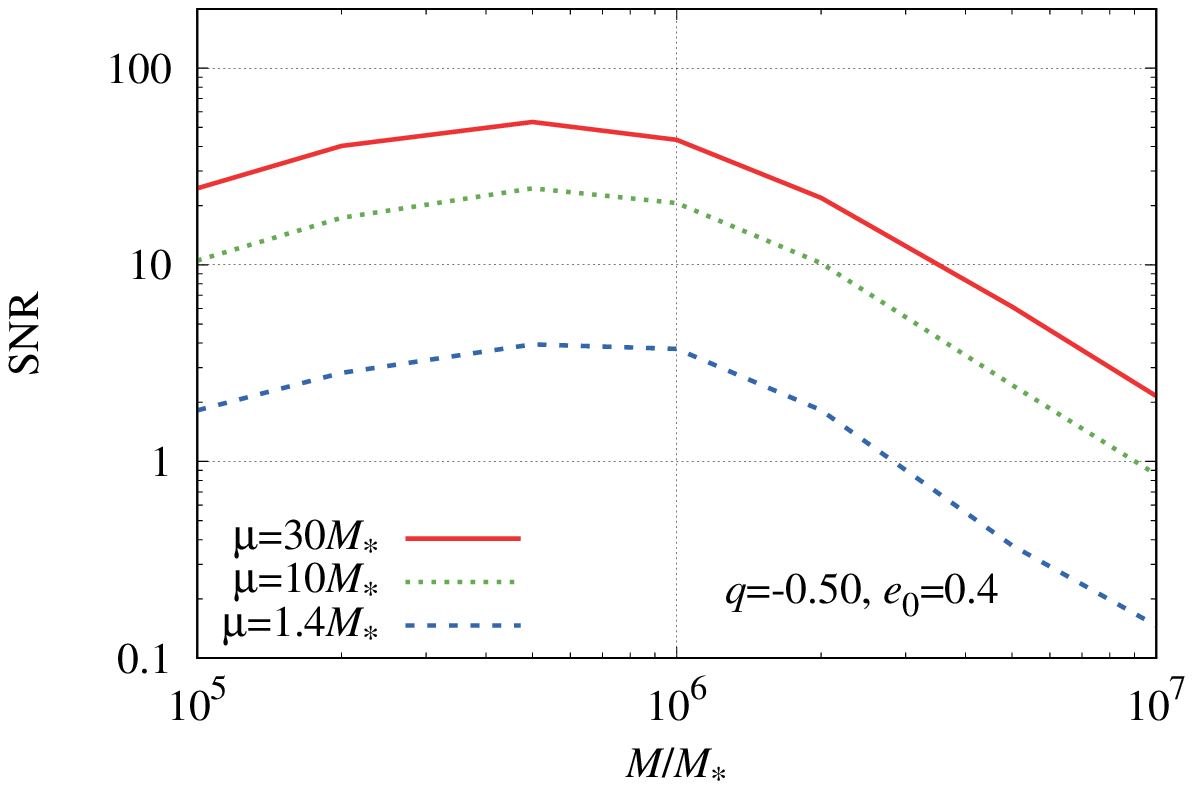}%
\includegraphics[width=59mm]{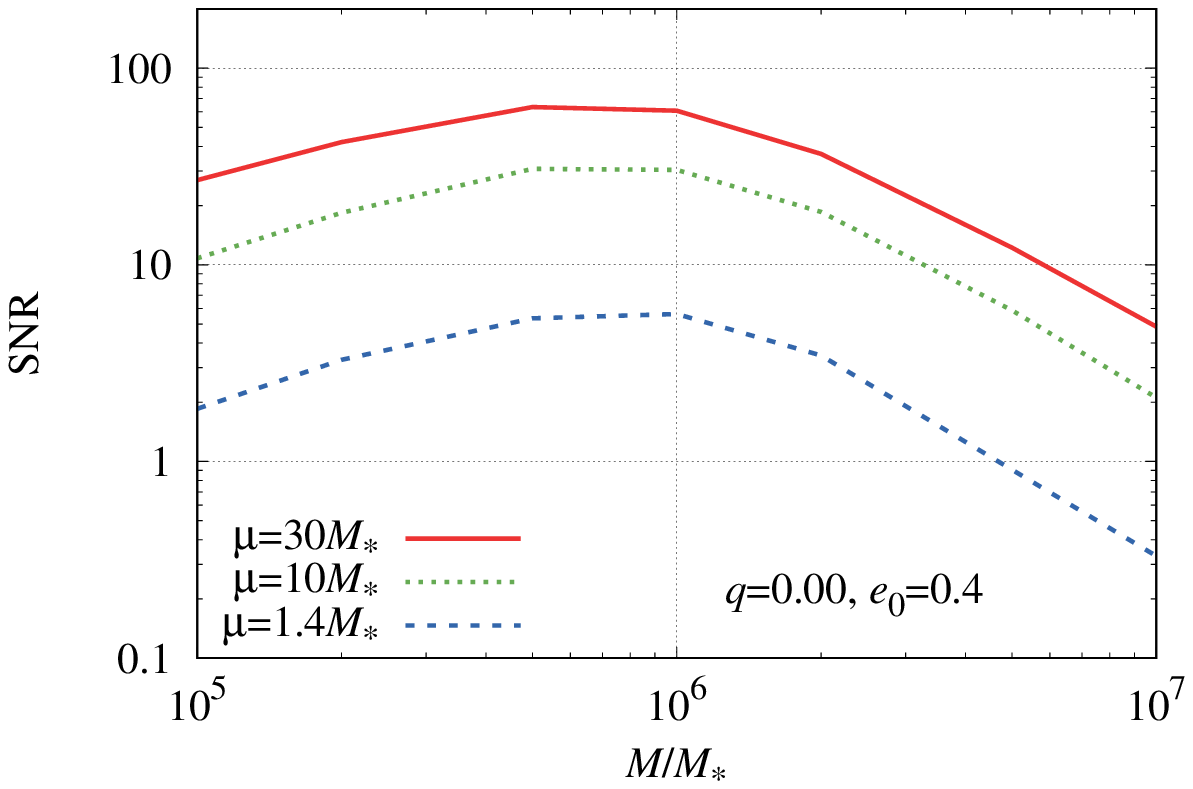}%
\includegraphics[width=59mm]{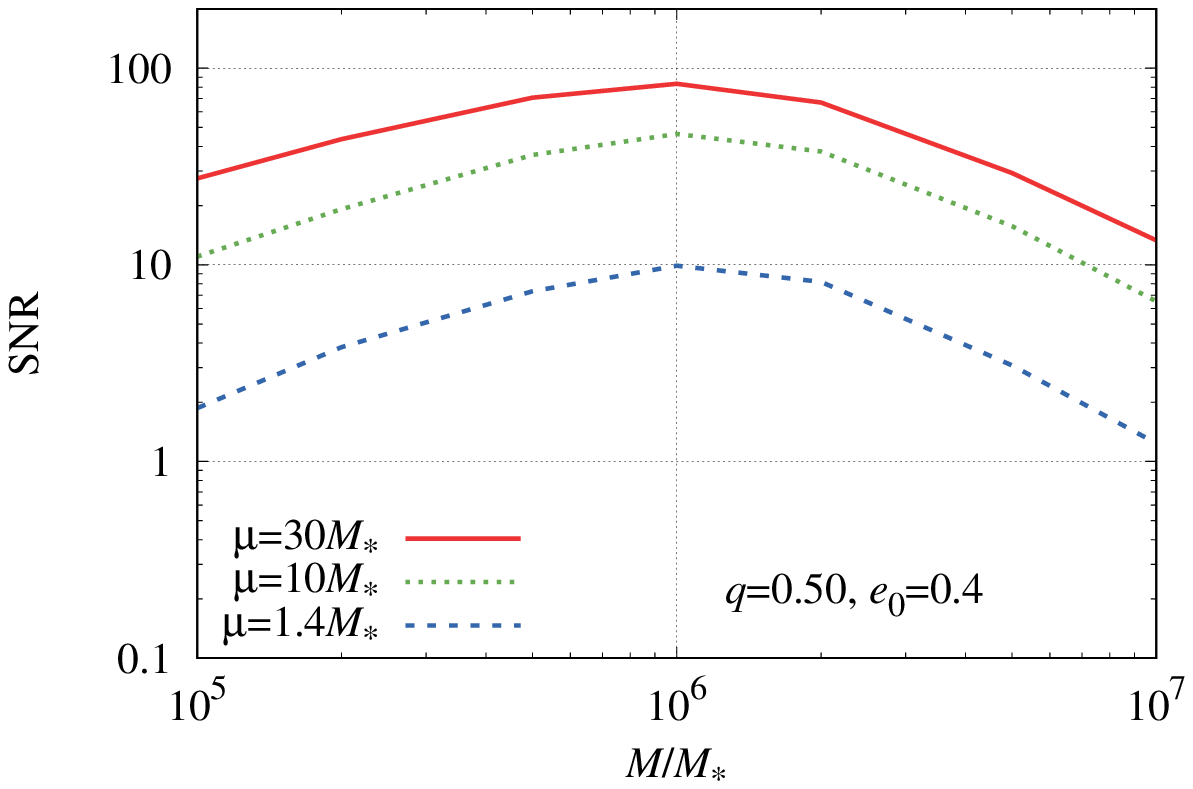}%
\caption{ SNR including the $\ell=2$--$4$ modes as a function of $M$
  for $q=-0.5$ (left), $0$ (middle), and $0.5$ (right) with $e_0=0.4$ 
  and $\mu=(30,10,1.4)M_\odot$ for the last 3-year inspiral
  before plunge.  }
\label{fig:snr_e0.6}
\end{figure*}

\begin{figure*}[t]%
\includegraphics[width=59mm]{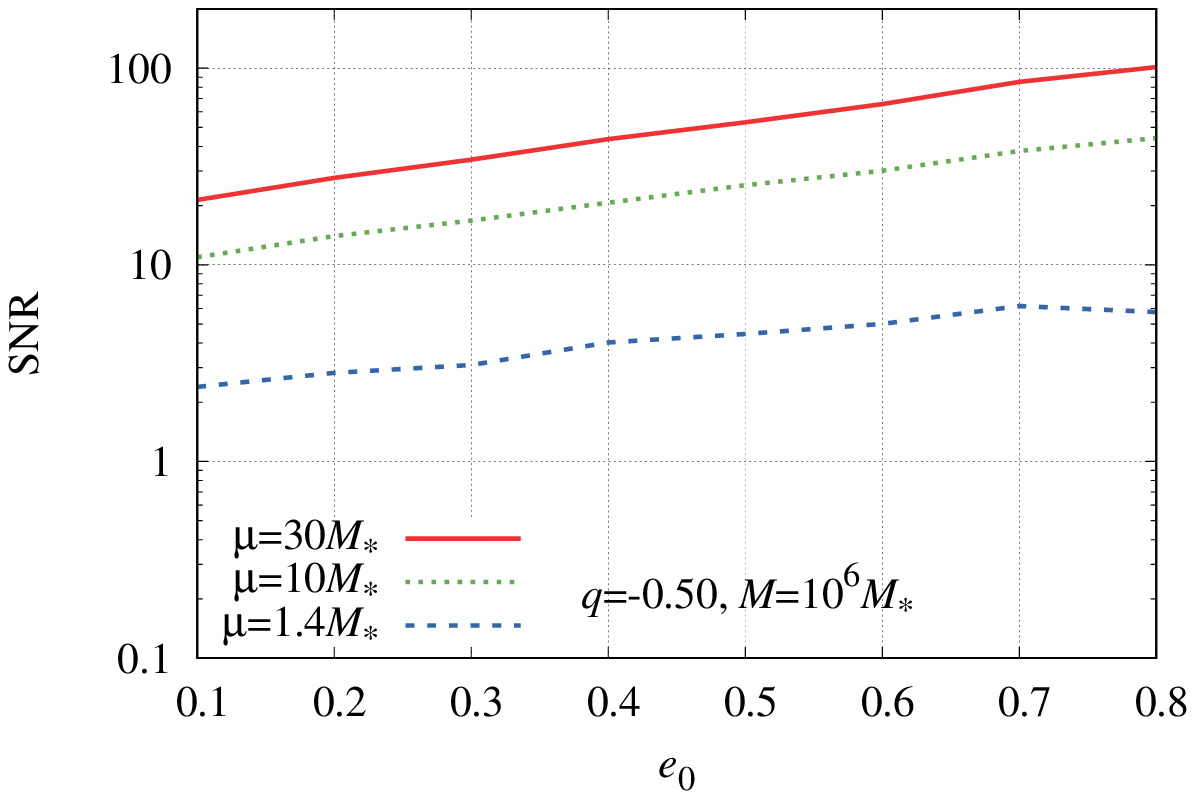}%
\includegraphics[width=59mm]{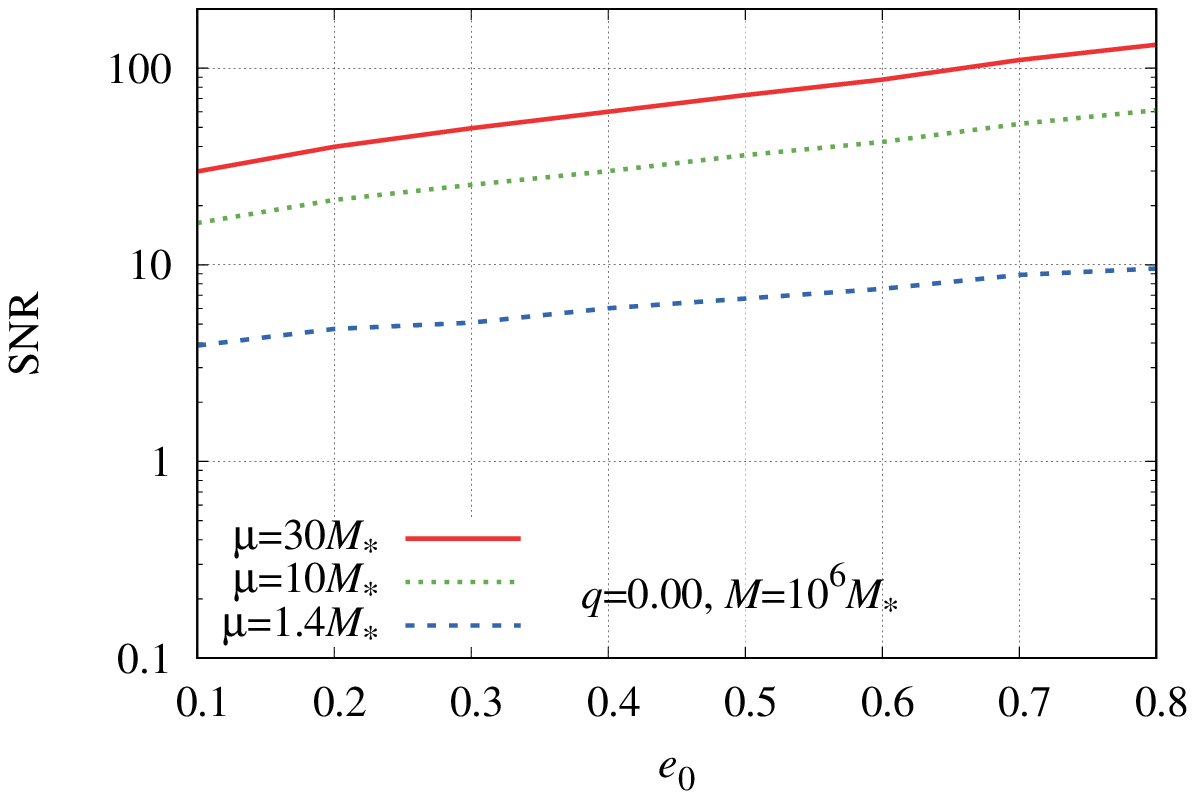}%
\includegraphics[width=59mm]{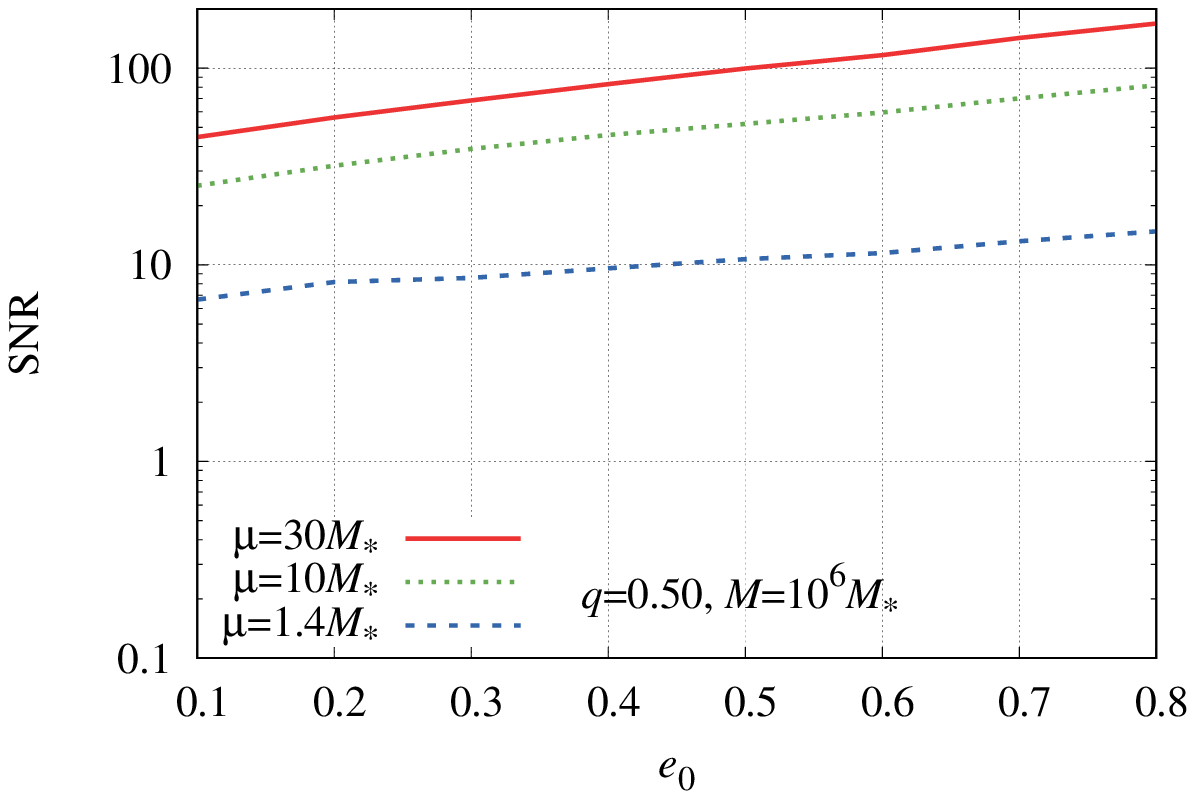}%
\caption{ SNR including the $\ell=2$--$4$ modes as a function of
  $e_0$ for $q=-0.5$ (left), $0$ (middle), and $0.5$ (right) with
  $M=10^6M_\odot$ and $\mu=(30,10,1.4)M_\odot$ for the last
  3-year inspiral before plunge.  }
\label{fig:snr_M6}
\end{figure*}

\begin{figure*}[th]%
\includegraphics[width=59mm]{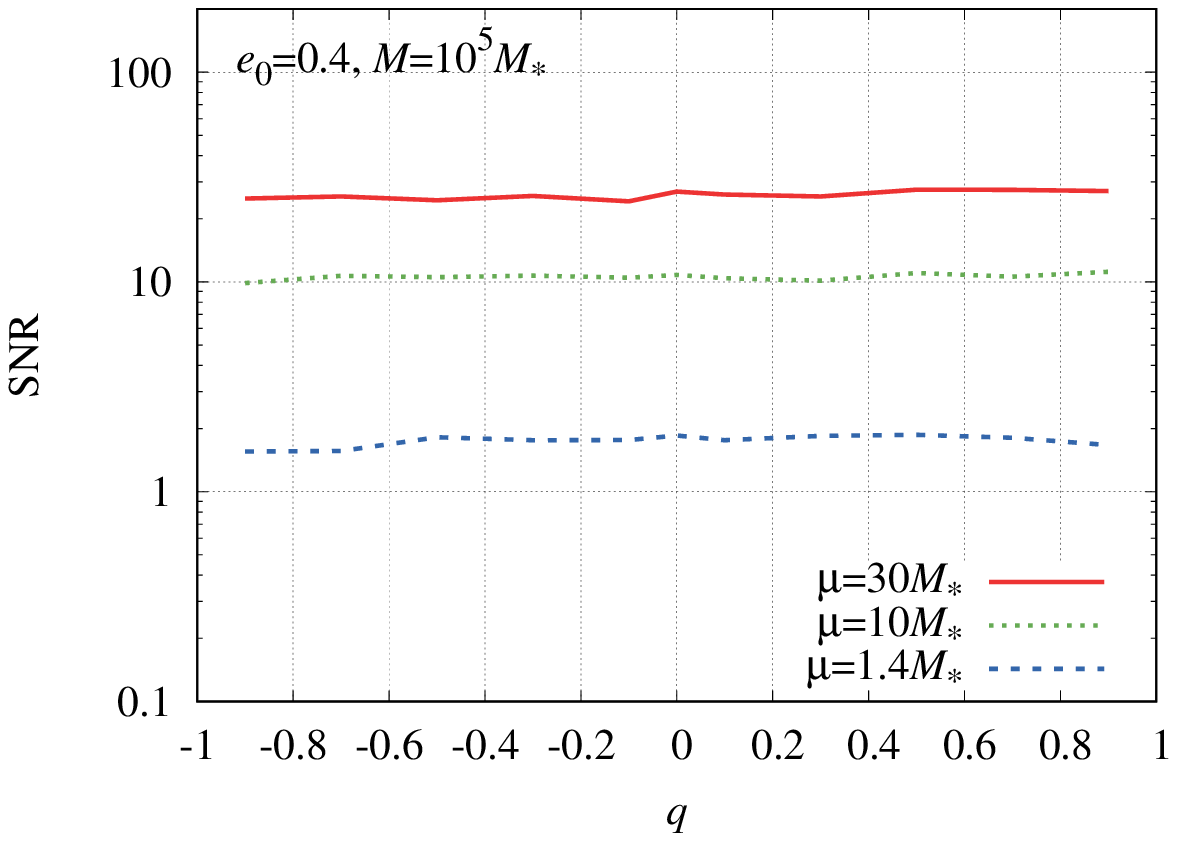}%
\includegraphics[width=59mm]{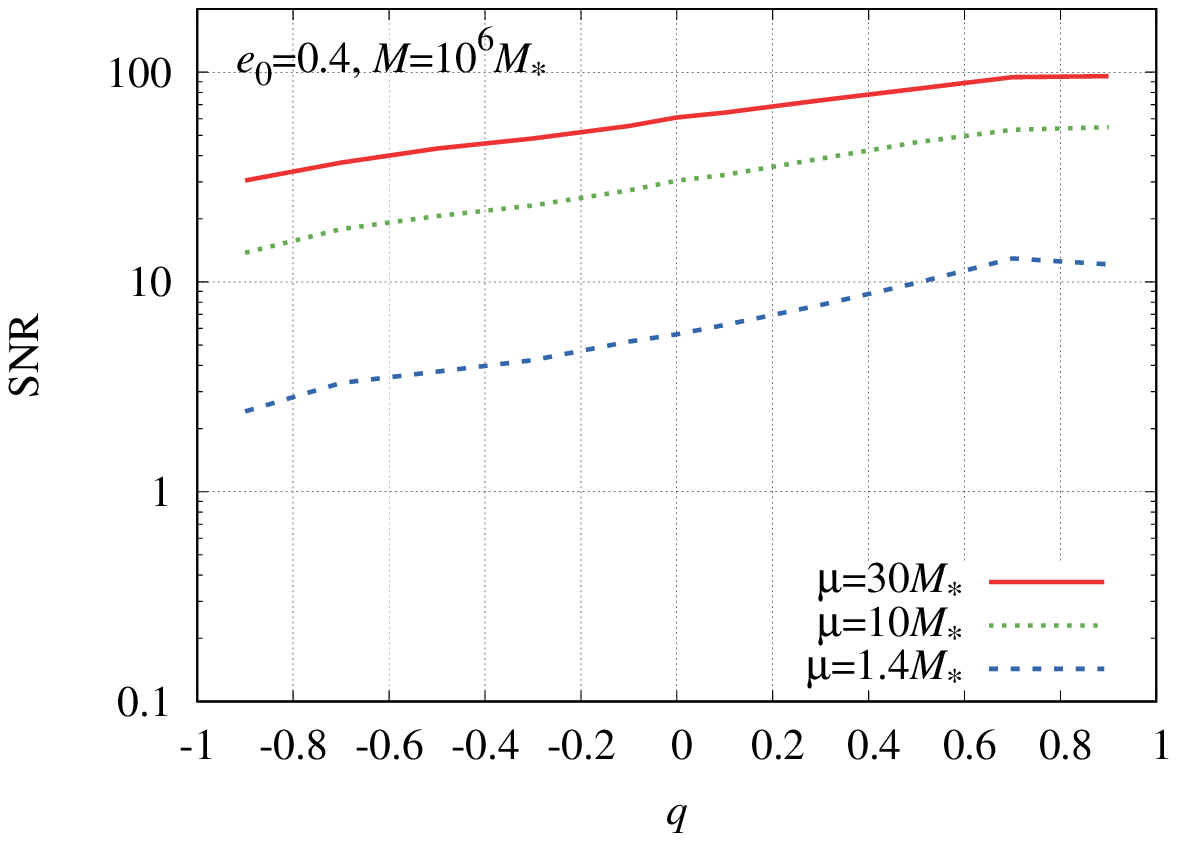}%
\includegraphics[width=59mm]{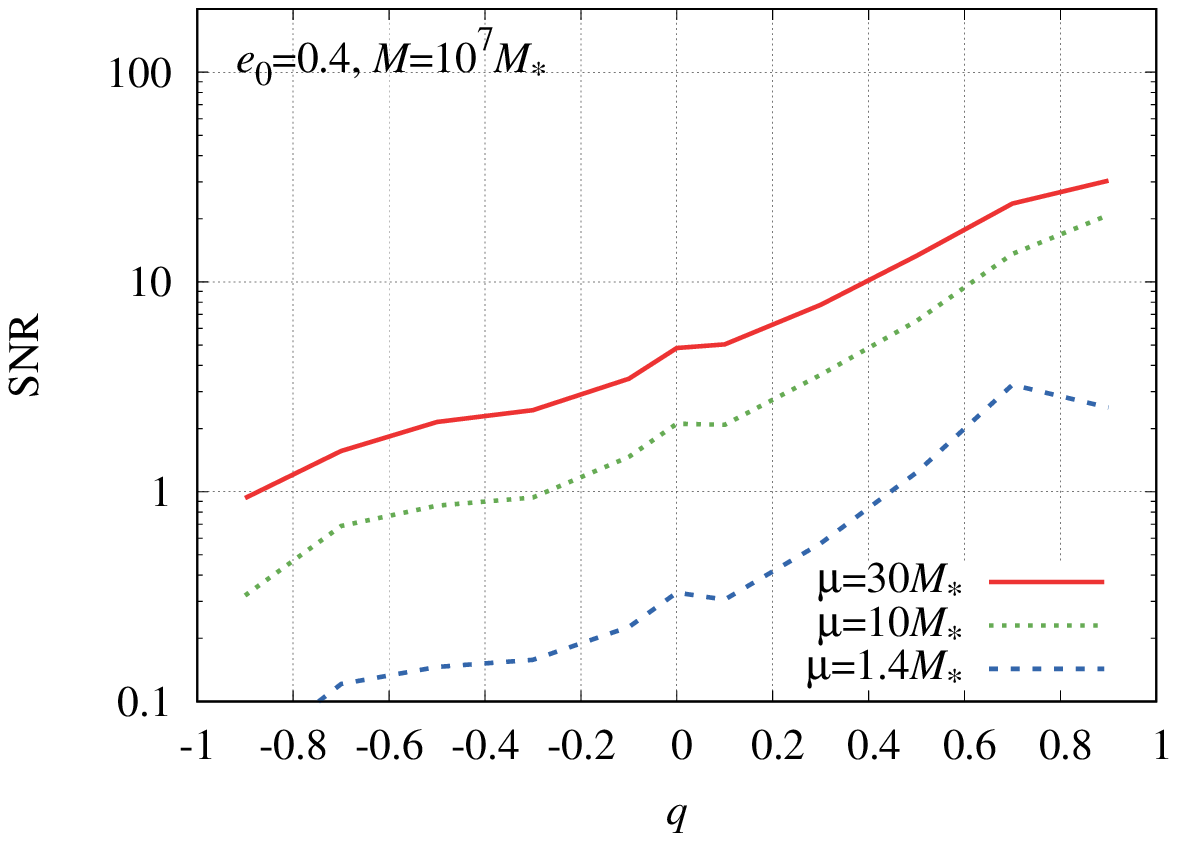}%
\caption{SNR including the $\ell=2$--$4$ modes as functions of $q$
for $M=10^5M_\odot$ (left), $10^6M_\odot$ (middle), and $10^7M_\odot$ (right) 
with $\mu=(30,10,1.4)M_\odot$ for the last 3-year inspiral before plunge.  
}
\label{fig:snr_q_e0.6M6Tobs3}
\end{figure*}

In Fig.~\ref{fig:heff_kludge}, we compare the power spectrum for
$\ell=2$ derived from our numerical results with those obtained by
kludge models~\cite{Barack:2003fp,Gair:2005ih,Babak:2006uv,Chua:2017ujo} as a
consistency check.  The power spectra for the kludge models are
obtained in the following manner. First, we compute the time domain
gravitational waveforms using the EMRI Kludge Suite in the Black Hole Perturbation Toolkit~\cite{BHPToolkit}. The public code enables us to compute the
inspiral orbits, the time domain waveforms, and the SNR for given
parameters such as $(q,p_0,e_0,\mu,M,D,T,\Delta t)$, where $T$ is the
duration of the waveform and $\Delta t$ is a time step.  The waveforms
are given by LISA response functions $h_I$ and $h_{II}$, which are
transformed from the waveform polarizations $h_+$ and $h_\times$ as
\begin{align}
h_I &=\frac{\sqrt{3}}{2}(F_I^{+}h_++F_I^\times h_\times),\cr   
h_{II} &=\frac{\sqrt{3}}{2}(F_{II}^{+}h_++F_{II}^\times h_\times),
\end{align}
where $F_I$ and $F_{II}$ are the antenna pattern
functions~\cite{Apostolatos:1994mx}.  We choose $\Delta t=63$\,s to
compute the inspiral orbits and the time domain waveforms, 
which are constructed from $\sqrt{h_I^2+h_{II}^2}$. We then
perform Fourier transformation for the time domain waveforms of the
kludge models into the frequency domain, and smooth
$h_\textrm{eff}(f)$ by using $100$ frequency bins. 

The power spectra are computed for a $10M_\odot$ compact object
inspiraling around a $10^6M_\odot$ SMBH of spin $q=0.5$ at $D=1$\,Gpc
during the last 3-year inspiral before plunge.  For this setting, the
inspiral starts from $p_0=10.3M$ with $e_0=0.2$.
We compute the power spectra $h_\textrm{eff}(f)$ for the
numerical kludge (NK) model~\cite{Gair:2005ih,Babak:2006uv} and the
augmented analytic kludge (AAK) model~\cite{Chua:2017ujo}.  Here, in
the NK, the orbital motion is determined by solving the geodesic
equation with the gravitational radiation reaction based on
post-Newtonian (PN) formulas~\cite{Blanchet:2013haa}, which are fitted
to the BH perturbation theory.  In the AAK, the orbital motion is
determined by solving the geodesic equation with the gravitational
radiation reaction based on PN formulas in the BH perturbation
theory~\cite{Sago:2015rpa}.  Gravitational waveforms are
determined from the resulting orbits using the quadrupole
formula~\cite{Peters:1963ux} both in the NK and AAK models.  

Figure ~\ref{fig:heff_kludge} shows that the power spectrum for
the $\ell=2$ mode derived in our calculation is closer to the one by
the NK model than that by the AAK model.  This is consistent because
the NK model should be more accurate than the AAK model.  In
Fig.~\ref{fig:heff_kludge}, the LISA's designed sky-averaged
sensitivity written in an analytic form~\cite{Babak:2017tow} is also
shown. It is found that for EMRIs with $\mu=10M_\odot$ together with an
SMBH of $M=10^6M_\odot$ and of $q=0.5$ at $D=1$\,Gpc, the SNR is of
the order of $10$ (more details on the SNR will be presented below).

The power spectra for $\ell=3$ and $4$ modes derived in our
calculation are also shown in Fig.~\ref{fig:heff_kludge}.  This
illustrates that the modes with $\ell=3$ and $4$ have the amplitude
approximately by 40\% and 20\% as large as that for the $\ell=2$
mode, respectively (see also Fig.~\ref{fig:snr_q050_p11.4} for the SNR
associated with the $\ell=3$ and $\ell=4$ modes).  This is reasonable
because the orbit which we consider here is very general relativistic,
and hence, the orbital velocity can be $\approx 50\%$ of the speed of
light, resulting in the enhancement of the higher multipole modes.
We note that the amplitudes for the $\ell\ge 5$ modes are 
less than 10\% of that for the $\ell=2$ mode. 
We ignore the amplitudes for the $\ell\ge 5$ modes because 
they are smaller than $h_{\textrm{n}}(f)$ for $e_0\lessapprox 0.8$.

To explore the dependence of the spectrum feature on the initial
orbital eccentricity and the BH spin, we generate
Figs.~\ref{fig:heff_q050_p11.4} and \ref{fig:heff_p11.4_e0.4}.  These
figures show the spectra of gravitational waves emitted by a
$10M_\odot$ compact object inspiraling around a $10^6M_\odot$ SMBH at
$D=1$\,Gpc for the last 3-year inspiral before plunge for a variety of
$q$ and $e_0$.  In Fig.~\ref{fig:heff_q050_p11.4}, the BH spin is
fixed to be $q=0.5$, while $e_0$ is varied from $0.2$ to $0.8$.  The
values of $p_0$ take $10.33M$, $10.28M$, $10.11M$, and $9.58M$ for
$e_0=0.2$, $0.4$, $0.6$, and $0.8$, respectively.  As the initial
eccentricity increases, the maximum frequency of gravitational waves
becomes higher, because with the large eccentricity, the minimum
value of $r_{\textrm{min}}$ is smaller resulting in the excitation of
the higher frequency modes (see Fig.~\ref{fig:orbit_pe_p0} and
Ref.~\cite{Barack:2003fp}): e.g., for $\ell=m=2$, the $n=0$ mode is
dominant for $e_0 \alt 0.1$, while the $n=0$ and $1$ modes are
equally dominant for $e_0\approx 0.3$ and the $n=2$ and $3$ modes
become dominant for $e_0\approx 0.7$. 

We find that the maximum value of $h_{\textrm{eff}}(f)$ 
increases as the value of $e_0$ increases (see also 
Fig.~\ref{fig:snr_q050_p11.4} for the SNR as a function of $e_0$). 
The reason for this is that $r_{\textrm{min}}$ decreases with the
increase of $e_0$, resulting possibly in the enhancement of the
gravitational-wave amplitude. 
Moreover, we need to sum over larger number of $n$-modes as the value of $e_0$ increases, 
and hence, the maximum value of $h_{\textrm{eff}}(f)$ at a peak frequency 
increases. Indeed, the power spectrum becomes broader in frequency 
as the value of $e_0$ increases. 

In Fig.~\ref{fig:heff_p11.4_e0.4}, the initial orbital eccentricity is
fixed to be $e_0=0.4$, while the BH spin is varied from $q=0$ to $0.9$.
The values of $p_0$ take $11.38M$, $11.15M$, $10.71M$, $10.28M$,
$9.83M$, and $9.41M$ for $q=0$, $0.1$, $0.3$, $0.5$, $0.7$, and $0.9$,
respectively.  The frequency of gravitational waves at plunge becomes
higher as the BH spin increases because the minimum value of
$r_{\textrm{min}}$ becomes smaller (see Fig.~\ref{fig:orbit_pe_e0}).
We also note that the maximum value of the power spectrum becomes
larger for the larger BH spin (see Fig.~\ref{fig:snr_q050_p11.4} for
the SNR as a function of $q$).  The reason for this is that for the
larger BH spin, the more compact orbits with smaller values of
$r_{\textrm{min}}$ is allowed, and for such orbits, gravitational
waves of the high amplitude can be emitted due to the more
relativistic motion.

Figure~\ref{fig:heff_M6} shows the power spectra for a compact object
of mass $\mu=(1.4,10,30)M_\odot$ inspiraling around a SMBH of mass
$10^6M_\odot$ at $D=1$\,Gpc during the last 3-year inspiral with
$e_0=0.6$.  The BH spin is varied from $q=-0.7$ to $0.7$.  For $q=0.7$
($-0.7$), the values of $p_0$ take $6.3M$ ($10.8M$), $9.6M$ ($12.9M$),
and $12.5M$ ($15.2M$) for $\mu=1.4M_\odot$, $10M_\odot$, and
$30M_\odot$, respectively.  Both the frequency of gravitational waves
and the power spectra become higher as the BH spin increases.
The power spectra increase as $\mu$ increases, and 
the maximum amplitudes are approximately proportional to $\sqrt{\mu}$.

In Fig.~\ref{fig:heff_mu10}, the spectra are shown for a $10M_\odot$
compact object inspiraling around an SMBH of mass
$(10^5,10^6,10^7)M_\odot$ at $D=1$\,Gpc during the last 3-year inspiral
with $e_0=0.6$.  The BH spin is again varied from $q=-0.7$ to $0.7$.  For
$q=0.7$ ($-0.7$), the values of $p_0$ are $16.8M$ ($18.9M$), $9.6M$
($12.9M$), and $6.0M$ ($10.6M$) for $M=10^5M_\odot$, $10^6M_\odot$,
and $10^7M_\odot$, respectively.  The maximum amplitude of the spectra
increases as the BH mass increases because $p_0/M$ becomes smaller and
thus the inspiral orbits are in more highly general relativistic
regions for a longer duration (see Fig.~\ref{fig:cycle}). The
frequency of gravitational waves at plunge becomes higher as the BH
mass decreases and the BH spin increases. As a result, gravitational
waves from a compact object around a $10^7M_\odot$ SMBH with $q<0$ are
not well in the LISA sensitivity band.  In addition, only
gravitational waves in an early part of the inspiral of a compact
object into a $10^5M_\odot$ SMBH is above the LISA sensitivity curve.
However, for larger values of $e_0$, the low-frequency tail of
gravitational waves (due to the contribution of low-$n$ modes) is 
above the LISA sensitivity curve and gravitational waves will be
detectable by LISA irrespective of $q$ for $M \approx 10^5M_\odot$
(see also Fig.~\ref{fig:snr_q_e0.6M6Tobs3}).

In the left panel of Fig.~\ref{fig:snr_q050_p11.4} we show the SNR of
gravitational waves with respect to the LISA sensitivity curve for
$\ell=2$--4 modes for a $10M_\odot$ compact object inspiraling around
a $10^6M_\odot$ SMBH at $D=1$\,Gpc during the last 3-year inspiral
before plunge. The SNR is plotted as a function of $e_0$ for $q=0.5$.
It is found that the SNR increases as $e_0$ increases 
because larger number of the $n$-modes could contribute to the SNR. 
The SNR for the $\ell=2$ mode with $e_0=0.8$ is $\approx 80$, 
which is about 4 times larger than that with $e_0=0.1$, $\approx 20$.
Thus, for $M = 10^6M_\odot$, highly eccentric EMRIs could dominate 
the detection by LISA. 
The right panel of
Fig.~\ref{fig:snr_q050_p11.4} shows the SNR as a function of $q$ with
$e_0=0.4$.  As the BH spin increases, the SNR is significantly
increased because the value of $r_{\textrm{min}}$ near the separatrix
decreases and general relativistic effects are enhanced. For example,
the SNR for the $\ell=2$ mode with $q=-0.9$ is $\approx 12$ which is
about $24\%$ of the one with $q=0.9$, $\approx 50$. This indicates
that rapidly spinning SMBHs could be more subject to the detection by
LISA.  However, this is the special feature for $M \agt 10^6M_\odot$
(see also Fig.~\ref{fig:snr_q_e0.6M6Tobs3}).

We also note that the SNR for the $\ell=3$ and $4$ modes is about
$40\%$ and $20\%$ of that for the $\ell=2$ mode, respectively. Thus, the
detection rate with a template that includes up to the $\ell=3$ and
$4$ modes of gravitational waves becomes $1.4^3\approx 2.7$ and
  $1.6^3\approx 4.1$ times larger than that using only the $\ell=2$
mode, respectively. Obviously, it is crucially important to take into
account the high-multipole modes in the waveform modeling. 

Figures~\ref{fig:snr_e0.6}--\ref{fig:snr_q_e0.6M6Tobs3} show the SNR
of gravitational waves including the $\ell=2$--$4$ modes with respect
to the LISA designed sensitivity curve for a compact object of mass
$\mu$ into an SMBH of mass $M$ at $D=1$\,Gpc during the last 3-year
inspiral before plunge.  In Fig.~\ref{fig:snr_e0.6}, the SNR is shown
as a function of $M$ for $\mu=(1.4,10,30)M_\odot$ with $e_0=0.4$, and
$q=-0.5$ (left), $0$ (middle), and $0.5$ (right).  It is found
  that the SNR is largest for $M\sim 10^6M_\odot$ irrespective of $q$
  and $\mu$, reflecting the sensitivity curve of LISA.  

Figure~\ref{fig:snr_M6} shows the SNR as a function of $e_0$ 
for $M=10^6M_\odot$ with $q=-0.5$ (left), $0$ (middle), and $0.5$ (right). 
As illustrated in Fig.~\ref{fig:snr_q050_p11.4}, 
the SNR is a monotonically increasing function of $e_0$, 
that increases by a factor of several for the change from $e_0=0.1$ to 0.8 with $M=10^6M_\odot$. 
This indicates that highly eccentric EMRIs for this SMBH mass would 
increase the detection rate in the LISA observation by a factor of several.

Figure~\ref{fig:snr_q_e0.6M6Tobs3} shows the SNR as a function of $q$
for $e_0=0.4$ with $M=10^5M_\odot$ (left), $10^6M_\odot$ (middle), and
$10^7M_\odot$ (right).  The SNR increases as the BH spin increases
for $M\agt 10^6M_\odot$ (see Fig.~\ref{fig:snr_q050_p11.4}), but
the SNR for $M=10^5M_\odot$ depends weakly on $q$ because the late part
of the inspirals can be below the LISA frequency band  for
  larger values of $q$ (see Fig.~\ref{fig:heff_mu10}).  As the
  right panel of Fig.~\ref{fig:snr_q_e0.6M6Tobs3} illustrates, the
  detection rate of the EMRIs from an SMBH of $M\sim 10^7M_\odot$
  depends strongly on the BH spin: for this SMBH mass, a higher spin
  BH will be much more frequently detected.  

\subsection{Limitation of post-Newtonian formulas}
\label{sec:pn_err}
\begin{figure*}[t]
\centering 
\includegraphics[width=59mm]{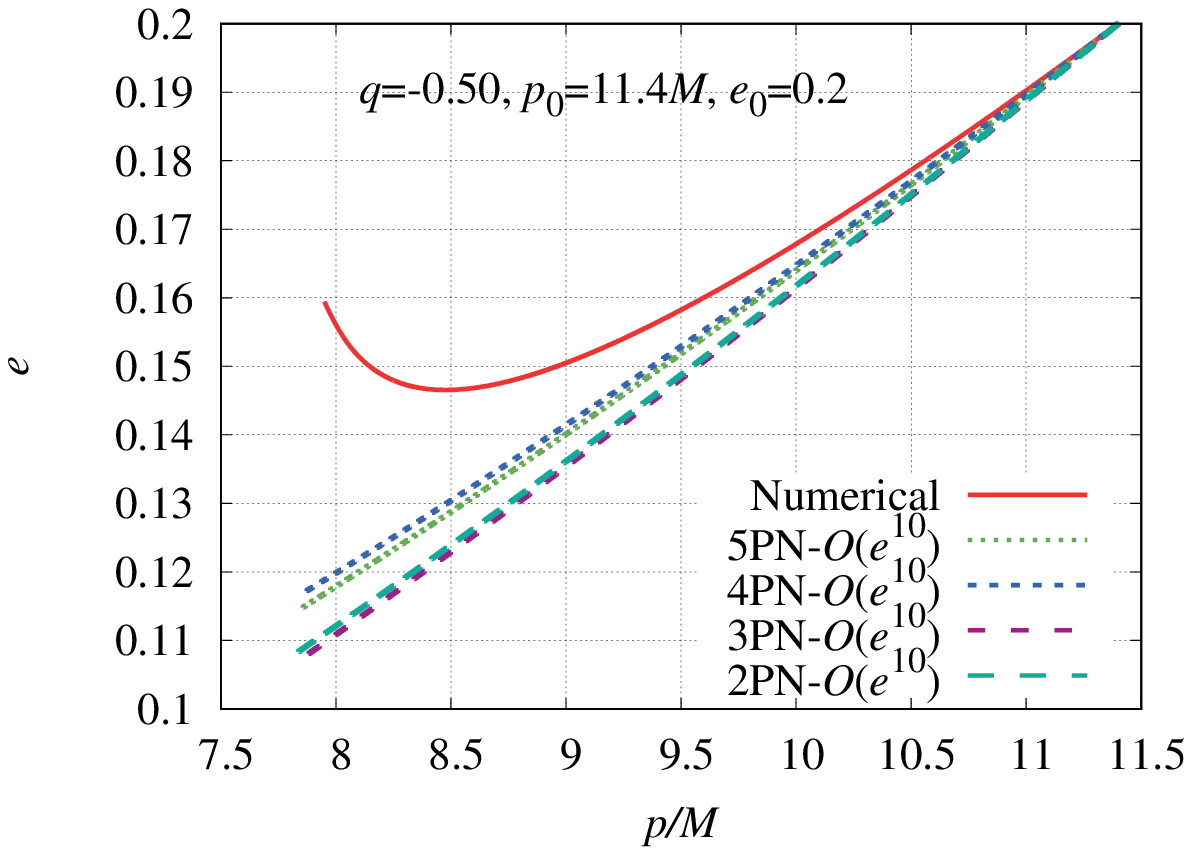} %
\includegraphics[width=59mm]{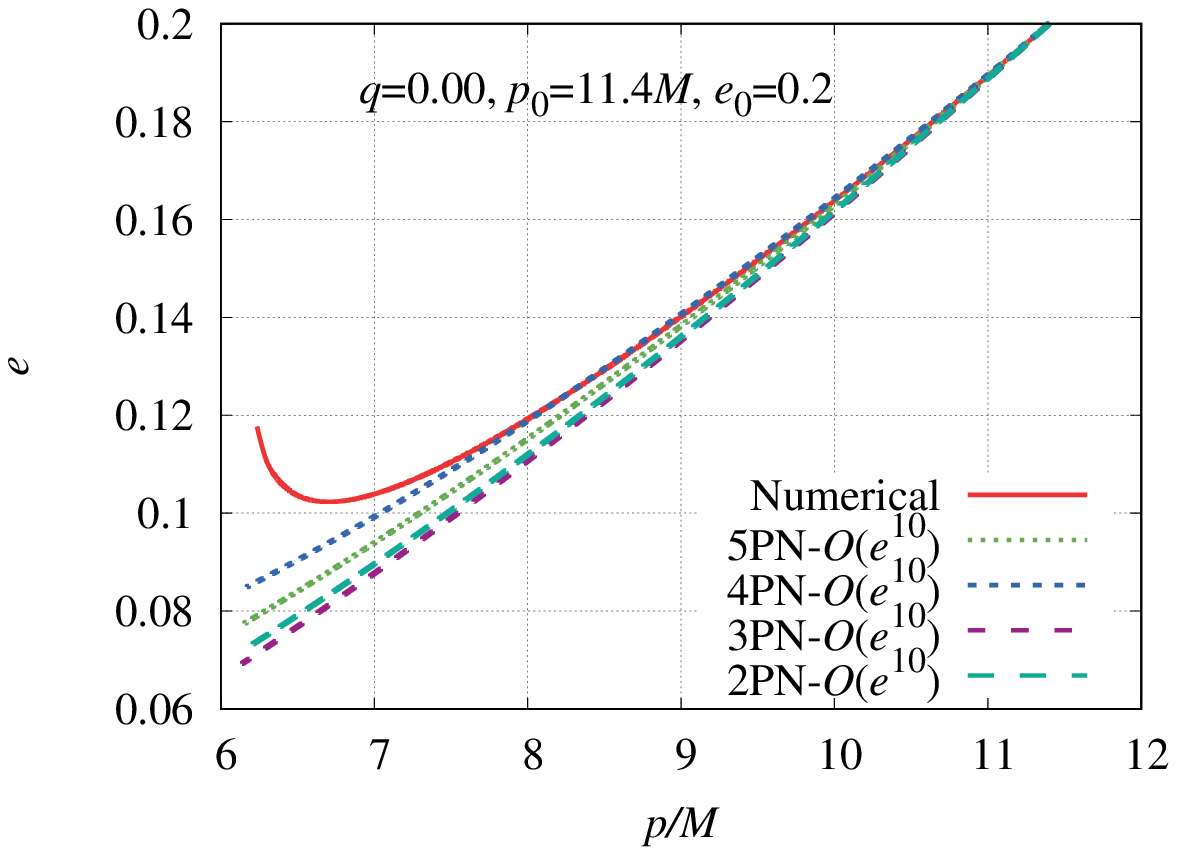}%
\includegraphics[width=59mm]{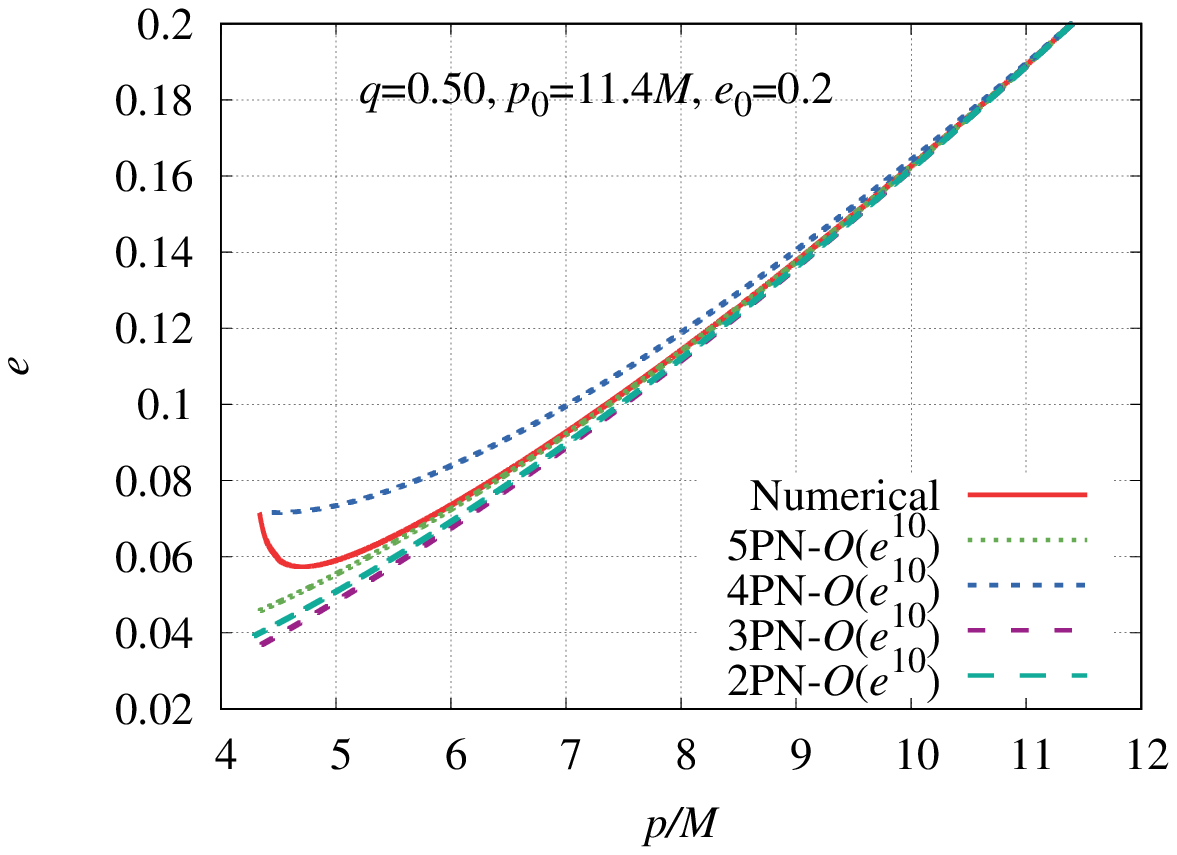}\\
\includegraphics[width=59mm]{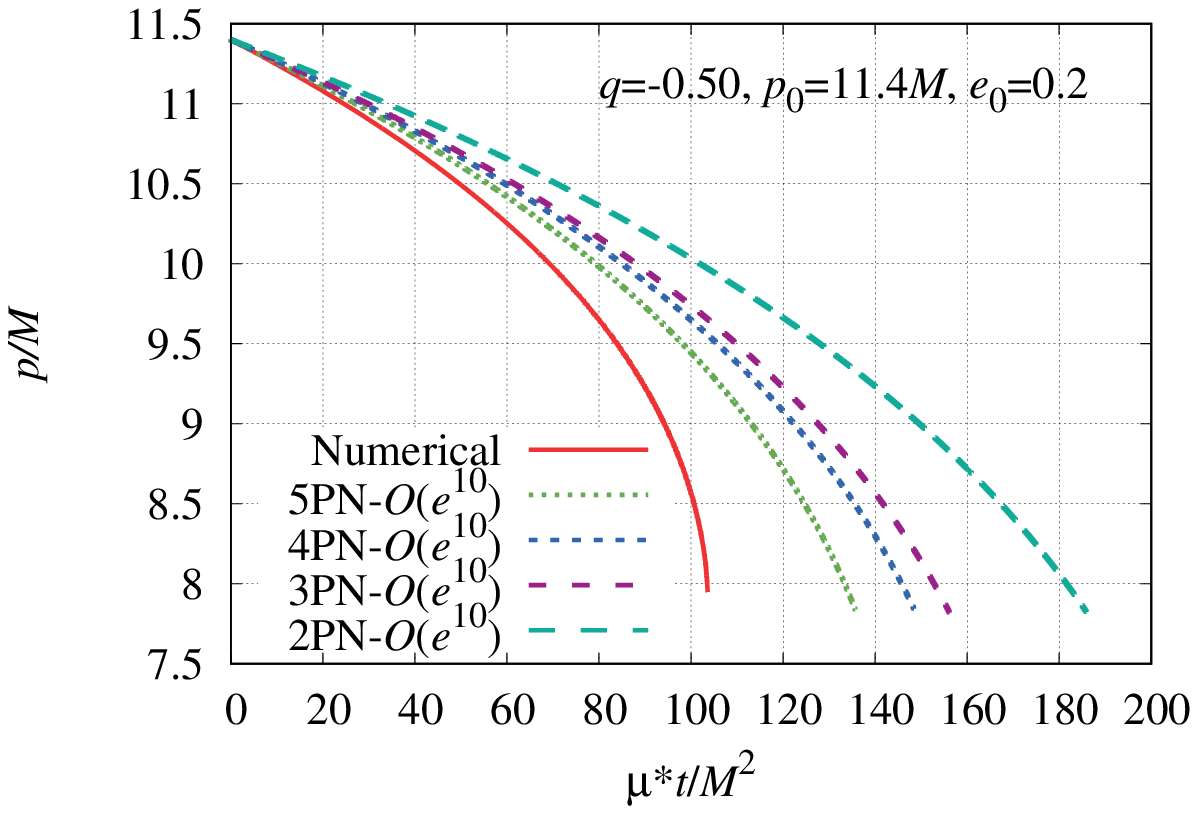} %
\includegraphics[width=59mm]{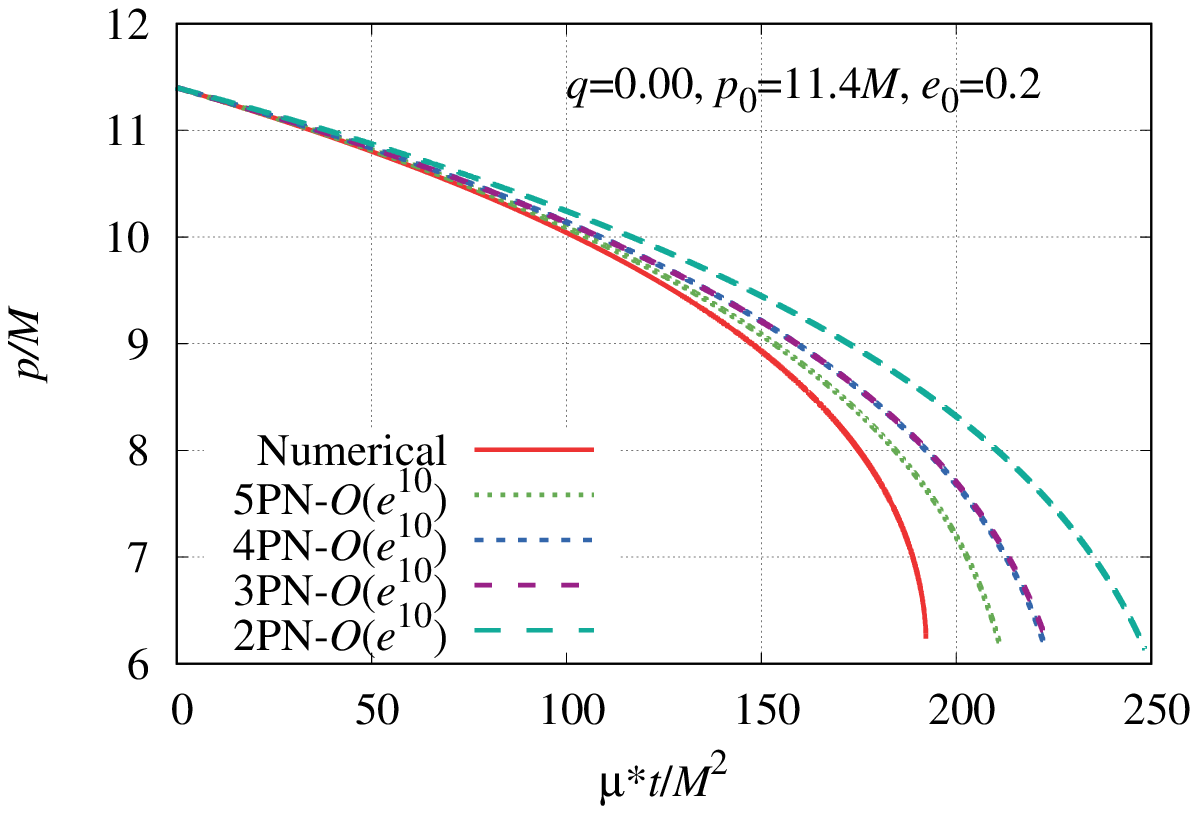}%
\includegraphics[width=59mm]{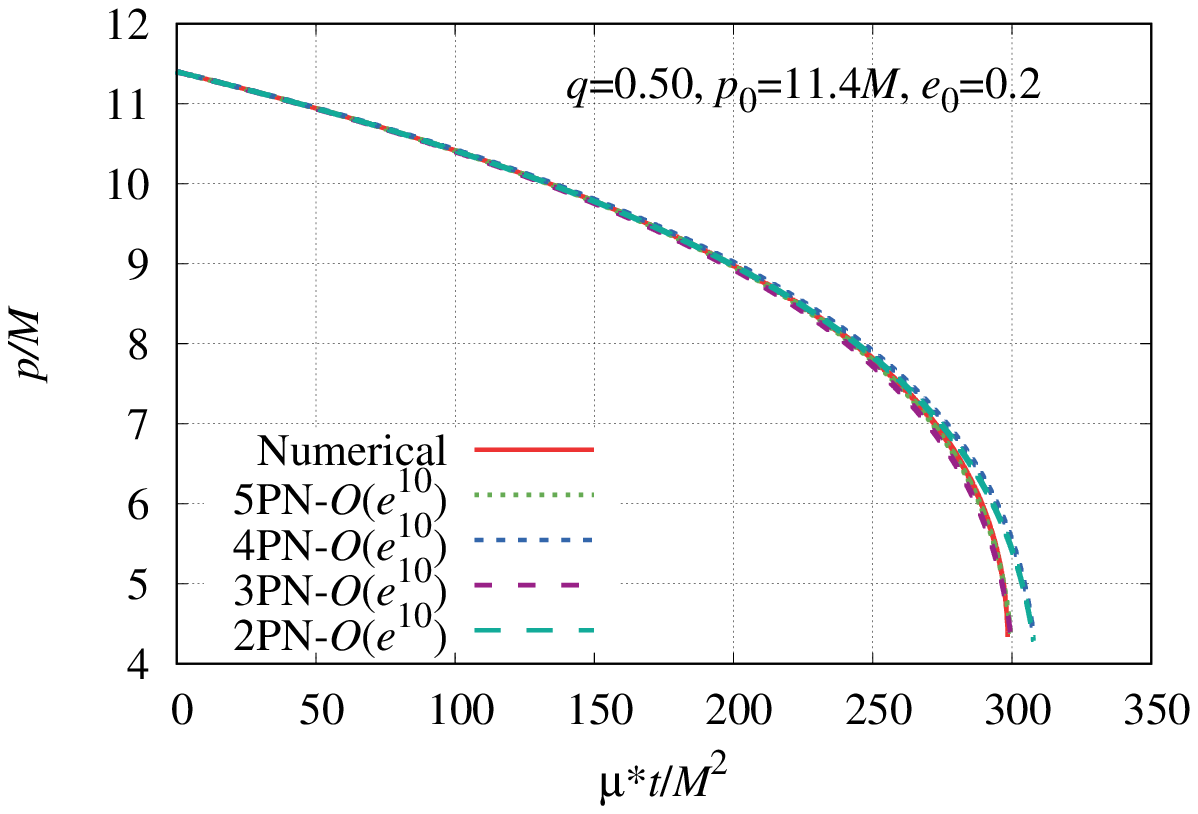}\\
\includegraphics[width=59mm]{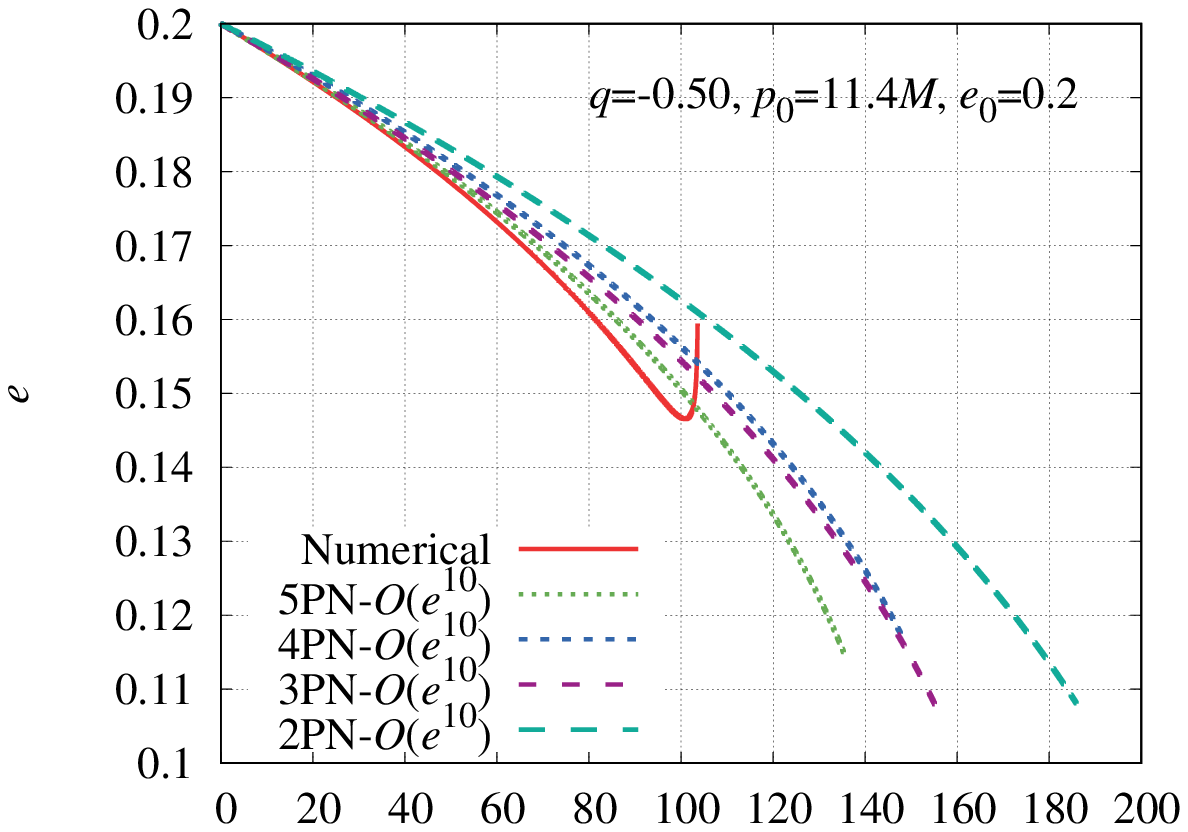} %
\includegraphics[width=59mm]{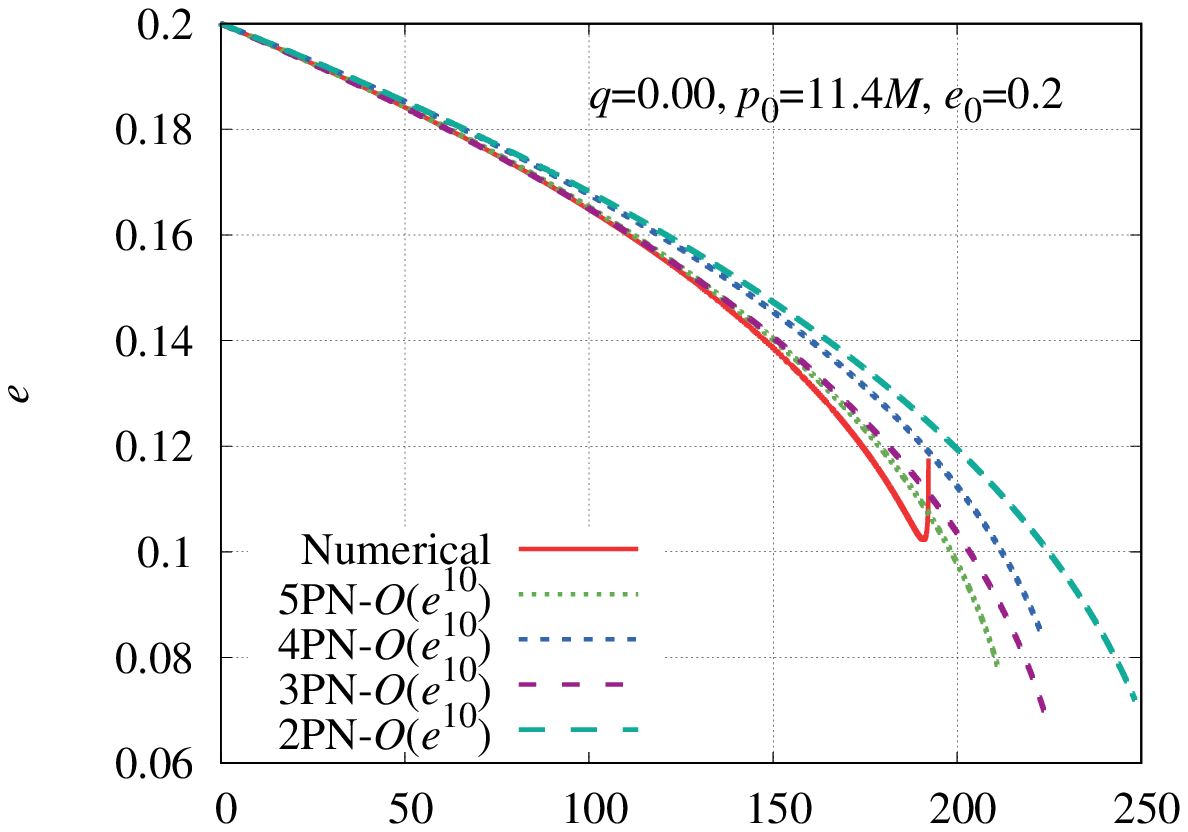}%
\includegraphics[width=59mm]{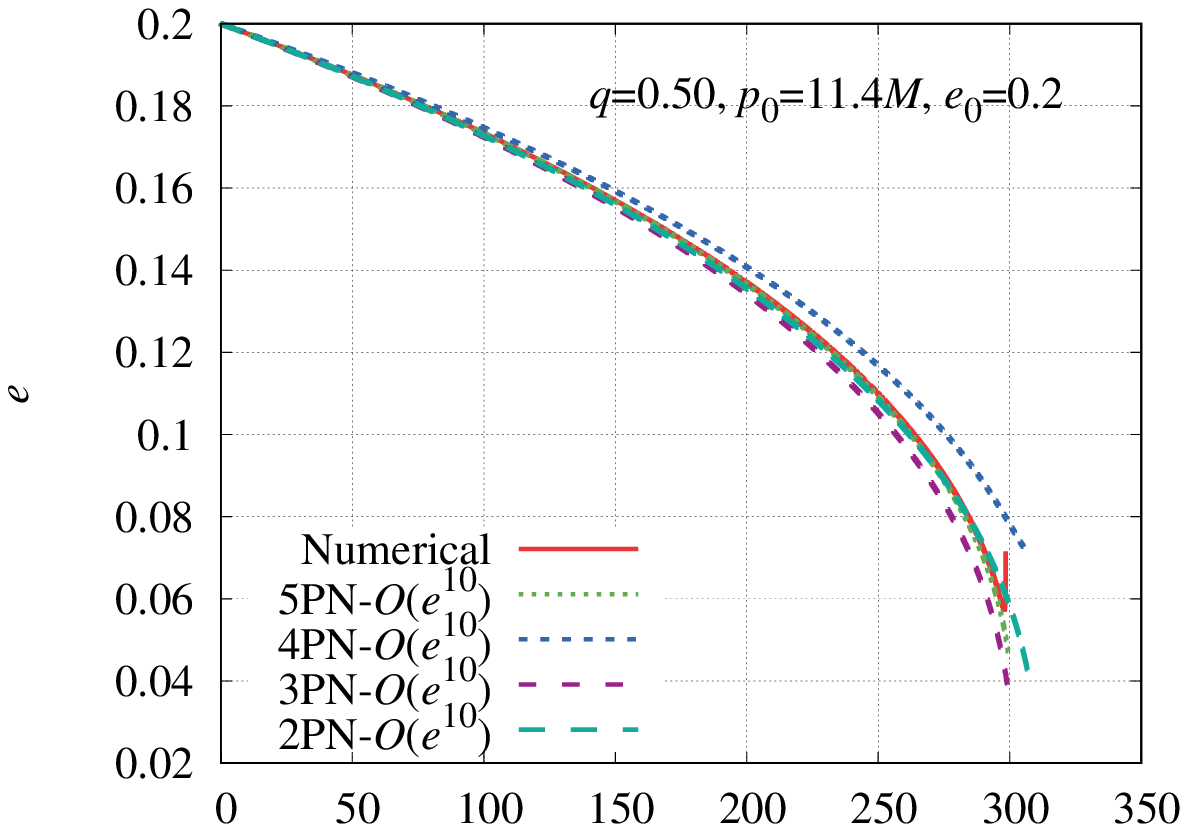}
\caption{Orbital evolution using numerical fluxes and PN fluxes for 
$q=-0.5$ (left), $0$ (middle), and $0.5$ (right) with $p_0=11.4M$ and $e_0=0.2$. Top 
panels show $e$ as a function of $p/M$, middle 
panels show $p/M$ as a function of $\mu t/M^2$, 
and bottom panels show $e$ as a function of $\mu t/M^2$. 
}
\label{fig:orbit_pn}
\end{figure*}

\begin{figure*}[htbp]
\centering 
\includegraphics[width=59mm]{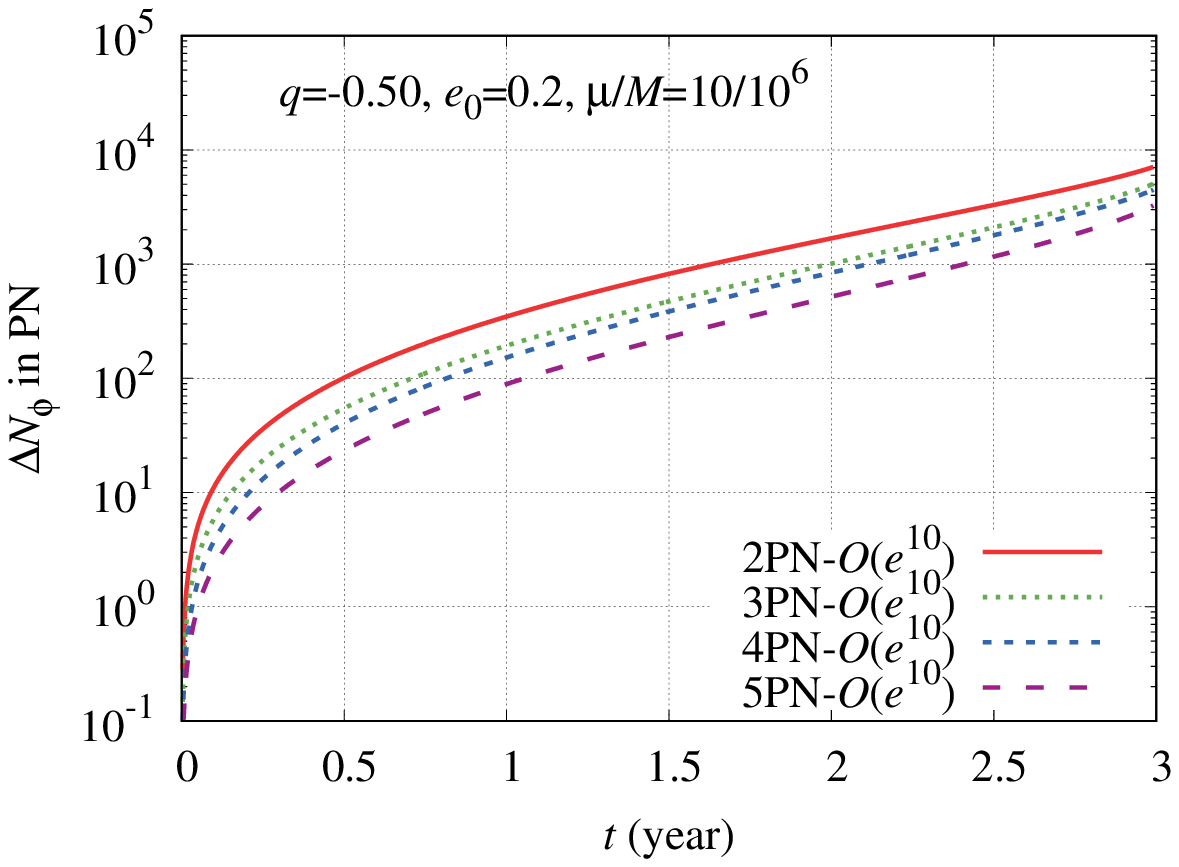}
\includegraphics[width=59mm]{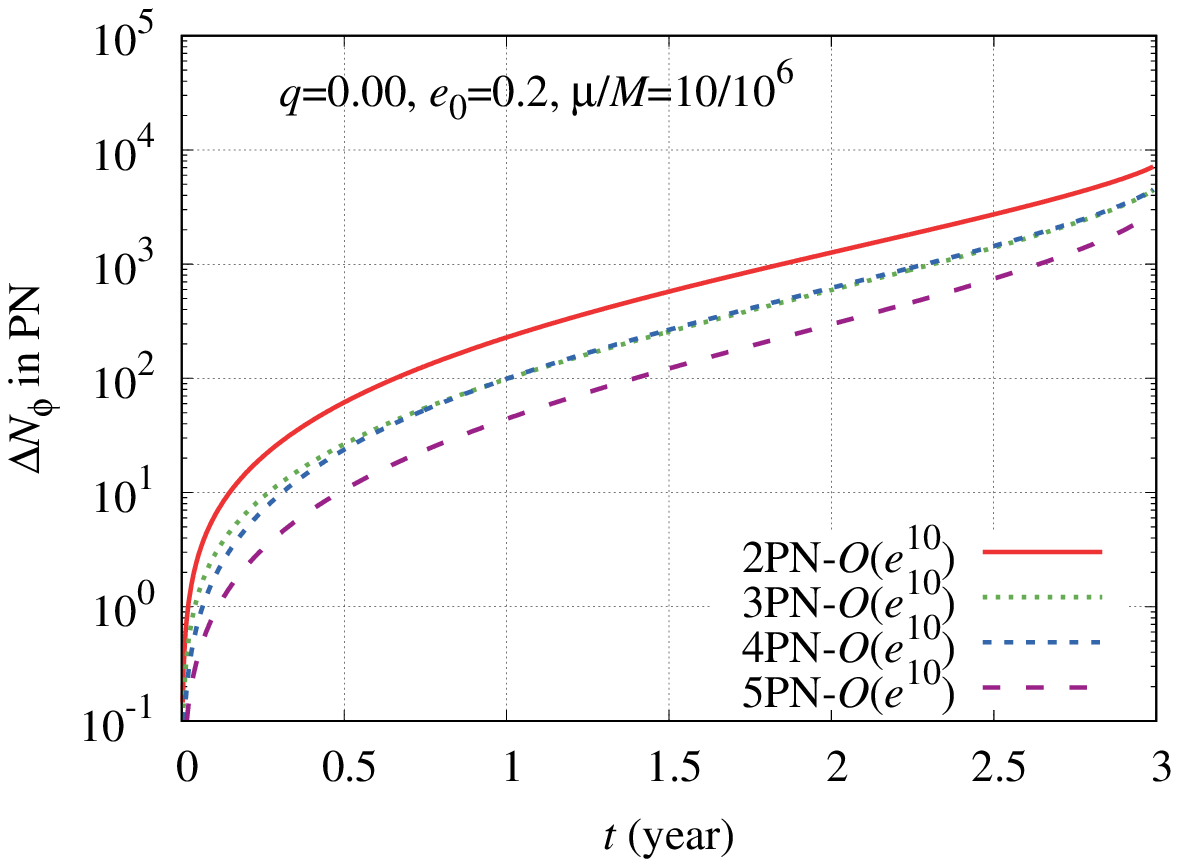}%
\includegraphics[width=59mm]{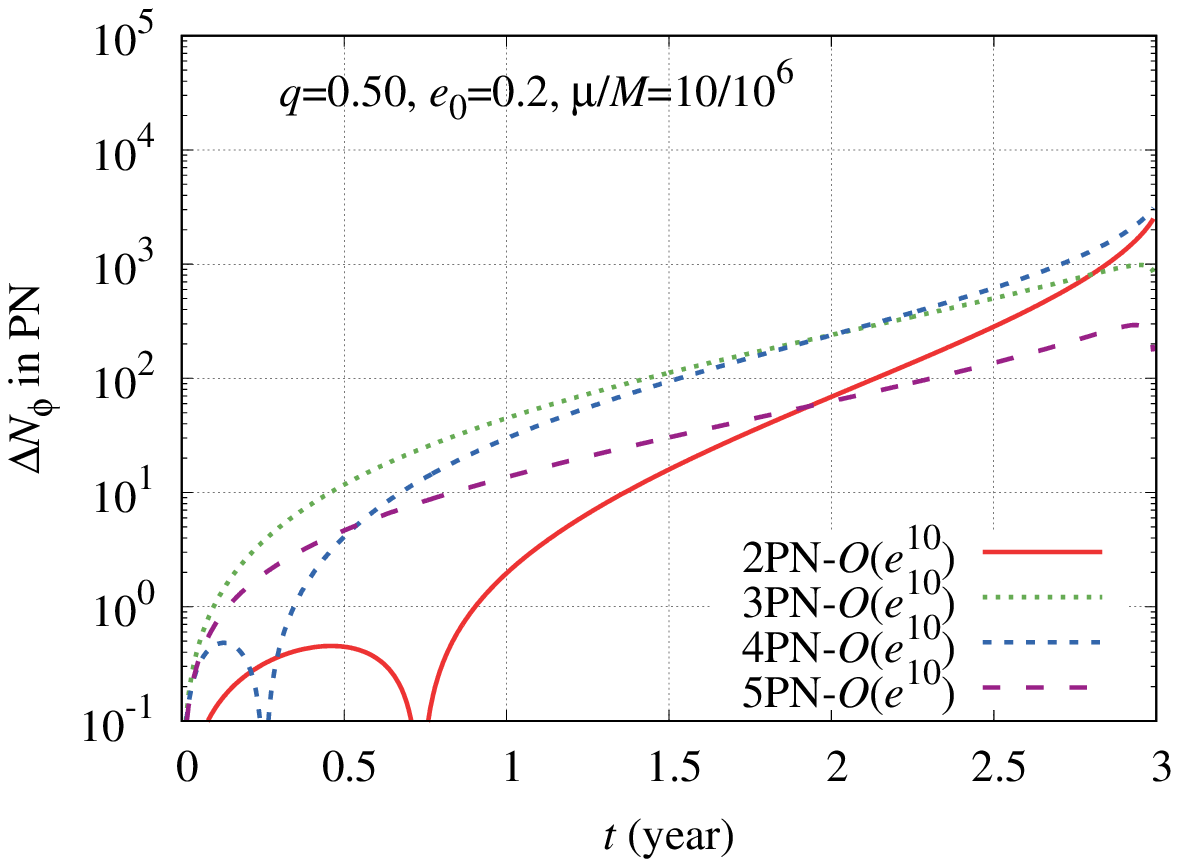}\\
\includegraphics[width=59mm]{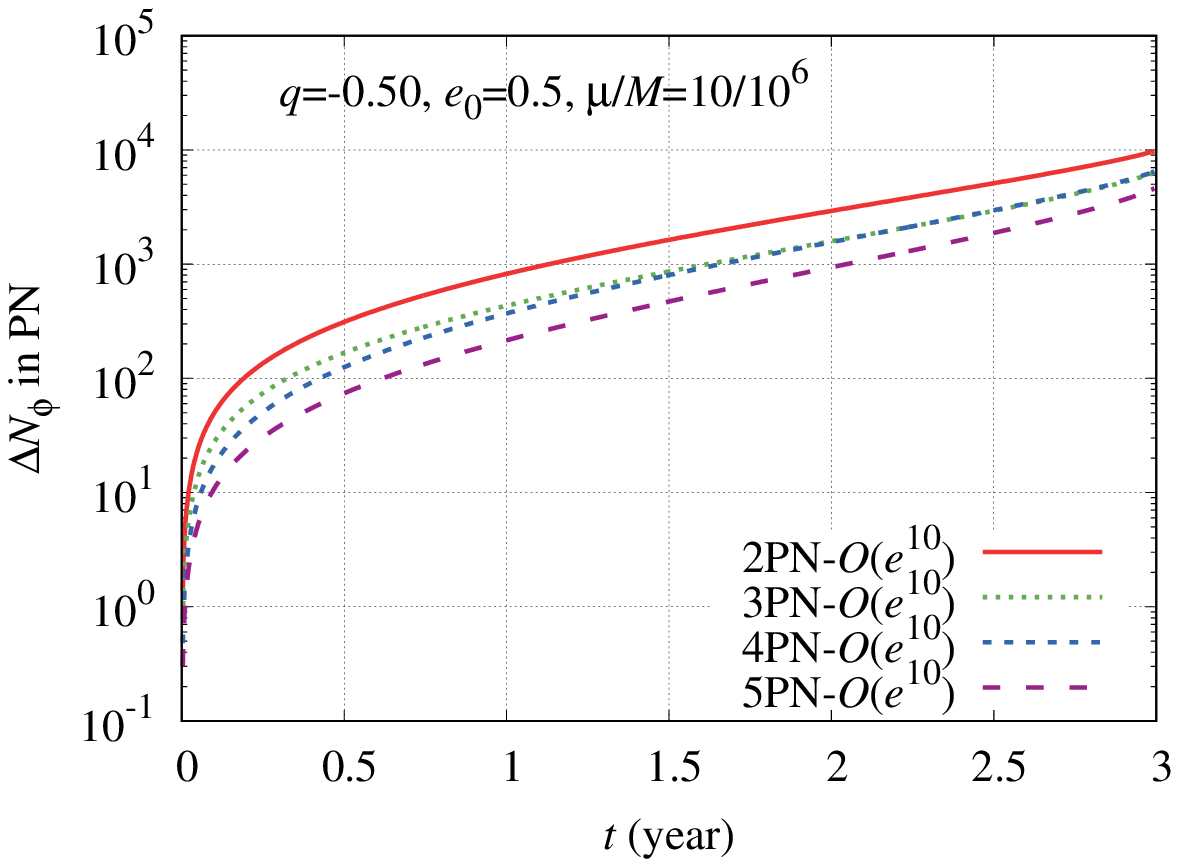}
\includegraphics[width=59mm]{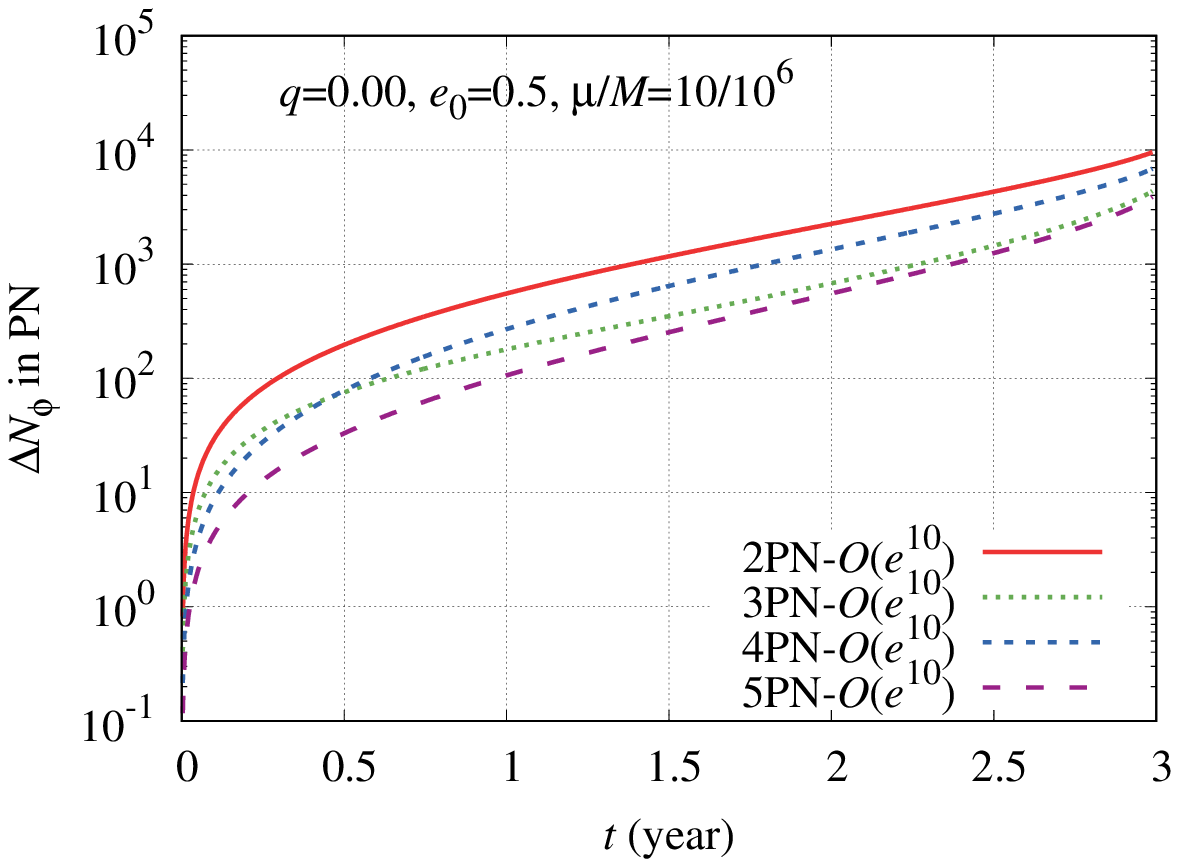}
\includegraphics[width=59mm]{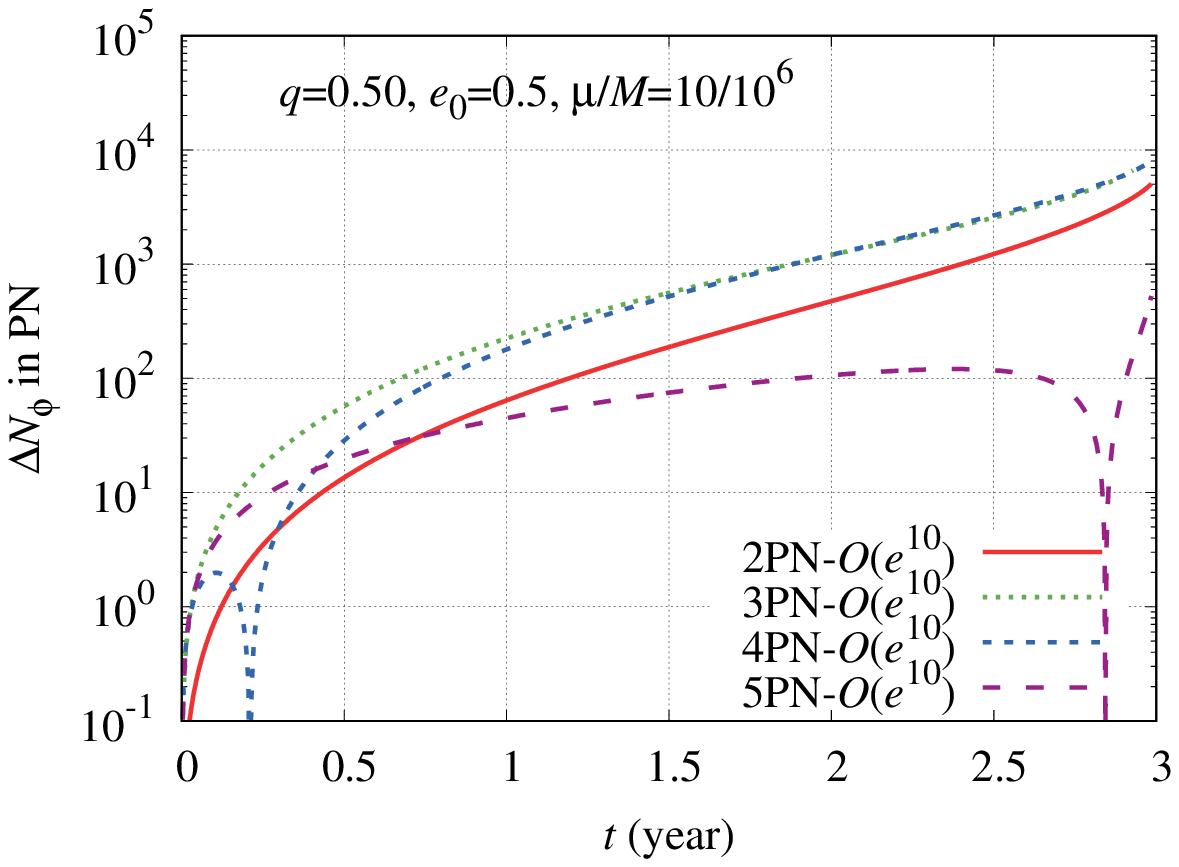}\\
\includegraphics[width=59mm]{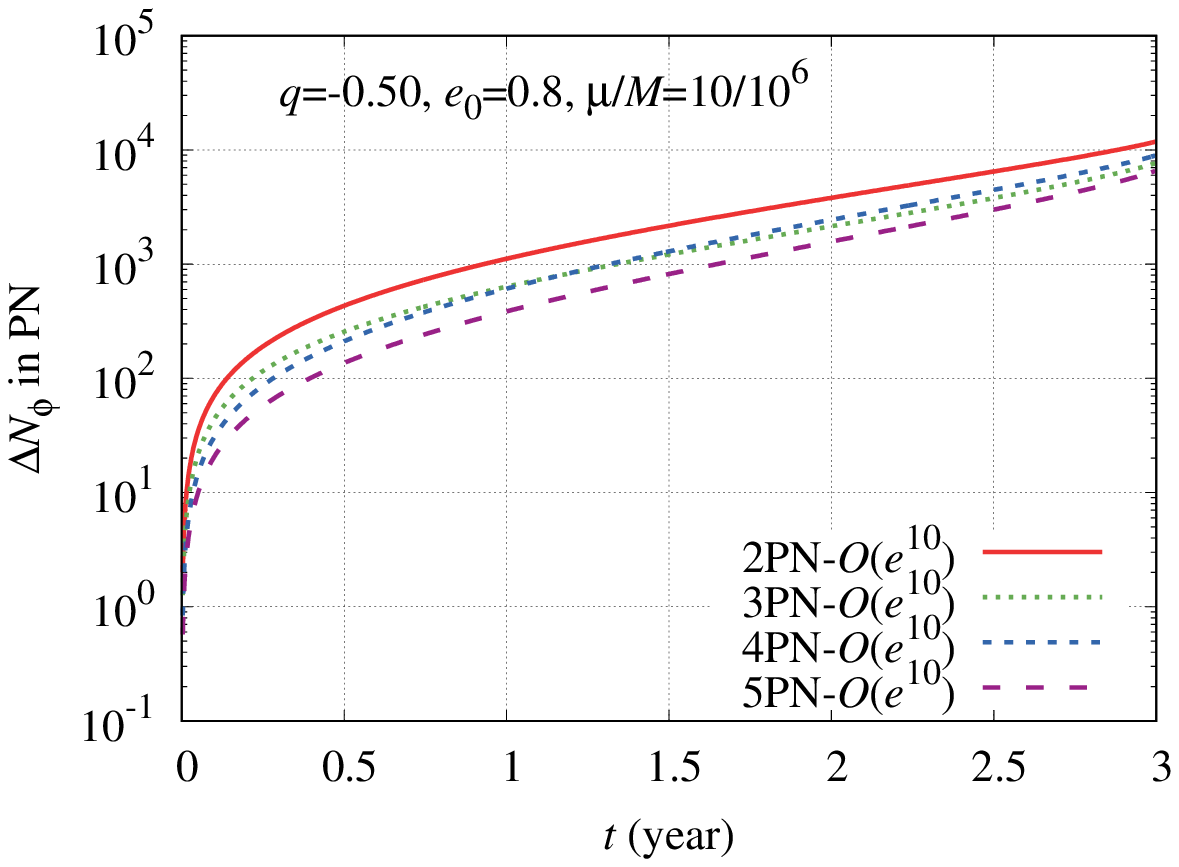}
\includegraphics[width=59mm]{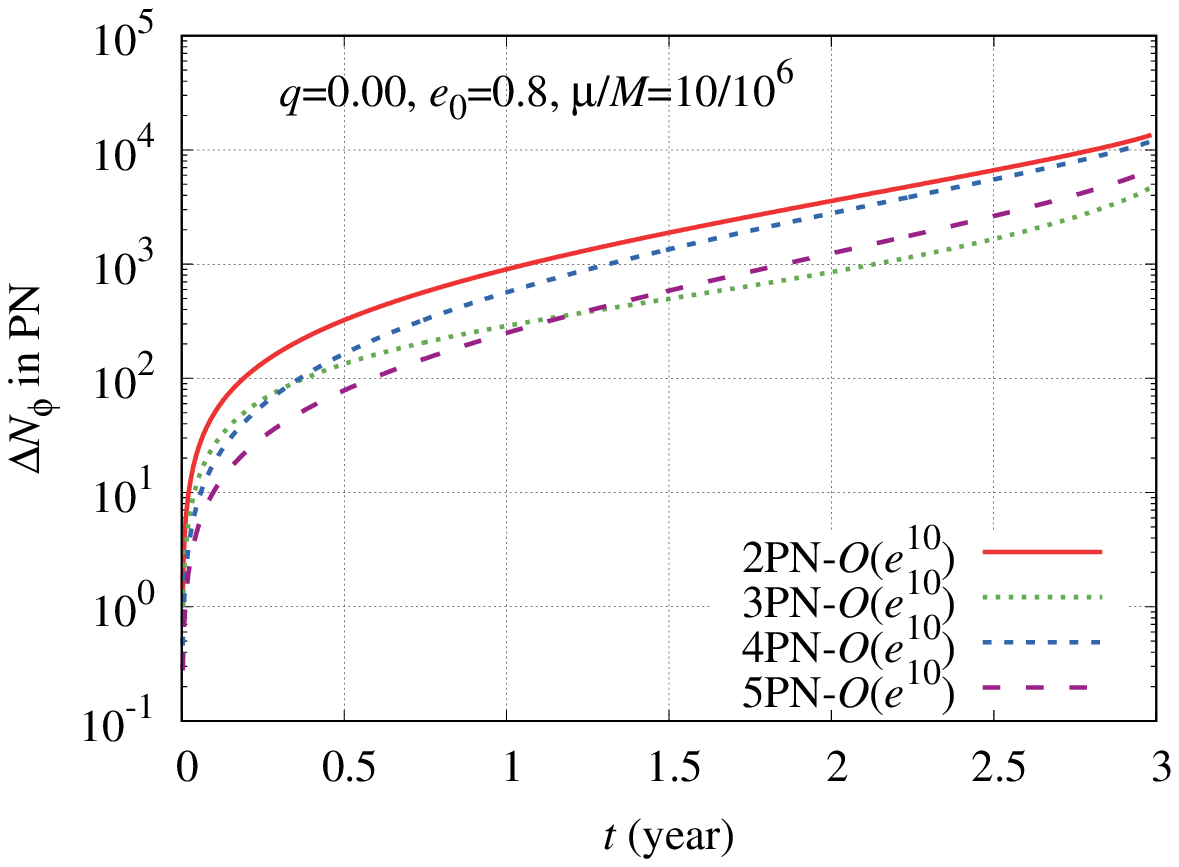}
\includegraphics[width=59mm]{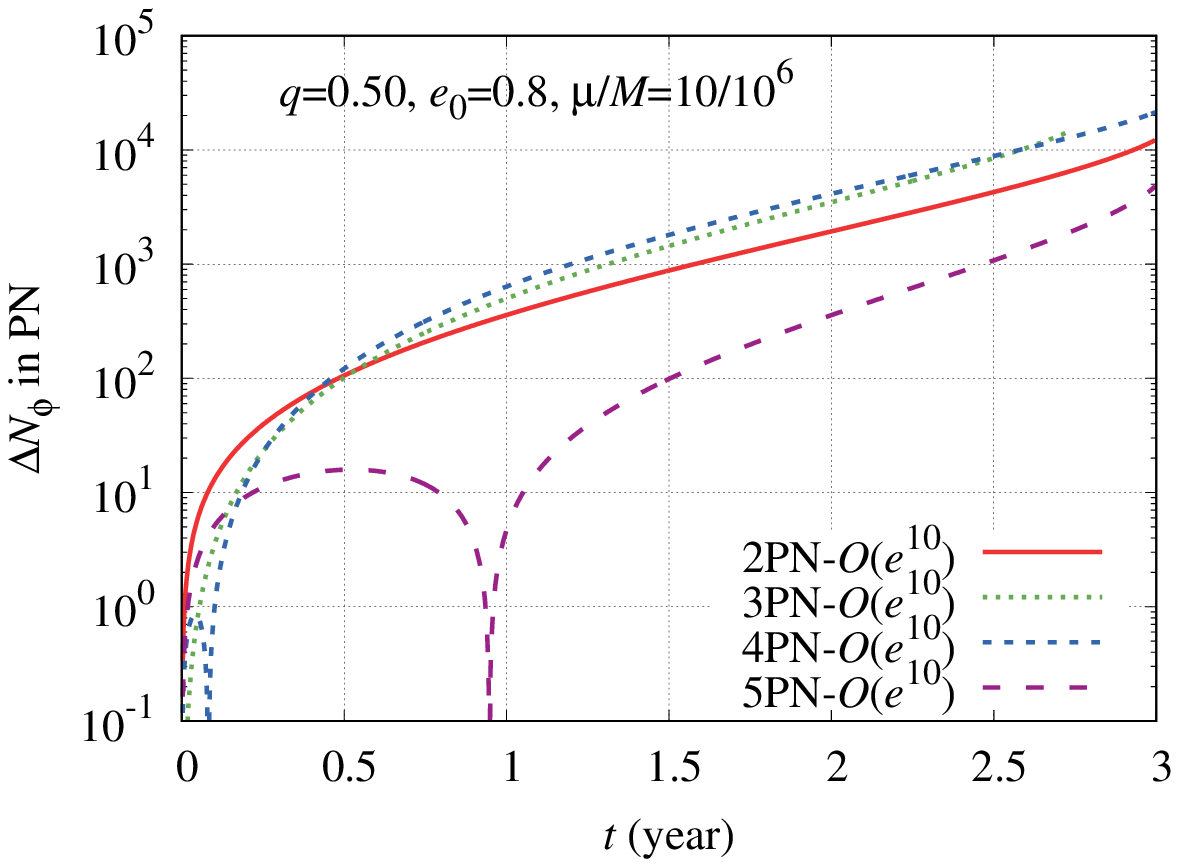}
\caption{Difference in orbital cycle for the last 3-year inspirals
  computed using numerical and PN fluxes for $q=-0.5$ (left), $0$
  (middle), and $0.5$ (right) with $e_0=0.2$ (top), $0.5$ (middle),
  and $0.8$ (bottom), and $(M,\mu)=(10^6,10)M_\odot$.  }
\label{fig:dephase_pn}
\end{figure*}

Before closing Sec.~\ref{sec:results}, we show the poor accuracy of inspiral orbits
determined in the PN approximation by comparing with our numerical
results.  We use a newly developed PN formula of $dI^i/dt$ that takes
into account the PN correction up to 5PN order and the tenth order in
eccentricity.  This new formula is the extension of the 4PN formula 
derived in Ref.~\cite{Sago:2015rpa}.

In Fig.~\ref{fig:orbit_pn}, we assess the accuracy of the 2PN, 3PN,
4PN, and 5PN formulas using inspiral orbits for $q=-0.5$, $0$, and
$0.5$ with $p_0=11.4M$ and $e_0=0.2$.  Disagreement between the PN and
numerical results is obviously non-negligible. Moreover, the
convergence of the PN expansion is quite slow, although with the
increase of the PN order the results gradually approach the numerical
results.  For the numerical inspirals, it takes $\sim 100M^2/\mu$ for
$q=-0.5$, $\sim 190M^2/\mu$ for $q=0$, and $\sim 300M^2/\mu$ for
$q=0.5$ until the plunge.  On the other hand, for the 5PN inspirals,
it takes $\sim 140M^2/\mu$ for $q=-0.5$, $\sim 210M^2/\mu$ for $q=0$,
and $\sim 310M^2/\mu$ for $q=0.5$ until the plunge.  The error in the
orbital eccentricity at the plunge between the numerical and 5PN
results is about $30\%$.  Thus, the 5PN formulas are not at all
accurate enough for gravitational-wave data analysis. 

It is interesting to note that the PN formulas work relatively well
for $q=0.5$ accidentally. In addition, the 4PN results often become
worse models than the 3PN results.  These facts illustrate that the PN
expansion has not only a poor-convergence property but also an
irregular convergence property~\cite{Sago:2016xsp,Fujita:2017wjq}.

We also note that the convergence of the eccentricity expansion 
in the PN formulas becomes slower if the eccentricity becomes higher~\cite{Sago:2015rpa}. 
For $q=0.9$ and $p=6M$, the relative error of the 5PN formula in $dE/dt$ 
determined by the comparison with our numerical results becomes $10^{-3}$ 
for $e=0.1$, $10^{-2}$ for $e=0.7$, and $10^{-1}$ for $e=0.9$, 
although the error is about an order of magnitude smaller than 
that of the 4PN and the tenth order in the eccentricity.
Thus, the accuracy of inspirals with higher eccentricity becomes 
worse than that with lower eccentricity. 

In Fig.~\ref{fig:dephase_pn}, we show the difference in orbital cycles
using the PN and numerical results.  This clearly illustrates the poor-convergence 
property and limitation in the PN formulas because the
difference in the orbital cycles from the numerical results is of the
order of $10^3$ even for the 5PN formula, in spite of the fact
that the required accuracy is within 0.1 orbital cycles. These results agree with
those in Refs.~\cite{Fujita:2012cm,Fujita:2014eta}, which study
quasicircular inspirals.
To conclude, the 5PN formulas, which are currently the best analytic
ones, cannot be used in the original forms for the data analysis of
gravitational waves. 

To derive an accurate analytic or semianalytic formula, the PN
formula combined with other methods such as resummation methods and
numerical fitting methods of higher PN order coefficients are
inevitable~\cite{Damour:1997ub,Buonanno:1998gg,Buonanno:2000ef,Damour:2000zb,Damour:2007xr,Damour:2007yf,Damour:2008gu,Buonanno:2009zt,Yunes:2009ef,Yunes:2010zj,Isoyama:2012bx}.
However, a significant improvement is required. In addition, new ideas
would be necessary for eccentric orbits because we need to perform a
resummation or fitting with respect not only to the PN expansion
parameter (e.g., $(M/p)^{1/2}$) but also to the eccentricity, $e$. In
particular, no idea for an efficient resummation with respect to the
eccentricity has been proposed.  We encourage the readers to perform a
careful analysis of our numerical data for developing a novel scheme
of a resummation/numerical fitting.  Our numerical data and 5PN formulas are published
in a web site~\cite{BHPC}.

\section{Summary}
\label{sec:summary}
We computed gravitational waves from a stellar-mass compact object
inspiraling around an SMBH.  The inspiral orbits were determined by
taking into account the adiabatic change of the constants of motion,
$dI^i/dt$, due to the emission of gravitational waves. In our
procedure, we first obtained $dI^i/dt$ for $\approx 2 \times 10^4$
data points in the parameter space of $(p,e)$ for each value of $q$.
Then, accurate interpolation was used to derive gravitational-wave
fluxes at arbitrary points within the region of the parameter space
computed in advance.  The relative error in the interpolated values of
$dI^i/dt$ is typically $\lesssim 10^{-6}$, which is smaller than the inverse of
the gravitational-wave phase for EMRIs during the last 3-year observation in LISA,
for most of the parameter space except for $r_{\textrm{min}}\lesssim
3M$ (see Sec.~\ref{sec:method}).

In Sec.~\ref{sec:results}, we derived the inspiral orbits and associated
gravitational waves. We then computed the spectrum of gravitational
waves and the SNR for several values of mass of a binary, the BH spin,
and the initial orbital eccentricity during the 3-year LISA observation
before final plunge.  We found that the SNR increases by a factor of
several as the BH spin and the mass of the compact object increase for
$M\agt 10^6M_\odot$.  The SNR as a function of the BH mass has a
maximum around $M=10^6M_\odot$ for fixed values of $q$ and $e_0$.  The
SNR as a function of $q$ is weakly dependent on $M$ around
$M=10^5M_\odot$ because only an early part of the inspirals can be
observed in the LISA frequency band for the larger BH spin.  
The SNR as a function of the initial orbital eccentricity 
for $M=10^6M_\odot$ is a monotonically increasing function, 
that increases by a factor of several for the change from $e_0=0.1$ to 0.8.
We also found that the SNR
for the $\ell=3$ ($\ell=4$) modes is about $40\%$ ($20\%$) of
that for the $\ell=2$ mode. This shows that taking account of the
higher multipole modes of gravitational waves is important for
increasing the detection rate in the LISA observation by a factor 
of 3--4.

In Sec.~\ref{sec:pn_err}, the limitation of the PN formulas is shown
by comparing the orbital cycles between the numerical and PN
inspirals.  The difference in the orbital cycles becomes larger than
$10^3$ even for the 5PN formula, which is much larger than the
required accuracy in the LISA data analysis, $\lesssim 1$\,rad in phase.  
This illustrates that we need much higher-order PN formulas or to develop a
special prescription such as resummation to improve the accuracy in
the PN formula.

In our present numerical computation, the numerical accuracy of the gravitational fluxes for 
compact orbits with $r_\textrm{min} \alt 3M$ is not high enough. Such compact 
orbits are possible for a high value of $q\agt 0.6$. As we showed in Sec.~\ref{sec:snr}, 
the SNR is higher for higher spin SMBHs with mass $M \approx 10^6$--$10^7M_\odot$, 
and hence, the detectability of EMRIs for the relatively high-mass SMBHs will be higher 
for the higher spin SMBH. This indicates that it is important to develop accurate gravitational-wave 
models for the high values of $q$. As we discussed in Sec.~\ref{sec:method},
the accuracy could 
be straightforwardly improved if we could perform the computation with higher numerical 
precision. A question is how high numerical precision is required for each value of $q$. 
This is one of our next issues to be clarified. 

In general, orbital inspirals of a compact object into an SMBH are not
only eccentric but also inclined from the equatorial plane of the
SMBH.  Thus, it is necessary to extend our approach to eccentric and
inclined inspirals.  The semilatus rectum at separatrix becomes
larger for larger orbital inclination angle with fixed orbital
eccentricity and BH spin.  This implies that orbital inclination
effectively reduces the effects of the BH spin and the frequency of
gravitational waves at separatrix.  We expect the power spectra of
gravitational waves and SNR for eccentric and inclined inspirals in
LISA observation would be smaller than those for equatorial inspirals
studied in this paper.  To check this quantitatively, we need to
compute gravitational waves for a large set of parameter space in the
BH spin, the semilatus rectum, the eccentricity, and the inclination
angle from the equatorial plane of the BH.  It would take about a
  year to derive gravitational waves for $\sim 10^6$ points in
  $(q,p,e,\theta_{\textrm{inc}})$ with $q\lesssim 0.9$ and $e\lesssim
  0.9$ using a $\sim 10$ Tflops machine if it takes 10 times longer to
  compute gravitational waves for a nonequatorial orbit than the one
  for an equatorial orbit (see Sec.~\ref{sec:method}).  However, it
is not clear how many data points are necessary to accurately derive
inspiral orbits for the generic case by interpolation methods.  
We are currently working on this issue, and the results will be 
published in future. 

\acknowledgments
We would like to thank the anonymous referee for useful comments and suggestions.
This work was in part supported by JSPS/Ministry of Education, Culture, Sports, Science 
and Technology (MEXT) KAKENHI Grants No.~JP16H02183, No.~JP18H04583, and No.~JP20H00158.


\end{document}